\documentclass[reqno]{amsbook}
\usepackage{amsmath}
\usepackage{calrsfs,dsfont,t1enc}
\usepackage[dvips]{graphicx}
\DeclareGraphicsExtensions{ps,eps}
\input cyracc.def
\input FEYNMAN
\font\tencyrbib=wncyr10 at 8pt
\def\cyrbib{\tencyrbib\cyracc}

\font\tencyritbib=wncyi10 at 8pt
\def\cyritbib{\tencyritbib\cyracc}

\font\goth=ygoth at 12pt
\def\got{\goth}
\newcommand{\GC}{{\got C}}
\newcommand{\GG}{{\got G}}
\newcommand{\GM}{{\got M}}
\newcommand{\Gg}{{\got g}}

\newcommand{\GF}{{\got F}}

\font\tenfrak=eufm10

\font\sevenfrak=eufm7
\font\fivefrak=eufm5

\newfam\frakfam
\textfont\frakfam=\tenfrak
\scriptfont\frakfam=\sevenfrak
\scriptscriptfont\frakfam=\fivefrak

\newcommand{\CH}{\ensuremath{\mathcal{H}}}
\newcommand{\CL}{\ensuremath{\mathcal{L}}}
\newcommand{\CB}{\ensuremath{\mathcal{B}}}
\newcommand{\CM}{\ensuremath{\mathcal{M}}}

\newcommand{\CE}{\ensuremath{\mathcal{E}}}
\newcommand{\CA}{\ensuremath{\mathcal{A}}}

\newcommand{\BA}{\ensuremath{\mathbf{A}}}
\newcommand{\BBeta}{\ensuremath{\boldsymbol{\eta}}}
\newcommand{\BV}{\ensuremath{\mathbf{V}}}
\newcommand{\BW}{\ensuremath{\mathbf{W}}}
\newcommand{\BD}{\ensuremath{\mathbf{D}}}
\newcommand{\BP}{\ensuremath{\mathbf{P}}}
\newcommand{\BU}{\ensuremath{\mathbf{U}}}
\newcommand{\BR}{\ensuremath{\mathbf{R}}}
\newcommand{\BJ}{\ensuremath{\mathbf{J}}}
\newcommand{\Balpha}{\ensuremath{\boldsymbol{\alpha}}}
\newcommand{\Btau}{\ensuremath{\boldsymbol{\tau}}}
\newcommand{\Bpi}{\ensuremath{\boldsymbol{\pi}}}
\newcommand{\Bxi}{\ensuremath{\boldsymbol{\xi}}}
\newcommand{\Br}{\ensuremath{\mathbf{r}}}
\newcommand{\Bx}{\ensuremath{\mathbf{x}}}
\newcommand{\Bq}{\ensuremath{\mathbf{q}}}
\newcommand{\Bp}{\ensuremath{\mathbf{p}}}
\newcommand{\Bv}{\ensuremath{\mathbf{v}}}
\newcommand{\Ba}{\ensuremath{\boldsymbol{a}}}
\newcommand{\BPhi}{\ensuremath{\boldsymbol{\Phi}}}
\newcommand{\Bell}{\ensuremath{\boldsymbol{\ell}}}
\newcommand{\Bsigma}{\ensuremath{\boldsymbol{\sigma}}}

\newcommand{\Z}{\ensuremath{\mathds{Z}}}
\newcommand{\NN}{\ensuremath{\mathds{N}}}
\newcommand{\R}{\ensuremath{\mathds{R}}}
\newcommand{\SU}{SU(2)}
\newcommand{\su}{{su(2)}}
\newcommand{\e}{\mathrm{e}}
\newcommand{\rt}{\tilde{r}}
\newcommand{\di}{\mathrm{d}}
\DeclareMathOperator{\Tr}{Tr}
\DeclareMathOperator{\mycup}{\cup}
\DeclareMathOperator{\diag}{diag}

\renewcommand{\thechapter}{\Roman{chapter}}
\renewcommand{\theequation}{\thechapter.\thesection.\arabic{equation}}

\usepackage{amsmath}
\begin{document}
%\baselineskip=1.5\normalbaselineskip
\frontmatter
\thispagestyle{empty}
\begin{center}
{\Large INSTITUTE OF  THEORETICAL PHYSICS AND ASTRONOMY}
\vspace{4cm}

{\bf \large Art\=uras Acus}
\vspace{0.99cm}

{\LARGE \bf BARYONS AS SOLITONS IN QUANTUM SU(2) SKYRME MODEL
\vphantom{\vbox to 0.6cm{}}}
\vspace{5cm}

{\large Doctoral Dissertation}
\vspace{0.3cm}

Physical Sciences, Physics (02P)
\vspace{6cm}

{\Large Vilnius, 1998}
\end{center}
\newpage
\thispagestyle{empty}
\begin{flushleft}
{\large
\noindent The dissertation has been 
accomplished in 1993-1998 at the Institute of Theoretical Physics and 
Astronomy, Vilnius, Lithuania.\\
\vspace{0.5cm}

\noindent  Institute of Theoretical Physics and Astronomy shares the joint
doctoral degree-granting authority with Vilnius University, 
following the Resolution No 457 of the Government of Republic of 
Lithuania, April 14, 1998.\\
\vspace{1.5cm}

\noindent {\bf The Doctorate Committee}
\vspace{0.5cm}

\noindent Chairman and scientific adviser:
\begin{itemize}
\item Egidijus NORVAI\v SAS, Dr. (Institute of Theoretical Physics and 
Astronomy, Physical Sciences, Physics, 02P);
\end{itemize}
\vspace{0.3cm}

\noindent Members:
\begin{itemize}
\item Sigitas ALI\v SAUSKAS,~Dr.~hab. (Institute of 
Theoretical Physics and Astronomy, Physical Sciences, Physics, 02P);
\item Adolfas BOLOTINAS,~Dr.~hab., Prof. (Vilnius University,
Physical Sciences, Physics, 02P);
\item Kazimieras PYRAGAS,~Dr.~hab., Prof.  (Vilnius Pedagogical  
University, Physical Sciences, Physics, 02P);
\item Zenonas RUDZIKAS,~Dr.~hab., Prof. (Institute of 
Theoretical Physics and Astronomy, Physical Sciences, Physics, 02P);
\end{itemize}
}
\end{flushleft}
\newpage
\thispagestyle{empty}
\begin{center}
{\Large TEORIN\.ES FIZIKOS IR ASTRONOMIJOS INSTITUTAS}
\vspace{4cm}

{\bf \large Art\=uras Acus}
\vspace{0.99cm}

{\LARGE \bf BARIONAI KAIP KVANTINIO SU(2) SKYRME'O MODELIO SOLITONAI
\vphantom{\vbox to 0.6cm{}}}
\vspace{5cm}

{\large Daktaro disertacija}
\vspace{0.3cm}

Fiziniai mokslai, fizika (02P)
\vspace{6cm}

{\Large Vilnius, 1998}
\end{center}
\newpage
\thispagestyle{empty}
\begin{flushleft}
{\large
\noindent Darbas atliktas 1993-1998 metais Teorin\.es fizikos ir 
astronomijos institute.\\
\vspace{0.5cm}

\noindent Doktorant\=uros ir daktaro mokslo laipsnio teikimo
teis\.e Teorin\.es fizikos ir astronomijos 
institutui suteikta kartu su Vilniaus 
universitetu 1998~04~14  Lietuvos Respublikos Vyriausyb\.es nutarimu 
Nr.~457.
\vspace{2cm}

\noindent {\bf Doktorant\=uros komitetas}
\vspace{0.5cm}

\noindent Pirmininkas ir darbo vadovas
\begin{itemize}
\item Dr.~Egidijus NORVAI\v SAS (Teorin\.es fizikos ir astronomijos 
institutas, fiziniai mokslai, fizika, 02P);
\end{itemize}
\vspace{0.3cm}

\noindent Nariai:
\begin{itemize}
\item Habil.~dr. Sigitas ALI\v SAUSKAS (Teorin\.es fizikos ir astronomijos 
institutas, fiziniai mokslai, fizika, 02P);
\item Prof.~habil.~dr. Adolfas BOLOTINAS (Vilniaus universitetas, fiziniai 
mokslai, fizika, 02P);
\item Prof.~habil.~dr. Kazimieras PYRAGAS (Vilniaus 
pedagoginis universitetas, fiziniai mokslai, fizika, 02P);
\item Prof.~habil.~dr. Zenonas RUDZIKAS (Teorin\.es fizikos ir astronomijos 
institutas, fiziniai mokslai, fizika, 02P);
\end{itemize}
}
\end{flushleft}
\tableofcontents
\listoftables
\listoffigures

\chapter*{Notations and conventions}
\begin{itemize}
\item Bold letters indicate multiple quantity  
structure, which may vary from case to case. For example,
\Balpha\ denotes triple of {Euler} angles $(\alpha^1,\alpha^2,\alpha^3)$,
\Br\ denotes spatial vector $\vec r$ and $\BD^j$ denotes {Wigner} matrix 
in representation $j$.
\item Calligraphic letters $\CA,\CH,\CL,\CM$ etc. denote densities.
\item Appearance of carets (\,\,$\hat{\!\!\!\CB},\hat V^a_t,\hat \BJ$, etc.)
indicate an operator or its component. Note that we do not follow this 
convention for coordinate $\Bq$ (also sometimes for operators which are 
functions of $\Bq$ only) and momentum $\Bp$ operators in order to keep 
notations simpler.
\item Special attention should be paid to $\hat\BJ,\hat\BJ^\prime$ and
$\BJ,\BJ^\prime$ operators. Operators with hats $\hat\BJ,\hat\BJ^\prime$ 
are dynamical operators (introduced in place of momentum operator $\Bp$
and depend on $\Bq,\dot\Bq$), 
whereas operators without hats $\BJ,\BJ^\prime$ are abstract \SU\ group 
generators. All group generators enter under $\Tr$ symbol, therefore, they 
yield representation dependence. Dynamical operators acting on appropriate 
states  provide spin $\ell$ dependence.
\item Dot over the symbol denotes full time derivative.  
\item The metric tensor $g^{\mu\nu}$ is 
$g^{00}=1,g^{0,i}=0,g^{i,j}=-\delta^{i,j}$ for spatial 
indices $i,j=1,2,3$. The four derivative 
$\partial^\mu=\frac{\partial}{\partial x_\mu}$ has 
components $(\partial /\partial t,-\nabla)$. The sign of 
totally anti-symmetric 
tensors ({Levi-Cevita} symbols) 
$\epsilon_{ijk},\quad \epsilon^{\mu\nu\sigma\gamma}$ are fixed by
$\epsilon_{123}=-\epsilon^{123}=1,\quad 
\epsilon_{0123}=-\epsilon^{0123}=1$, respectively.
\item Isovector of Pauli 
isospin matrices $\Btau$ in Cartesian coordinates have a 
form 
$\Btau_{1}=\bigl(\begin{smallmatrix}0&1\\1&0\end{smallmatrix}\bigl),\quad 
\Btau_{2}=\bigl(\begin{smallmatrix}0&-i\\i&0\end{smallmatrix}\bigl),\quad 
\Btau_{3}=\bigl(\begin{smallmatrix}1&0\\0&-1\end{smallmatrix}\bigl)$.
\item 
$*$ is a complex conjugation mark.\item \SU, \su\ denotes the group and the 
group algebra, respectively.
\item The indices $\alpha, \beta, \gamma\dotsc$ 
at the head of the Greek alphabet usually represent Euler angles. The 
middle Greek letters $\mu,\nu,\lambda,\rho,\eta,\dotsc$ represent axes in 
{Minkowski} space, whereas $i,j,k,l,o,p,r,s$ usually indicate spatial 
dimensions. Indices $a,b,c,d,e,f,g,h$ and $m,n$ are reserved for SU(2) 
values. Index $t$ denotes time component.
\item The curly bracket $\bigl\{\ ,\ \bigr\}$ and the square bracket
$\bigl[\ ,\ \bigr]$ denotes the anti-commutator and the commutator, 
respectively.
\end{itemize}
We assume summation convention under repeated (dummy) indices. Dummy 
indices sometimes can involve phase factors. In this case three indices are
required for summation convention. For example, we assume summation in
$(-1)^a A_a B^{(-a)}$, but not in $(-1)^a A_a$.

Natural units $c=\hbar=1$ are used in the work. 
Mass/energy, momentum then are usually measured in MeV (or fm${}^{-1}$; 
$1$~MeV$=\frac{1}{197.3}\mathrm{fm}^{-1}$), length and time in 
MeV${}^{-1}$. Unitary field $\BU$ and model parameter $e$ are 
dimensionless, right $\BR_\mu$ (left $\mathbf{L}_\mu$) {Maurer-Cartan} 
forms, Lagrange function, pion and sigma fields, pion decay constant $f_\pi$ 
have dimensions of MeV. Lagrange function density  is proportional to 
MeV${}^4$. Vector, axial-vector and baryon current densities are 
proportional to MeV${}^3$, whereas Lagrange/Hamilton function 
density\footnote{If we use densities integrated over spherical angles 
$\varphi$ and $\vartheta$, then it is natural to multiply them by $r^2$ 
--- the rest part of the Jacobian. This introduces additional dimension  
factor MeV${}^{-2}$.} to MeV${}^4$.

\subsection*{Note} Notations in the first chapter differ from notations in 
the rest chapters, whereas notations are the same in second and third 
chapters.

\chapter*{List of publications}
\begin{enumerate}
\item  A.~Acus, E.~Norvai\v sas, and D.~O.~Riska, 
"Stability and Representation Dependence of the Quantum Skyrmion",
Phys. Rev. C, V.57, Nr. 5, p.2597-2604 (1998)

\item A.~Acus, E.~Norvai\v sas, and D.~O.~Riska,
"The Quantum Skyrmion in Representation of General Dimension",
Nucl. Phys. A, V.614, p.361-372 (1997)

\item A.~Acus and E.~Norvai\v sas,
"Stability of SU(2) Quantum Skyrmion and Static Properties of Nucleons",
Lithuanian Journal of Physics, 1997, V.37, Nr. 5, p.446-448

\item  E.~Norvai\v sas and A.~Acus, "Canonical quantization of SU(2)
Skyrme model", Physical Applications and Mathematical Aspects 
of Geometry, Groups, and Algebras, Editors: H.~D.~Doebner, W.~Scherer, 
P.~Nattermann, World Scientific, Singapure, Vol.1, p.456-460, (1997)

\item A.~Acus and E.~Norvai\v sas,
"New  quantum corrections in Skyrme model for baryons,"
Proceedings of the International Workshop on Quantum Systems: new trends and 
methods, Editors: Y.~S.~Kim, L.~D.~Tomil'chik, I.~D.~Feranchuk, A.~Z.~Gazizov
World Scientific, Singapure,  p253-258, (1997)

\item A.~Acus, "Barion\k u  SU(2) Skyrme'o modelio tyrimas",
Report to Lithuanian State Science and Studies Foundation,
supervisor: dr E.~Norvai\v sas,   Nr.97-103/2F
\end{enumerate}
\vspace{2cm}

Results of the investigation have been reported in the following
conferences:

\begin{itemize}

\item "BARYONS `98", $\mathrm{8}^{\text{th}}$ International Conference on the
Structure of Baryons, Bonn, September 22--26, 1998.

\item 32${}^{\text{th}}$ Lithuanian National Conference of Physics, 
Vilnius, October 8--10, 1997.

\item "XVII${}^{\text{th}}$ UK Institute for Theoretical High Energy 
Physicists", Durham, August~26--September~13, 1996.

\item "XXI International Colloquium on Group Theoretical Methods in 
Physics", Goslar, July~15--20, 1996.

\item "Origin of masses", International workshop, Tartu, June~19-22, 1996.

\item "QS-96 Quantum Systems: New Trends and Methods", International 
workshop, Minsk, June~3-7, 1996.

\item 31${}^{\text{th}}$ Lithuanian National Conference of Physics, 
Vilnius, February 5--7, 1996.

\item "International Europhysics Conference on High Energy Physics",
Brussels, July~27--August~2, 1995.

\item "Nordic School in Particle Physics Phenomenology", Solvalla, 
June~11--17, 1994.
\end{itemize}

\chapter*{Preface}
This thesis is a compendium of our work on extension 
of basic Skyrme model to arbitrary representations of \SU\ group, hoping 
that higher representations would be helpful for more adequate 
description of static baryon properties~\cite{acus97,acus96a,acus97a} .
\newline
{\bf General ideas and historical remarks.}
The idea that the ordinary proton and neutron might be 
solitons\footnote{Soliton history begins in 1834, from {\it D.S.~Rassel's}
(1808--1882) "great solitary wave". There have been, however, no more than 
twenty scientific works during the period 1845--1965, directly related to 
solitons~\cite{filipovbook}.} in nonlinear model has a long history. The 
first suggestion was made by {\it T.H.R.~Skyrme} about 40 years 
ago~\cite{skyrme61a}. The essential feature of the theory is the 
representation of the fundamental field quantities in terms of angular 
variables rather than linear ones. Realistic three-dimensional model is 
possible only when there are also three angular variables. The condition is 
satisfied by the pion fields of nature. The periodicity of angular variables 
introduces a new constant of motion, which measures the number of times that 
space (three dimensions) is mapped by the fields onto the elementary volume 
of angular space and which can be interpreted as a baryon number. The origin 
of the new constant of motion is related to topological features of the 
Skyrme model. In contrast, conservation of energy and momentum follows from 
space-time symmetry, as usually. 

The mathematical construction outlined above is to ensure possibility of 
particle-like states "of a kind that cannot be reached by perturbation 
theory and which cannot necessarily be discounted by general 
arguments"~\cite{skyrme62}.
To provide readers a link between fundamental theory of strong interactions 
(QCD) and Skyrme model we need to consider briefly the chiral symmetry 
concept and, therefore, the idea of isospin.

The concept of isospin was introduced in 1933 by {\it W.~Heisenberg}, who 
considered proton and neutron as different projections of single 
state\footnote{{\it W.~Heisenberg} even suggested to explain interaction 
between proton and neutron by particle exchange~\cite{rekalob}. The 
existence of pion, however, was predicted by {Yukawa} theory in 1934. 
The particle was discovered by {\it G.~Lattes, H.~Muirhead, G.~Occhialini} 
and {\it S.F.~Powell} in 1947.}. From contemporary 
point of view {\it W.~Heisenberg} actually  assumed \SU\ (flavour) symmetry 
of nuclear interactions. In 1962 {\it M.~Gell-Mann} succeeded much more in 
suggesting very predictive SU(3) (flavour) symmetry of strong interactions 
(The Eightfold Way) and the concept of isospin was extended to all baryons. 
The SU(3) symmetry had enormous influence in becoming of QCD: it was 
realized that each basis element, i.e. a product of quark and anti-quark 
functions, can be identified with some hadron state. The entire basis, 
therefore, is interpreted as multiplet of hadrons, belonging to irreducible 
representation of (flavour) SU(3) group.

A revival of interest in the Skyrme model~\cite{skyrme61a,skyrme62}, 
begins from the work \cite{adkins83} of
{\it G.S.~Adkins} et al., who demonstrated  that 
this model could fit observed properties of the baryons to an accuracy of 
about 30\%. This rebirth of attention was stimulated by the belief that 
some such model is the long-wavelength limit of QCD, as reviewed, e.g., in 
Refs.~\cite{witten84,aitchison87}. The interplay between various 
phenomenological models and QCD is still open problem~\cite{dorey95}.
\newline
{\bf Extensions and modifications of basic Skyrme model.}
In the intervening period, there has been a large number of works 
extending the range of applications, modifying and extending the model,    
and improving the way in which consequences are drawn from it. This work 
also serves as an extension of basic Skyrme model to arbitrary \SU\ 
representation. Among the further applications, the most prominent have 
been to pion-nucleon scattering~\cite{mattisphd,mattis84} and the 
two-nucleon problem~\cite{manton95,wambach97}, and, very recently, to 
multi-soliton "chemistry"~\cite{sutcliffe97c,carson91,eisenberg93}. The 
basic Skyrme model has been extended in various directions. Excluding the 
mention of models that contain the quarks explicitly, one encounters in 
the literature models with higher-order terms involving the same fields 
\cite{marleau90,marleau92} or even higher unitary groups 
\cite{walliser92}, models in curved space~\cite{manton86,sutcliffe97b} and 
models in which vector mesons have been added 
\cite{mattis86,mattis88,kaiser93}, as well as extensions to include strange 
\cite{pari91} and even charmed mesons~\cite{rho90}. Related but more 
fundamental extension of the basic Skyrme model is the incorporation 
of the Wess-Zumino term into this theory. This term eliminates an extra 
discrete symmetry that is not a symmetry of QCD~\cite{witten83} and, 
therefore, has far reaching consequences and, of course, no lack of 
attention~\cite{rabinovici84}.

All these models are first presented as classical field theories, since one 
can do much physics using only selected classical solutions. The need to 
address the problem of quantization is, however, manifested in the intrinsic 
properties of the classical solution.

The capability of extracting interesting physics from the Skyrme model is 
grounded on the existence of a special solution of the classical field 
theory, the hedgehog skyrmion. Like all interesting classical (or mean 
field) solutions it breaks some of the symmetries of the underlying 
Lagrangian. The hedgehog skyrmion violates translation, spatial rotation, 
and iso-spatial rotation symmetry. The restoration of these symmetries 
requires, at the very least, the quantization of the generators of the 
symmetry transformations and of the associated canonically conjugate 
collective coordinates. As a consequence, maximum attention has been paid 
to this aspect of the problem of quantization~\cite{asano91}. In 
addition, to study pion-baryon 
scattering~\cite{mattisphd,karliner86,hayashi91}, it is 
necessary to discuss quantization of the small oscillations of the pion 
field \cite{braaten85} (for different approach see~\cite{hayashi92}).
There have also been some discussion on 
quantization of radial oscillations~\cite{kostyuk95,sawada91,weigel91} in 
connection with problems of stability. Other quantization 
methods applied to Skyrme model include 
{\it cutoff quantization}~\cite{balakrishna92},
(which uses short-distance cutoff $\epsilon:\quad F(\epsilon )=\pi$), the 
{\it general covariant Hamiltonian 
method}\/~\cite{dewittbook1,dewitt52,dewitt57} (which preserves the original 
symmetry of classical Lagrangian), {\it Kerman-Klein quantization 
procedure}~\cite{cebula93} (based on formal quantization of entire classical 
field). Due to rich and beautiful mathematical structure the model has 
numerous applications.
\newline{\bf Applications of Skyrme model.}
Despite the original model has been introduced to describe strongly 
interacting particles, there are attempts to apply similar gauged
construction to describe weak interactions ("electroweak 
skyrmions")~\cite{balachandranbook}~(p.250). Apart from high energy physics 
the model proved to be useful in cosmology~\cite{rho95} and solid state 
physics\footnote{The number of applications of the Skyrme model to quantum 
Hall effect have greatly increased during the last two years (1996--1998).}
\cite{sondhi93,fradkin88}. 
Before brief review of the manuscript organization we explicitly state main 
tasks this thesis is intended to solve.\newline
{\bf Main tasks:}
\begin{enumerate}
\item Investigate representation dependence of the quantization procedure
\cite{fujii87} applied to the SU(2) Skyrme model.
\item Numerically evaluate obtained expressions for physical quantities and 
compare the results with experimental data.
\end{enumerate}
{\bf Scientific novelty.}
This work demonstrates the new possibilities to extend basic Skyrme model
to arbitrary representations. Quantization of the Skyrme model (in collective
coordinate approach from the outset) yields different quantum Lagrangian
density for each SU(2) group representation $j$. The classical limit
of these quantum Lagrangian
densities is the same original Skyrme Lagrangian density.
For the first time it has been shown that stable quantum solitons exist
{\it both} for spin, isospin $\ell=\frac12$ and $\ell=\frac32$ states.
These quantum solitons possesses Yukawa asymptotic and, therefore, imply
non-vanishing pion mass. Noether currents, magnetic momenta etc., operators
have been calculated and numerically evaluated in this approach for 
self-consistent quantum chiral angles in various representations $j$.

The generalization considered in the work has far-reaching consequences 
and, we believe, can readily be extended to other models and theories.
\newline
{\bf Manuscript organization.}
The manuscript is organized into three chapters plus appendices, containing 
numerous tables and illustrations. Chapters and sections (if structure of 
the latter is complicated enough) have short information about its 
content and, therefore, not need to be repeated here. We find useful, 
however, briefly to describe what purposes each chapter is intended to 
serve.
{\bf Chapter~\ref{chapone}} contains mathematical formulation and physical 
motivation of the Skyrme model in background level. Apart from few 
presentation details it contains no new results.
We give formulation of classical Skyrme 
model in group theory terms and introduce mathematical apparatus which is 
convenient for model formulation in arbitrary representation 
in {\bf Chapter~\ref{chaptwo}}. This chapter includes results of 
Ref.~\cite{norvaisas94}.
{\bf Chapter~\ref{chapthree}} contains main new results and deals with 
the quantization of the Skyrme model.

Despite the thesis has no lack of references when investigating concrete 
problems we found useful to provide a list of sources about the entire
model. These are books 
\cite{balachandranbook,mahankovbook} and review articles
\cite{balachandran86,holzwarth86,zahed86}.
Literature on solitons currently is untraceable\footnote{We have found over 
1100 sources under single keeword {\it skyrmion[s]} in data basis "WOS" 
{\bf http:$\backslash\backslash$wos.isitrial.com/wos},
(starting from year 1983).}, but we still mention few books, 
namely~\cite{filipovbook,rajaramanbook,rebbibook} to begin with.
\newline
{\bf Acknowledgments}
\begin{itemize}
\item First of all I am indebted to my teacher and collaborator {\it 
Egidijus NORVAI\v SAS}. It has been a joy and privilege these five years to 
have benefited from his gently guidance, clear ideas and informal 
communication.

\item I tender thanks to my wife {\it Janina} and our sons  {\it Algirdas} 
and {\it K\k estutis} for everything, most of all keeping me sane and 
making me happy through it all.

\item I also would like to thank my {\it doctorate committee} for 
consultations and support, many {\it teachers} both at the University and 
school, as  their help and education have played an important role in my 
further investigations.

\item Many thanks to {\it Vytautas \v SIMONIS} and {\it Rimantas \v SAD\v 
ZIUS} for careful reading of the manuscript, also {\it Tadas KRUPOVNICKAS} 
in helping me with illustrations.

\item I am grateful to {\it the administration of Institute of Theoretical 
Physics and Astronomy} for taking care of our (constantly improving) work 
conditions.

\item Many thanks to staff of library of Institute of Theoretical 
Physics and Astronomy for quick and comprehensive service. 

\item I am indebted to {\it Gediminas VILUTIS, Vygandas LAUGALYS, Gintaras 
VALIAUGA, Gytis VEKTARIS, Gintautas GRIGELIONIS, Edvardas 
DU-\penalty-10000KO,
Art\=uras KULIE\v SAS} for help 
and guidance in sideless jungle of enormous (and still very rapidly 
expanding) Computerland.

\item Finally I would like to thank many others at Institute of Theoretical 
Physics and Astronomy, and especially our coffee team ({\it Gediminas 
JUZELI\= UNAS, Bronislovas KAULAKYS, $\dots$}) for inspiration, 
companionship and support.

\item This study was supported by Lithuanian Government, Lithuanian State
Science and Studies Foundation, by Grant N~LA5000 from
the International Science Foundation (in part), and
by Joint grant N~LHU100 from Lithuanian Government and ISF (in part).
\end{itemize}
\mainmatter

\chapter[Introduction]{Introduction to the Skyrme model}
\label{chapone}
This chapter is intended to provide very short but more or less 
consistent introduction to the Skyrme model. 
From this point of view it is essential to clear out
difference between models realized linearly and nonlinearly.
The simplest examples are linear and nonlinear $\sigma$ models.  Skyrme 
model then arises naturally by adding the fourth-order term in field 
functions to nonlinear $\sigma$ model Lagrangian. This term enables 
existence of stable soliton in three spatial dimensions (skyrmion) and, 
therefore, is called a stabilizing term. 
\section{Linear $\sigma$  model in two spatial dimensions}
\setcounter{equation}{0}
We start with linear $\sigma$ model. When physical 
boundary conditions are imposed, all model solutions fall into disconnected 
classes regardless of what equations of motion are. It is this property 
which is peculiar to nonlinear models only and play an important role 
in the Skyrme model particularly.
\label{linsigmod}
\subsection{The Lagrangian}
\begin{figure}
\begin{center}
\includegraphics*[140,585][310,670]{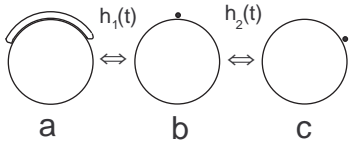}
\end{center}
\caption{Example of zero winding number maps.}
\label{fig1}
\end{figure}

Let's take $\BPhi $ to be a scalar doublet of real
fields $(\Phi_1 ,\Phi_2 )$ and consider the Lagrangian
\begin{equation}
\CL=-\frac 12\partial _\mu \Phi _r \partial ^\mu \Phi _r
-(\left| \BPhi \right|^2-a^2)^2;\qquad  r=1,2;\quad a^2>0.
\label{pir}
\end{equation}
Here we understand $\left|\BPhi \right|^2=\Phi _r \Phi_r$ and $\mu =0,1,2$
($0$ denotes time component). 
Assume that for $\Bx\rightarrow \infty $ field configuration tends to some 
constant state\footnote{Only these states are interesting from physical 
point of view: energy at infinity should be zero.}
\begin{equation}
\Phi _1^2+\Phi _2^2\rightarrow a^2,\qquad \text{for}\left| \Bx\right|
\rightarrow \infty
\label{artvac}
\end{equation}
with approach rate, which guarantees finiteness of total system energy
$E$. Then the set $\BPhi _\infty =\left( \Phi _{\infty 1},
\Phi _{\infty 2}\right) $ of all fields at spatial infinity make up
a circle $S^1$. Spatial infinity in argument plane also can be imaged
as a circle $\tilde S^1$, with infinitely large radius
\begin{equation}
x^2_1+x^2_2=R^2 ,\qquad R\rightarrow \infty.
\end{equation}
Field $\BPhi _{\infty}$ maps a circle $\tilde S^1$ to a circle $S^1,\quad
\BPhi _{\infty}:\quad \tilde S^1 \rightarrow S^1.$
Identification of infinities with a circle of infinite radius does not
involve any topology change. The hint is similar to coordinate
system change. We have here spaces --- both argument and function space
--- flat. Moreover, these  spaces are vector spaces as
well. Consider first a vacuum (or trivial) solution $\BPhi _{\rm
vac}(x_1,x_2)\equiv (a,0)$. At spatial infinity the field $(\BPhi
_\infty )_{\rm vac}=\lim\limits_{R\rightarrow \infty }\BPhi _{\rm
vac}$ maps all points of $\tilde S^1$ circle (argument infinity) to
the same vacuum point $(a,0)$ of function space $S^1$. Thus $(\BPhi
_\infty )_{\rm vac}$ is characterized by zero winding number: $(\BPhi
_\infty )_{\rm vac}\in Q^\infty _0$. To $(\BPhi _\infty )_{\rm vac}$ we
can, therefore, associate all maps $\BPhi ^{(0)}_\infty $ which are
homotopic to $(\BPhi _\infty )_{\rm vac}$. The set of all maps $\BPhi
^{(0)}_\infty $ makes up the trivial sector $Q_0^\infty $ of system
configuration space $Q^\infty $ (see Fig.~\ref{fig1}).
Of course, one can choose any other vacuum state simply redefining
$\BPhi _{\rm vac}(\Bx)=(a/\sqrt{2},a/\sqrt{2})$ or more generally
$\BPhi _{\rm vac}(x_1,x_2)=(a\cos \chi, a\sin \chi )$, where $\chi $
is fixed from $[0,2\pi ]$ (see Fig.~\ref{fig1}~$\!{}_c$). All vacuuma
belong to trivial $n=0$ sector and are homotopic to
$\BPhi _{\rm vac}=(a,0)$. The homotopy can be defined as $h_2(t)=(a\cos
t\chi ,a\sin t\chi )\quad t\in  [0,1]$ (see Fig.~\ref{fig1}~$\!{}_{b,c}$). 
One could think about gauge freedom corresponding to transformation from
one vacuum to another. With any vacuum choice model SO(2) symmetry becomes
spontaneously broken. Physical motivation comes from tunnelling
possibility: vacuuma aren't separated by any potential barrier,
therefore, in the case of such vacuuma interference fields must rotate
everywhere in space. This involves infinite rotational energy
\cite{balachandranbook}~(p.107). Of course such interference also
restores SO(2) symmetry of the ground state and, therefore, should be 
forbidden.
\subsection{Soliton sectors and invariants of linear $\sigma$ model}
\begin{figure}
\begin{center}
\includegraphics*{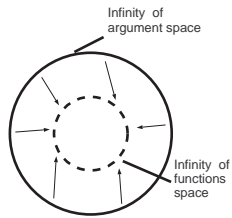}
\end{center}
\caption{Example of winding number one map in linear $\sigma$ model.}
\label{fig2}
\end{figure}

A field $\BPhi _\infty^{(1)}(\theta )$ of the sector $Q^\infty _1$
can be defined as $\BPhi _\infty ^{(1)}(\theta)=(a\cos \theta,a\sin \theta
)$. When $\theta $ runs from $0$ to $2\pi $ all points on $S^1$ are
covered once and only once. $\BPhi _\infty ^1$ is a typical winding
number one map (see Fig.~\ref{fig2}). The equivalence class of maps homotopic
to $\BPhi ^{(1)}_\infty $ forms the winding number one sector $Q^\infty
_1$. The winding number $n$ map can be defined as $\BPhi _\infty
^{(n)}(\theta)=(a\cos n\theta,a\sin n\theta )$, where $n$ is an
integer, due to our requirement of $\BPhi $ single valuedness  on
$\tilde S^1:\quad \BPhi _\infty ^{(n)}(0)=\BPhi _\infty ^{(n)}(2\pi)$.
$Q^{(\infty )}_n$ consists of all $\BPhi ^{(n)}_\infty $  homotopic
maps. We remind that $\Phi ^{(n)}_\infty $ with any $n$ (not only
$n=0$) satisfy \eqref{artvac}. From this picture it becomes clear that it
is not possible to deform a field $\BPhi^{(n)}_\infty $ to a field $\BPhi 
^{(m)}_\infty $ {\sl continuously}, if $n\neq m$. Indeed, for the action we 
need to cut the mapping curve and take off
$|n-m|$ twists. Thus the fields in sector $Q_n^{(\infty )}$ are not
homotopic to fields in $Q_m^{(\infty )}$ for $n\neq m$. As a
consequence, system configuration space $Q^\infty $ falls into an
infinite number of disconnected components $Q^{\infty }_n =\{\BPhi
^{(n)}_\infty \}$, $Q^\infty $ being a union $\mycup\limits_n Q^\infty _n$. 
The same is then true for the space $Q$ of fields defined over all space: 
$Q=\mycup\limits_n Q_n$. Here $Q_n$ is the space of all configurations 
$\BPhi ^{(n)}$ whose limit as $|\Bx |\rightarrow \infty$ is an element 
$\BPhi _\infty ^{(n)}$ of $Q_n^{(\infty )}$. For example, $\BPhi ^{(n)}$ at 
given time $t_0$ can be defined by
\begin{equation}
\BPhi ^{(n)}(\Bx,t_0)=f(|\Bx|)\, \BPhi _\infty ^{(n)}\Bigl(\frac{\Bx}{|
\Bx|}\Bigr),
\end{equation}
where $f(|\Bx|)$ is any smooth function, such that $f(\infty)=1,\quad 
f(0)=0$.

The physical significance of the integer $n$ associated with the field
$\BPhi ^{(n)}(\Bx,t)$ is that it is an integral of
motion. The integer is just the label of homotopy classes of the fields
at a fixed time. Sectors $Q_n \quad (n\neq 0)$ are topologically stable
in the sense that a field in $Q_n$ will not evolve in
time to the vacuum (an element of $Q_0$) or to a field in any other 
sector $Q_m\quad (m\neq n)$, since evolution in time is a continuous 
deformation.

The existence of inequivalent topological sectors leads to additional
invariants in the theory. These new invariants of quite different origin
is the most interesting and important point in such models. As a 
consequence, we have two kinds of invariants:
\begin{itemize}
\item {Invariants which are closely related to the symmetry of the system,
under simultaneous coordinate frame and fields change (corresponding
to this frame transformations). We can find all these invariants by
Noether theorem. Examples of the invariants are: energy, momentum, angular 
momentum, electromagnetic charge.\footnote{
The Lagrangian~\eqref{pir} is symmetric under rotations of SO(2).
The SO(2) is known to be isomorphic to U(1), therefore, the Lagrangian
$\CL=-\frac 12\partial _\mu \Phi \partial ^\mu \Phi ^{*}
-(\Phi \Phi ^{*}-a^2)^2;\quad  \mu=0,1,2;\quad a^2>0$ can be chosen
instead of~\eqref{pir}. Conserved Noether currents exist corresponding
to continuous symmetries (SO(2) or U(1)) of these Lagrangians. 
In the case of one complex field the conserved electromagnetic current
has a simple form $J^\mathrm{e}_\mu \sim
i(\Phi ^{*} \partial _\mu \Phi -\partial _\mu \Phi ^{*} \Phi )$. These
currents have nothing to do with the conserved topological 
current~\eqref{topcurrent1}. Also there is one interesting difference
between two real and one complex field case.
Namely, there is no real vector, which is invariant under SO(2) rotation, 
but there is a pair of complex eigenvectors $(1\pm i)$ with eigenvalues 
$\e^{\pm i\phi}$ in complex plane (group U(1)).}}
\item{Invariants, involving boundary conditions in one or another
way. Conservation of the number of particles in classical mechanics 
(in conservative systems) is an example.}
\end{itemize}
Let us concentrate on the second type of invariants. It has been shown
that solutions from one sector cannot evolve in time to the solutions of
any other sector. Consequently, we need discrete quantity to label
each sector. The most natural choice seems to be the number of twists,
describing mapping of one circle onto another. The number is called a
topological charge. One could introduce a conserved topological current 
density, corresponding to the charge
\begin{equation}
J^{\rm top}_\mu  \sim \epsilon _{\mu \sigma \rho }\epsilon ^{rs
}\partial^\sigma\Phi ^r\partial^\rho \Phi ^s\quad \mu ,\sigma
,\rho =0,1,2;\quad r,s=1,2.
\label{topcurrent1}
\end{equation}
It is easy to check that expression~\eqref{topcurrent1} has a
divergence zero and meets our requirements. We stress that the current 
density is conserved irrespective of what the equations of motion 
are (because of the antisymmetric properties of $\epsilon _{\mu \sigma \rho 
}$) and thus it is topological. The corresponding charge density is
\begin{equation}
J^\mathrm {top}_0={\rm div}\BP,\qquad \text{where}\qquad
P_k\sim \epsilon _{kl}\epsilon ^{rs}\Phi ^r \partial_l \Phi
^s;\quad k,l=1,2.
\label{divp}
\end{equation}
The charge $Q^\mathrm{top}\sim  \int J^\mathrm{top}_0\di^2 \Bx$ is nonzero 
only for fields $\BPhi $ with non-vanishing asymptotic\footnote{
We have in mind configurations with $(\Phi ^1_\infty )^2+(\Phi 
^2_\infty )^2=a^2$. Indeed, using Gauss theorem from \eqref{divp} and 
\eqref{topcurrent1} we obtain $\int J^\mathrm{top}_0\di^2 \Bx \sim \int 
\mathrm{div} \BP\di^2 \Bx \sim \int _{R\rightarrow \infty} (\BP\cdot 
\di\Bell )\sim 0$, if $\lim _{|\Bx|\rightarrow \infty } \BPhi 
^r(x_1,x_2,t)=0$.}~\cite{fadeev76}, which is realized, for example, by the 
Higgs mechanism.
\subsection{Derrick theorem}
\label{derrictheorem}
The existence of topologically stable sectors and conservation of 
topological current are independent on the Lagrangian form and, thus, on 
equations of motion. The presence  of such sectors, therefore, does not 
guarantee that equations of motion actually have solutions in each sector. 
It is known that Lagrangian \eqref{pir} does not lead to nontrivial 
stable static solutions of equations of motion if only the spatial 
dimension is $D\neq 1$. This can be shown by simple 
scaling argument of Derrick~\cite{balachandranbook,derrick64}. Suppose that 
$\BPhi _{\rm cl}$ is static solution and the energy of the solution consists 
of terms $E=E_1+E_2$:
\begin{equation}
E_1=\frac 12\int \di^D\Bx (\partial _i(\BPhi _{\rm cl })_s)^2\quad
\text{and}\quad E_2=\int \di^D\Bx U[((\BPhi _{\rm cl })_s)^2]
\quad i=1,2,\dotsc, D.
\end{equation}
Under a scaling transformation $\BPhi _{\rm cl }(\Bx,t)\rightarrow
\Phi _{\rm cl }(\lambda \Bx,t)$ these terms scale as
\begin{equation}
E \equiv E(1) \rightarrow E(\lambda )=
\lambda ^{2-D}E_1 +\lambda ^{-D}E_2.
\label{enesca}
\end{equation}
Requiring that $\lambda =1$ corresponds to energy $E$ minimum yields the
condition
\begin{equation}
\frac{\di E(\lambda )}{\di \lambda }\Bigl| _{\lambda =1}=0\quad
\text{or}\quad (2-D)\,E_1=D\, E_2.
\end{equation}
Since $E_1,E_2\ge 0 $, it follows that $E_1=E_2=0$ when $D>2$.
This implies that $\BPhi _{\rm cl }$ must be the vacuum solution for $D>2$.
In the case $D=2$, we have $E_2=0$, so that
$((\Phi _{\rm cl })_1)^2 +((\Phi _{\rm cl })_2)^2=a^2$ for all $\Bx$.
This requires that $\BPhi _{\rm cl }$ (as $|\Bx|\rightarrow \infty $)
has zero winding number and hence is in $Q_0^{(\infty )}$. We can
prove this result as follows. Let $r,\theta $ denote polar coordinates
in the plane. For any nonzero $r$, $\BPhi _{\rm cl }$ defines a map of the
circle (with coordinate $\theta $) to a circle (because of the
condition $((\Phi _{\rm cl })_1)^2 +((\Phi _{\rm cl })_2)^2=a^2$). The
winding number $n$ of this map cannot depend on $r$, as changing $r$
is a continuous change. When $r\rightarrow 0$, all values of $\theta $
represent the same spatial point, so that $n\rightarrow 0$. Hence n is
identically zero which proves the result.
\section{Simplest nonlinear topological model}
\begin{figure}
\begin{center}
\includegraphics*{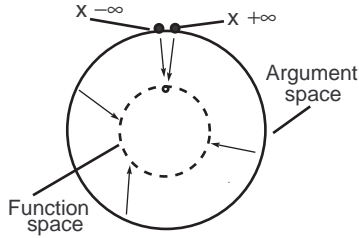}
\end{center}
\caption{Example of one-dimensional model map in a simple nonlinear model.}
\label{fig3}
\end{figure}

\setcounter{equation}{0}
The simplest model which modifies space topology can be found in 
one-dimensio-\penalty10000nal field theory. Consider the set of all 
mappings $\alpha $ from the real line $R^1$ (argument space) onto the circle 
$S^1$ (function space). $S^1$ can be parametrized by two real variables 
$\BPhi =(\Phi _1,\Phi _2)$, whose squares add up to one: $\Phi _1^2+\Phi 
_2^2=1$. Note that function space is not flat (circle $S^1$). To prevent the 
escape of interesting structures at infinity we consider only the class of 
functions on $S^1$, such that $\BPhi (\infty )=\BPhi (-\infty )$. The 
restriction of functions class allows us to identify argument space (line 
$R^1$) with a circle. In other words it makes possible compactification of 
$R^1$ to a circle\footnote{One point compactification theorem 
\cite{matuzeviciusbook}~(p.86). Two-dimensional analog of the 
compactification is known as a stereographic projection.}.
A field $\BPhi^{(1)}$ from $n=1$ sector may be illustrated pictorially by a
strip familiar with M\"obius strip, except that the M\"obius strip has a
twist through $\pi $ whereas the sector $n=1$ field has a twist through
$2\pi $. $\BPhi^{(1)}(x)$ then specifies twist angle of
the strip about its center line at a given point $x$. Note 
quite different meaning we give to the external circle in this 
(see Fig.~\ref{fig3}) and linear 
case (Fig.~\ref{fig2}${}_b$). Since by classical field we mean a field 
that is single-valued under the action of the rotation group, it follows 
that particles involved must be bosons. Quantization of such a classical 
field introduces a quantity that can be interpreted as a particle number. 
It is known that after quantization the states, corresponding to $n=1$ 
classical configuration, are, in fact, fermion states. Dynamics and quantum 
mechanical operators can be introduced\footnote{Lagrangian of the 
toy model: $\CL =\partial _\mu \Phi_s \partial ^\mu \Phi_s ;
\quad \mu=0,1;\quad s=1,2$,  where $\BPhi$, in addition, is
subject to the constraint $\Phi _1^2+\Phi_2^2=1$.} into this 
theory~\cite{skyrme61a,williams70}. 

To summarize, the simple nonlinear model fields are subject to nonlinear 
constraint, when linear $\sigma$ model fields are not. This explains the 
need to reduce one argument space dimension 
in order to have the same global symmetry group for both models. In other 
words, function space of the first model is a flat space (vector space as 
well), when function space of the second one is a compact manifold (not 
a vector space at all).
\section{Nonlinear $\sigma$ model in two spatial dimensions}
\setcounter{equation}{0}
In this section we formulate nonlinear $\sigma$ model in fundamental \SU\
representation using well known Pauli and rotation matrices technique. The
Skyrme model then is obtained by adding forth order term in field functions
which ensures stable soliton solution in 3D. As a consequence, Skyrme model 
inherits all essential features from nonlinear $\sigma$ model.
\label{nonlinsigmod}
\subsection{Formulation}
It was sufficient to look at $Q^\infty $ (the space of physical fields at 
spatial infinity) for the topological consideration in the linear $\sigma$ 
model. For solitons in nonlinear models it is often necessary to consider 
the topology of physical fields defined {\it over all space}. Physical 
fields in these models for all points $\Bx$ take values in a manifold \GM\
which generally is not a vector space.

By definition group \GG\ acts on manifold \GM\ transitively, if for 
any pair $p,p'\in \text{\GM }$ there exists an element $g\in \text{\GG 
}$, such that $T_g(p)=p'$. Assume, this is the case. Then \GM\ is called a 
homogeneous space for \GG. If $\text{\Gg}_p$ is the stability group of 
point $p\in \text{\GM}$:
\begin{equation}
\text{\Gg}_p=\{h\in \text{\GG}|hp=p\} ,
\end{equation}
and \GM\ is a homogeneous space for \GG\, then any two 
$\text{\Gg}_p,\text{\Gg}_{p'} \quad p,p'\in \text{\GM }$ are isomorphic. 
If $p\neq p'$ and $T_g(p)=p'$ the isomorphism $\text{\Gg}_p\rightarrow 
\text{\Gg}_{p'}$ can be defined\footnote{It is 
assumed that group \GG\ acts on \GM\ from the left. Homogeneous spaces
play an important role in ensuring uniqueness of solution of equations of 
motion. There is no homogeneous space problem in 3-dimensional $\sigma$
(Skyrme) model, because in this case there exists one-to-one correspondence 
between function space and \SU\ manifold, which, of course, is natural 
homogeneous space for the same group. The problem again arises in SU(3) 
Skyrme model, when \SU\ ansatz is employed.}: $h\mapsto 
ghg^{-1}$. Now we can identify \GM\ with space of left cosets \GG$/$\Gg\ 
by the following procedure. First let's fix point $p_0\in \text{\GM}$. With 
each class of left cosets $\{g\text{\Gg}\}$ we identify point $T_g(p_0)$, 
where $\text{\Gg}=\text{\Gg}_{p_0}$ is a stability group of point $p_0\in 
\text{\GM }$. The identification is in one-to-one correspondence and do 
not depend on particular $g$ in the class of left cosets.

The nonlinear $\sigma$ model in two spatial dimensions
has \GG\ $=$\SU$,\quad $ \Gg\ $=$U(1)
$\equiv \{ {\e}^{i\alpha {J_z}}, 0\leq \alpha \leq 2\pi \}$ and 
\GM$=$\GG$/$\Gg\  is a {\em two} sphere $S^2$. To show this, define
\begin{equation}
K\equiv gJ_zg^{\dagger }=n_{a'} J_{a'},\quad a'=(x,y,z)\equiv 
(1,2,3),\label{ndef}
\end{equation}
where $g\in \mathrm{SU(2)}$ and $J_{a'}$ are generators of SU(2).

$K$ is an invariant under transformation\footnote{And only under these 
transformations, because $[J_z,J_{x[y]}]\neq 0$.}  $\pi :g\rightarrow 
g\e^{i\alpha J_z}$. A map $\pi $ projects $g\e^{i\alpha J_z}\in \text{\GG}$ 
for all $\alpha $ to the same point of left cosets $\{ g\text{\Gg}\}$. 
Since $\alpha $ is a continuous parameter, $\{ g\text{\Gg}\}$ defines a 
two dimensional manifold, namely a two sphere $\mathrm{SU(2)/U(1)}=S^2$. 
Indeed, the  scalar product of $n_{a^\prime}$ is $1$,
\begin{equation}
\Tr K^2=n_{a'} n_{a'}\equiv (n_{a'})^2=1.
\end{equation}
Fields $\Phi _{a'}(x_1,x_2,t)$ of nonlinear 2-dimensional $\sigma$ model 
are subjects to the constraint: $\Phi _{a'}(\Bx,t)\Phi _{a'}(\Bx,t)=1$ and 
thus can be identified with $n_{a'}$. The action of \GG\ on these fields 
yields
\begin{equation}
\Phi '_{a'}=R_{a'b'}(g) \Phi _{b'},
\end{equation}
where $R_{a'b'}$ is the usual rotation matrix --- an element of adjoint 
representation of SU(2): $\mathbf{R}\in SO(3)$. Note that the constraint is 
invariant under this action of \GG. 

The Lagrangian density is chosen so that it is invariant under \GG.
\begin{equation}
\CL =-\frac{\beta }2 \partial _\mu \Phi _{a'} \partial ^\mu \Phi _{a'},\quad
\beta =\mathrm{const},\quad \mu =0,1,2.
\label{nonlinsiglagran}
\end{equation}   
Because of the constraint on $\BPhi _{a'} $, the Lagrangian
\eqref{nonlinsiglagran} does not describe a free system. Interactions of 
the field with itself are implicit. To see this, one can write $\CL $ in 
terms of two independent degrees of freedom, say $\Phi _1\equiv \Phi _x$ 
and $\Phi _2\equiv \Phi _y $
\begin{equation} 
\CL=-\frac{\beta }{2} \left(\frac{2-\Phi _k\Phi _k }{1-\Phi _k\Phi 
_k}\right)(\partial _\mu \Phi _k )^2,\quad \Phi _3\neq0,\quad k\ \text{is 
summed over\ } 1 \text{\ and\ } 2.
\end{equation} 
Note that the action of \GG\ is nonlinear in terms of two degrees of 
freedom.

The energy density associated with (\ref{nonlinsiglagran}) is
\begin{equation}
\CE (\Bx,t)=\frac\beta 2 \bigl((\partial _0 \Phi _{a'} )^2+
(\partial _i \Phi _{a'} )^2\bigr);\quad i=1,2. 
\end{equation}
\subsection{Topological structure}
The vacuum solution (which 
is subject to the constraint on $\BPhi$)
is $\BPhi _\mathrm{vac}=\mathrm{const}$. $\CE $ is invariant under
global SU(2) transformations. Thus $\BPhi _\mathrm{vac}$ can be reduced to 
$(0,0,1)$ by action of SU(2) without affecting the energy ($\CE 
\equiv 0\quad\text{for}\quad\BPhi _\mathrm{vac}$). After the choice only 
rotations about the third axis leave the vacuum invariant, consequently, 
the global SU(2) is spontaneously broken to a global U(1), by special 
vacuum choice.
\begin{figure}
\begin{center}
\includegraphics*{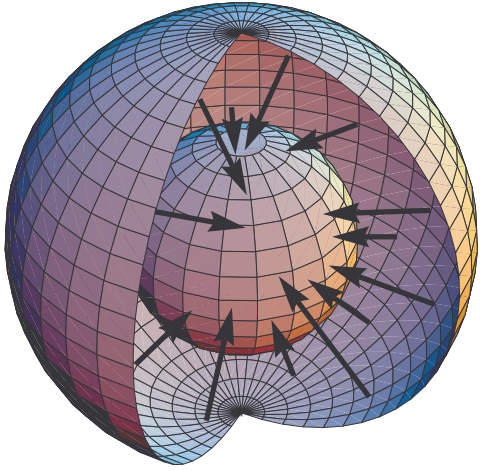}
\end{center}
\caption{Example of 
mappings: $\tilde S^2$ sphere $\rightarrow $ $S^2$ sphere.}
\label{fig6}
\end{figure}
Next, consider general configuration with nonzero, but finite energy. Let 
us first show that homotopic sector generation mechanism described in 
linear $\sigma$ model in Sec.~\ref{linsigmod}  now fails.
Indeed, for large $|\Bx|\equiv R$ the fields $\BPhi $ define a mapping 
from circle $S^1$ with radius $R$ to $S^2$. Since $\pi _1(S^2)=0$ this 
mapping is homotopic\footnote{
Even if the fundamental group  of the manifold \GM\ is nontrivial:
$\pi _1(\text{\GM})\neq 0$, the fields at $R$ would have to be homotopic to 
the vacuum solution $\BPhi _\mathrm{vac}$. This is so since $\BPhi $ defines 
a trivial mapping at $r=0$, and the topological index cannot change as $r$ 
is continuously varied from $r=0$ to $R$. (Constraint $\BPhi^2=1$ here is 
fulfilled at each point.  In linear $\sigma$ model, in contrast, we had this 
condition satisfied at infinity only.)} to $\BPhi _\mathrm{vac}$ at $|\Bx|= 
R$. Despite the fail there is a way out. Indeed, finiteness of energy 
requires $\BPhi (\Bx,t)$ to approach $\BPhi _\mathrm{vac }$ as $r\rightarrow 
\infty$ and that the rate of approach is fast enough to guarantee that the 
energy $E$ is finite. Again we choose $\BPhi _\mathrm{vac}=(0,0,1)$. Assume 
that $\BPhi $ approaches $\BPhi _\mathrm{vac}$ and there is no angle 
dependent limit at $r=\infty $. Thus, we may think of all points at spatial 
infinity as being a single point. Such a restriction of function class 
$\BPhi $ essentially converts (in topological but not metrical sense)
the plane $R^2\equiv \{(x_1,x_2)\}$ at a 
constant time $t_0$ to the surface of a two sphere $\tilde S^2 $. The fields 
$\BPhi =(\Phi _1,\Phi _2,\Phi _3)$ are well defined on this  $\tilde S^2$ in 
view of boundary conditions. Consider this in detail. Let $\xi _a,\quad 
a=1,2,3$ be the stereographic coordinates associated with $\Bx$
\begin{gather}
\xi _a(\Bx)=\frac{2 x_a}{r^2+1},\quad 
a=1,2;\quad \xi _3(\Bx)=\frac{r^2-1}{r^2+1}\notag \\
\xi _1^2(\Bx)+\xi _2^2(\Bx)+\xi _3^2(\Bx)=1,
\quad r^2=x_a x_a.\label{ksidef}
\end{gather}
Coordinates $\Bxi$ span a two-sphere $\tilde S^2$. They are not global 
valid coordinates for $R^2$ which unlike $S^2$ is not a compact 
manifold. Indeed, all the "points of infinity" of $R^2$ which 
correspond to $r\rightarrow \infty $ are mapped to one point of $S^2$, 
namely $(0,0,1)$. In reality the "points of infinity" are not points at
$R^2$ at all. Thus, to get a topologically accurate representation 
of $R^2$ we should remove the north pole $(0,0,1)$ from 
$\tilde S^2:\quad R^2=\tilde S^2\backslash \{(0,0,1)\}$.

The topological difference between $R^2$ and $S^2$ can make 
difference for some functions. For example, the function $f(\Bx)=|\Bx|$ is 
a continuous function on $R^2$, but the function obtained by the 
substitution $x_a =\xi _a/(1-\xi _3)$ is not a continuous function on 
$\tilde S^2$, becoming infinite at north pole $\xi _3 =1$. Another 
example is the function $\bar \Bx =\Bx/|\Bx|$ which is continuous on 
$R^2$, while its image function on $\tilde S^2$ has no well 
defined limit as the north pole is approached. However, for functions which 
approach a constant limit as $r\rightarrow \infty $, the change of variable 
$\Bx\rightarrow \Bxi$ does produce a well defined function on $\tilde 
S^2$. In this sense then, because of the boundary condition on $\BPhi $, we 
can imagine that the space on which the field $\BPhi $ is defined is 
$\tilde S^2$.

\begin{figure}
\begin{center}
\includegraphics*{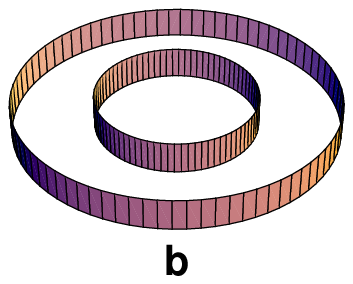}
\includegraphics*{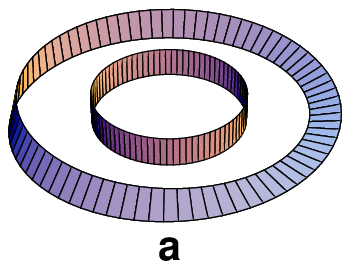}
\end{center}
\caption{Examples of 
mappings: cylinder $\rightarrow $ cylinder and 
cylinder $\rightarrow $ M\"obius strip.}
\label{fig4}
\end{figure}
Thus, the configuration space $Q$ of the nonlinear $\sigma$ model is made 
up of fields which map $\tilde S^2$ to $\text{\GM} =S^2\quad \BPhi:\tilde 
S^2\rightarrow S^2$ (see Fig.~\ref{fig6}). The situation, thus, is 
analogous (despite quite different origin of the map)
to the maps $\BPhi _\infty $ in linear $\sigma$ model, where we 
had $\tilde S^1\rightarrow S^1$. It is then plausible to expect that for 
nonlinear $\sigma$ model the configuration space $Q$ falls into an infinite 
number of disconnected components $Q_n$, with $Q=\mycup\limits_{n} Q_n$. 
This result is true. Here $n$ is generalization of the previous winding 
number associated with $\BPhi_\infty $ and is also called the winding 
number. It indicates the number of times the sphere $S^2$ is covered by 
sphere $\tilde S^2$ as $\Bxi$ runs over all values. Strictly speaking, 
one cannot transfer this very illustrative definition from 
the one-dimensional case (number of times one circle is covered by another 
circle) to higher dimensions (number of times one $S^k$ sphere is covered by 
another $S^k$ sphere). Mathematically irreproachable definition of mapping 
degree\footnote{The illustrative definition of the winding number is indeed 
correct for the one-dimensional case. The reason is that curve does not have 
internal structure. In general it is not true for higher dimensional 
manifolds. For example mappings cylinder $\rightarrow $ cylinder  and 
M\"obius strip $\rightarrow $ cylinder in Fig.~\protect\ref{fig4} obviously 
are not homotopic, but illustrative winding number definition does not allow 
us to clearly distinguish both cases.} $n$ on $m\ge 2$ dimensional manifold 
is based on very general triangulation concept~\cite{shashinbook}. Despite 
the criticism, it is known that for mappings $\tilde S^k\rightarrow S^k, 
\quad k=1,2,\dots $ the $k$-th homotopy group is $\pi _k(S^k)=\Z $, which in 
some sense justifies the illustrative idea of the winding number. After the 
generalization once again, equivalence classes $Q_n$ can be made into a 
group under a suitable product. This group is called the second homotopy 
group and is denoted by $\pi _2(\text{\GM})$. Here $\text{\GM}=S^2$. Like 
$\pi_1(S^1),\quad \pi_2(S^2)$ is isomorphic to the group \Z\ of the all 
integers under addition.

The equivalence class $Q_0$ contains the vacuum solution $\BPhi 
^{(0)}\equiv \BPhi _\mathrm{vac}=(0,0,1)$. $Q_0$ consists of all maps 
which are homotopic to $\BPhi _\mathrm{vac}$. An element $\BPhi ^{(1)}$ of 
$Q_1$ is obtained by simply setting $\Phi _{a^\prime} ^{(1)} (\Bxi  
,t)=\xi _{a^\prime} $, here $t$ being fixed and $\Bxi$ being 
coordinates defined in \eqref{ksidef}. A typical element 
$\BPhi ^{(n)}$ of $Q^{(n)}$ is
\begin{subequations}
\begin{alignat}{3}
&\Phi _1^{(n)}(\Bx,t)=\sin \vartheta \cos n\varphi,\quad&&
\xi _1(\Bx)=\sin \vartheta \cos \varphi,\quad&&
0\le \vartheta \le \pi ,\\
&\Phi _2^{(n)}(\Bx,t)=\sin \vartheta \sin n\varphi,&&
\xi _2(\Bx)=\sin \vartheta \sin \varphi,&&
0\le \varphi \le 2\pi \\
&\Phi _3^{(n)}(\Bx,t)=\cos \vartheta &&
\xi _3(\Bx)=\cos \vartheta && \qquad\qquad\qquad .
\end{alignat}
\label{finspcoord}
\end{subequations}
Here $\vartheta ,\varphi$ are spherical coordinates for the argument 
two-sphere $\tilde S^2$. $Q_n$ consist of all maps homotopic to the  $\BPhi 
^{(n)}$.

Again, the significance of the above classification (since time 
evolution is a continuous operation), is that 
the integer $n$ is an integral of 
motion. It is useful to have an explicit formula for this conserved quantum 
number. For this purpose consider the current
\begin{equation}
j^\mu (\Bx,t)=-\frac1{8\pi}\epsilon _{a'b'c'}\,\epsilon ^{\mu \nu \lambda }
\Phi _{a'} \partial _\nu \Phi _{b'} \partial _\lambda \Phi _{c'},
\label{nonlinsigmodcurrent}
\end{equation}
where $\epsilon ^{\mu \nu \lambda }$ is the totally antisymmetric 
tensor. This current is conserved regardless of the equations of 
motion. Taking its divergence we obtain
\begin{equation}
\partial _\mu j^\mu (\Bx,t)=-\frac1{8\pi}\epsilon _{a'b'c'}\,\epsilon ^{\mu 
\nu \lambda }\partial _\mu \Phi _{a'}\,\partial _\nu \Phi _{b'}\,\partial 
_\lambda \Phi _{c'}.
\label{nonlinsigmodcurrentdiv}
\end{equation}
The right hand side of (\ref{nonlinsigmodcurrentdiv}) contains the triple 
scalar product of the three tangent vectors $\partial _0 \BPhi,\ 
\partial _1 \BPhi $ and $\partial _2\BPhi$ defined at $(\Bx, t)$. 
When multiplied by $\di^2\Bx \di t$, it represents an infinitesimal volume 
element at $(\Bx,t)$. But because of the constraint on $\BPhi$, the 
tangent vectors at $(\Bx,t)$ are enforced to lie in a plane. 
Consequently, the volume element and the right hand side vanishes $\partial 
_\mu j^\mu =0$ and from \eqref{nonlinsigmodcurrentdiv} we have
$\displaystyle \frac{\di j^0}{\di t}=0$. 
It follows that the associated charge\footnote{Note that the charge density 
now is not a pure divergence $j^0\neq \mathrm{div} \BP$ 
(as it was in linear model) and has only integer values 
when properly normalized.}
\begin{equation}
B(\BPhi )=-\frac1{8\pi}\int \di^2\Bx \epsilon _{a'b'c'}\,\epsilon _{ij}
\Phi _{a'} \partial _i \Phi _{b'} \partial _j \Phi _{c'},
\label{topcharge}
\end{equation}
is a constant of motion. Its value is the conserved quantum number; it has 
the value $n$ when $\BPhi =\BPhi^{(n)}\in Q_n$. The factor $-\frac1{8\pi }$ 
is chosen so that $B(\BPhi )$ is, in fact, an integer.
To see that  $B(\BPhi )=n$, write $\BPhi$ in terms of spherical coordinates
\eqref{finspcoord}. Then
\begin{equation}
B(\BPhi )=\frac{n}{4\pi }\int \sin \vartheta (\Bx)\di\varphi(\Bx) 
\di\vartheta (\Bx).
\end{equation}
Since $\frac1{4\pi }\sin \vartheta \di\varphi\wedge \di\vartheta $ is the 
normalized volume element\footnote{The symbol $\wedge$ is an exterior
multiplication mark.} on two sphere, $B(\BPhi )$ indicates the number of 
times the sphere $S^2$ is covered as $\Bx$ runs over all values and is, 
therefore, an integer. Derrick scaling argument rules out (see 
Sec.~\ref{derrictheorem}) the possibility of having nontrivial static 
solution to  a linear scalar field in two (or greater) space dimensions. 
However, for the nonlinear $\sigma$ model with the Lagrangian 
\eqref{nonlinsiglagran}, Derrick's argument can only be used to rule out the 
existence of static solutions in all but two space dimensions. This is 
because the static energy contains only one term which we denote by 
$E_\sigma$. Under $\Bx \rightarrow \lambda \Bx$, it scales like $E_\sigma 
\rightarrow \lambda ^{2-D} E_\sigma$. The minimum value of the energy for 
this variation is zero in all except $D=2$ dimensions.

A lower bound on the energy (the "Bogomol'nyi bound")
\cite{rajaramanbook,rebbibook} for the classical solutions can be obtained 
from the identity
\begin{equation} 
(\partial _i \Phi _{a'} \pm  \epsilon _{a'b'c'}\epsilon 
_{ij}\Phi _{b'} \partial _j\Phi _{c'} )^2\ge 0.
\end{equation}
After completing the square, we can write 
\begin{equation}
\frac2\beta E_s=\int \di^2 \Bx(\partial _i \Phi _{a'})^2\ge 8\pi |n|.
\end{equation}
Here the $\beta$ is the same as in \eqref{nonlinsiglagran}.
The bound is saturated if
\begin{equation}
\partial _i \Phi _{a'}=\mp  
\epsilon _{a'b'c'}\,\epsilon _{ij} \Phi _{b'} \partial _j\Phi _{c'}.
\label{boucon}
\end{equation}
Here identities
\begin{equation}
\epsilon _{a'b'c'}\,\epsilon _{a'b''c''}=
\delta _{b'b''}\,\delta _{c'c''}-\delta _{b'c''}\,\delta _{c'b''}\quad
\text{and}\quad
\epsilon _{ij}\,\epsilon _{ij'}=\delta _{jj'}
\end{equation}
have been used.

A general solution to equation \eqref{boucon} was obtained by {\it 
A.A.~Belavin} 
and {\it A.M.~Polyakov} \cite{belavin75}. Here we shall only look for a 
spherically symmetric $n=1$ solution. Spherical symmetry in two spatial 
dimensions means $\epsilon _{ij} x_i\partial _j \Phi _{a'} =0$. This 
condition is consistent with the constraint on $\BPhi$. However, it has the 
undesired result that all fields satisfying it have $B(\BPhi )=0$. This is 
because the general solution to $\epsilon _{ij} x_i\partial _j \Phi _{a'} 
=0$ is\begin{equation} 
\Phi _{a'}(\Bx,t)=\tilde \Phi _{a'} (r,t),\quad \text{so that}\quad
\partial _i \Phi _{a'} =\bar x_{i} \frac{\partial \tilde \Phi 
_{a'}}{\partial r},\quad \bar x_{i}=\frac{x_i}r.
\end{equation}
Upon substituting this into \eqref{topcharge} we immediately obtain the 
result $B(\BPhi )=0$. After some symmetry requirement modification it is 
possible to obtain configurations with $B(\BPhi )\ne 0$. We refer for 
details to~\cite{balachandranbook}~(p.119-121).
\subsection{Going to 3D space}
\label{threedimcase}
The change of space dimension is a highly nontrivial action.  The existence 
of many objects and phenomena which are allowed in some dimensions are 
forbidden in another's. For example, two-dimensional creatures should have 
different digestive tract and blood circulation system, otherwise eating 
or blood circulation would divide them in two separate halves 
\cite{hawkingbook}(p.164). There also would be problems with more than 
three space dimensions, in particular with gravitational force. As a 
consequence, orbits of planets would be unstable. Here are 
theories more or less successfully describing phenomena when higher 
dimensions are introduced (string theories). The problem usually then 
becomes how to reduce these nonobservable dimensions. Our aim now is to 
construct realistic 3D nonlinear theory, with essential features inherited 
from two-dimensional $\sigma$ model.

What do we need in order to extend $\sigma$ model to real 3D space? First, 
we note that the compactification method used in nonlinear $\sigma$ model 
can be easily extended to the 3D case. Indeed, compactification $R^3$ at 
fixed time $t_0$ to $\tilde S^3$ leads to the mapping $\tilde S^3 
\rightarrow S^2$ with trivial homotopy group $\pi _3(S^2)=0$. Therefore in 
order to get nontrivial topological classes we should add one more field, 
satisfying
\begin{equation}
\Phi _1^2(\Bx,t)+\Phi _2^2(\Bx,t)+\Phi _3^2(\Bx,t)+\Phi _4^2(\Bx,t)
=1;\quad  \Bx \equiv (x_1,x_2,x_3).
\end{equation}
Then again we have $\pi _3(S^3)=\Z $. Additional field component ensures 
that field $\BPhi$ has values in the whole \SU\ manifold. Thus 
group manifold becomes natural homogeneous space for the group itself
and no identification of \GM\ with space of cosets is needed.
The additional field, however, does not eliminate the soliton stability 
problem. The simplest way to eliminate {Derrick} scaling argument (which 
excludes static stable nontrivial solution) in classical level of the 
theory\footnote{There exist stable solutions with only $\sigma$ term 
when coupling to vector mesons is included~\cite{meissner86}. There is 
discussion on the market, however, whether scale parameter or breathing 
mode quantization can stabilize the solution without the Skyrme term. For 
arguments see~\cite{jain89,kobushin95,bhaduri90}, for 
contra-arguments we refer to~\cite{asano91,iwasaki89,rajaraman86}.} 
is to add a new term in the 
Lagrangian density which would stabilize the energy~\eqref{enesca}. 

Skyrme succeeded in suggesting the following fourth-order term (ensuring  
stable soliton solution) to be added to Lagrangian density 
\eqref{nonlinsiglagran}:
\begin{gather*}
\begin{align}
\CL_{Sk}=&\frac1{32e^2}\Tr\bigl[ 
\BR_\mu,\BR_\nu\bigr]\bigl[\BR^\mu,\BR^\nu\bigr]\notag\\
=&\frac1{16e^2}\Tr 
\bigl[\BR_0,\BR_i\bigr]\bigl[\BR^0,\BR^i\bigr]+\frac1{32e^2}\Tr 
\bigl[\BR_i,\BR_j\bigr]\bigl[\BR^i,\BR^j\bigr],
\end{align}\\
\intertext{where} 
\BR_\mu=(\partial_\mu \BU )\BU^\dag;\qquad \BU(\Bx,t)\in\SU.
\end{gather*}
The contribution to {\it static soliton energy}\/ comming from the Skyrme 
term scales as
\begin{equation}
E_{Sk}(1) \rightarrow E_{Sk}(\lambda )=\lambda ^{4-D} E_{Sk}, 
\end{equation}
under a scaling transformation $\BU(\Bx)\rightarrow
\BU(\lambda \Bx)$. Requiring again that $\lambda=1$ corresponds to energy 
minimum yields the equation
\begin{equation}
E_\sigma=E_{Sk},
\label{enemincond} 
\end{equation}
in $D=3$ space dimensions. Assuming that soliton energy is proportional to 
its size $R$ and taking into account dimensions of $f_\pi$[MeV] and 
$e$[dimensionless] we conclude that the leading term ($E_\sigma$) is 
proportional to $\sim c_1 f_\pi^2 R$, whereas the Skyrme term to 
$\sim\frac{c_2}{e^2 R}$ ($c_1,c_2$ are positive constants). Thus, adjusting
soliton size $R$ equation \eqref{enemincond} can always be 
satisfied\footnote{Moreover, \eqref{enemincond} is satisfied for only one 
positive $R$ value due to the second-order algebraic equation $c_1 f_\pi^2 
e^2 R^2-c_2=0$, describing the energy extremum condition.}, for nonzero 
soliton size $R$.

We can also add terms involving more
than four derivatives (for example, $\CL_6$ and $\CL_6^\prime $ terms in 
Sec.~\ref{chaptwo}.\ref{higher}). 
There is no good argument to suggest that these terms 
are ignorable. For example, the so-called large $N_c$ limit of QCD 
\cite{witten79,thooft74} fails to show that higher derivative terms are down 
by powers of $N_c$ as compared to the leading terms. Despite these 
criticisms, we will approximate the action density by $\CL_\sigma +\CL_{Sk}$.

When the Skyrme term is also included in the Lagrangian density, there is an 
elegant lower bound to the energy of soliton. The bound is analogue of the 
Bogomol'nyi bound we considered earlier although it predates the latter by 
many years. It is based on the observation that 
\begin{equation}
\int \di^3 \Bx \Tr \Bigl( \frac{f_\pi}{2}\BR_i + \frac1{8e} 
\epsilon_{ijk}\bigl[\BR_j,\BR_k\bigr]\Bigr)^2\le0.
\end{equation}
To show this result, notice that $\BR_i$ and 
$\epsilon_{ijk}\bigl[\BR_j,\BR_k\bigr]$ are antihermitian matrices and that 
for any antihermitean matrix $\mathbf{G}\quad\Tr \mathbf{G}^2\le0$. From 
this we arrive at the bound 
\begin{equation}
-\int \di^3 \Bx \Tr\Bigl( \frac{f_\pi^2}4 \BR_i^2 +\frac1{32e^2}
\bigl[\BR_i,\BR_j\bigr]^2 \Bigr)\ge \frac{6\pi^2 f_\pi}{e} |B(\BU)| ,
\label{bound}
\end{equation}
due to Skyrme. Here $|B(\BU)|$ denotes a winding number, explicit expression 
of which is given in the next chapter (see \eqref{genbarcur}). The left 
hand side of 
\eqref{bound} is the potential energy of field $\BU(\Bx)$. The bound thus 
shows that in the presence of the Skyrme term, the soliton energy and mass 
are bounded from below. Although there is no nontrivial solution which 
saturates the bound \eqref{bound}, static solutions to field equations are 
known to exist for $|B|=1$. In Sec.~\ref{symetries} of Chapter~\ref{chaptwo}
we shall discuss the 
"spherically symmetric" static $B=1$ solution, which by $\sim 
23$\%~\cite{manton86} exceeds the bound\footnote{We shall see in 
Chapter~\ref{chapthree} that the negative quantum mass correction 
\eqref{newdefdeltam} can lower quantum Skyrmion mass. The question, however, 
can be asked, whether quantum Bogomol'nyi bound similar to \eqref{bound} can 
be defined when dynamical variables don't commute.}.

\section{QCD and the Skyrme model}
\setcounter{equation}{0}
Links between fundamental theory (QCD) and phenomenological theories of 
strong interactions (including the Skyrme model) are briefly considered here.
\subsection{Historical remarks}
{\it T.H.R.~Skyrme} proposed his model in 1961~\cite{skyrme61b}. For almost 
two decades the theory has been ignored and only in early 80-ies it has been 
realised that the model, as effective theory of mesons, may provide a link 
between QCD and the familiar picture of baryons interacting via meson 
exchange. Low energy domain of QCD becomes forbiddingly difficult due to the 
rising coupling constant which possess a major obstacle to a satisfactory 
description of the dynamical behaviour of the elementary quark and gluon 
fields of QCD at the relevant large distances. {\it R.~Rajaraman}'s 
\cite{rajaramanbook} and 
{\it E.~Witten}'s~\cite{witten79} results suggest that 
baryons may be regarded as soliton solutions of the effective meson theory 
without any reference to their quark content. This was precisely what Skyrme 
had suggested in his remarkable 
papers~\cite{skyrme61a,skyrme62,skyrme61b,skyrme58}. 
There are a lot of works analyzing one or another aspect  
of this extremely important and interesting problem. For overview we refer 
to Ref.~\cite{aitchison87} and references therein. Here we consider only 
general phenomenological requirements for effective theory of strong 
interactions and very briefly describe the $1/N_c$ expansion idea. 
Unfortunately, we completely escape chiral perturbation 
theory recently making a huge progress. This theory, however, explicitly 
involves baryon fields (when describing processes involving baryons) and is 
outside the Skyrme's idea that baryons are solitons of meson fields.
\subsection{General requirements for effective theory of 
strong interactions}
The starting point is an idealized world where $N_{\text{flavours}}=2$ or 
$3$ of the quarks are massless ($u,d$ and possibly $s$). In chiral limit 
the QCD Lagrangian exhibits a global symmetry
\begin{equation}
\underbrace{\mathrm{SU}(N_f)_L\times \mathrm{SU}(N_f)_R}_{\text{
chiral group } G}\times \mathrm{U}(1)_V\times \mathrm{U}(1)_A.
\end{equation}
At the effective hadronic level the quark number symmetry $\mathrm{U}(1)_V$ 
is realized as baryon number. The axial $\mathrm{U}(1)_A$ is not a symmetry 
at the quantum level due to the Abelian anomaly~\cite{thooft76,crewther77} 
that leads, for instance, to $M_{\eta^\prime}\neq 0$ even
 in the chiral limit.

There is compelling evidence both from phenomenology and from theory that 
the chiral group $G$ is spontaneously broken~\cite{ecker98}:
\begin{itemize}
\item Absence of parity doublets in the hadron spectrum.
\item The $N_f^2 -1$ pseudoscalar mesons are by far the lightest hadrons.
\item The vector and axial-vector spectral functions are quite different. 
\item In vector-like gauge theories like QCD (with the vacuum angle 
$\theta_{\text{QCD}})$, vector symmetries like the diagonal subgroup of 
$G$, $\mathrm{SU}(N_f)_V$, remain unbroken.
\end{itemize}
All these arguments together suggest very strongly that the chiral symmetry 
$G$ is spontaneously broken to the vector subgroup $\mathrm{SU}(N_f)_V$ 
(isospin for $N_f=2$, flavour $\mathrm{SU}(3)$ for $N_f=3$)
\begin{equation}
G\rightarrow H=\mathrm{SU}(N_f)_V.
\end{equation}
Then the Goldstone theorem tells us that there exist $N_f^2-1$ massless 
mesons. For two flavours, these Goldstone modes are identified with the 
three pions, while for three flavours, these modes are identified with the 
pseudoscalar octet. In chiral limit (when quarks have zero masses), the 
pseudoscalar mesons are exactly massless. They become massive when the 
interactions between the quark and Higgs fields are turned on, the quarks 
acquire mass and $G$ gets explicitly broken in the Lagrangian.

The effective Lagrangian emerges when we attempt to construct a model which 
describes the dynamics of these Goldstone modes. Let us list the properties 
we require for this Lagrangian in the zero quark mass 
limit~\cite{balachandran86}:
\begin{enumerate}
\item  The Lagrangian $L$ must be invariant under $G=\mathrm{SU}(N_f)_L\times
\mathrm{SU}(N_f)_R$, this property being the analogue of the $G$-invariance 
of the QCD Lagrangian. Thus $L$ is to be constructed from a multicomponent
field $\BPhi$ which is transformed by $G$, $L$ being invariant under these 
transformations.

\item Field $\BPhi$ should have exactly $N_f^2-1$ degrees of freedom per 
space-time point. This is a requirement of minimality: we want to describe 
the dynamics of the Goldstone modes and only of these modes. It is possible 
to improve effective theory by introducing vector or/and axial vector 
mesons~\cite{mattis86}, or even massive non-Abelian 
gauge bosons~\cite{adkins86}.

\item We require that the subgroup of $G$ which leaves any value of the 
field invariant is exactly (or isomorphic to) subgroup $H$ and no more. If 
this can be arranged, then we would have nicely built in spontaneous 
symmetry breakdown $G\rightarrow H$ in the geometry of the fields itself. 
\end{enumerate}
It is an easy task to check that the Skyrme model satisfies all these 
requirements~\cite{balachandranbook}. 

Chiral perturbation theory (CHPT) is also based on similar chiral symmetry 
principles. Generally, many more chiral invariant terms can be included 
into the Lagrangian. The Skyrme model, basically, takes only two of them 
$L_2$ and $L_4$. 
CHPT, on the other hand, provides us with a scheme which tells us which 
terms should be included and which ones should not. Roughly speaking, the 
essential idea of chiral perturbation theory is to realize that at low 
energies the dynamics should be controlled by the lightest particles, the 
pions, and the symmetries of QCD. Therefore, S-matrix elements, i.e. 
scattering amplitudes, should be expandable in Taylor-series of the pion 
momenta and masses\footnote{In the baryon sector one has an additional 
parameter --- the nucleon mass.}, which is also consistent with chiral 
symmetry. This scheme is valid until one encounters a resonance, such as the 
$\rho$-meson, which corresponds to a singularity of the S-matrix. It should 
be stressed, however, that chiral perturbation theory is not a perturbation 
theory in the usual sense, i.e. it is not a perturbation theory in the QCD 
coupling constant. In this respect, it is actually a nonpertubative method, 
since it takes infinitely many orders of the QCD coupling constant in order 
to generate a pion. In the meson sector CHPT is quite successful, whereas 
the precision achieved in heavy baryon CHPT is not comparable to the meson 
sector accuracy. For explanation we refer to lectures~\cite{ecker98}. 

\subsection{The $1/N_c$ expansion}
Assuming confinement, the asymptotic states of QCD are not the coloured 
quarks and gluons, but rather the observed colour singlet hadrons. In view 
of this, one might wonder whether in some way QCD itself could not be 
equivalently formulated in terms of these observed asymptotic degrees of 
freedom. Quite remarkably, the work of {\it G.~'t Hooft}~\cite{thooft74} and 
{\it E.~Witten}~\cite{witten79} 
shows that QCD is indeed equivalent --- in the full 
field theory sense --- to a theory of mesons and glueballs\footnote{There 
exist at least few effective field theories in four dimensions, as the 
number $N$ of fields of some type becomes large~\protect\cite{weinberg97}.}, 
with meson-meson coupling constant $\sim 1/\sqrt{{N_c}}$. From the 
first sight, there seems to be one very large gap in the equivalence
\begin{equation}
\text{confined QCD}\equiv \text{theory of mesons and glueballs},\notag
\end{equation}
namely, where are the baryons? It is here that the real interest of the 
$1/N_c$ idea lies. 
{\it E.~Witten} showed ~\cite{witten79} that for large $N_c$ 
baryon masses scale like $N_c$. This is reminiscent of the behaviour of 
solitons in a theory in which the coupling constant is $g$: the soliton mass 
is $\sim 1/g^2$, 
so that putting $g\sim 1/\sqrt{{N_c}}$ we find mass $\sim N_c$.
But this interpretation is exactly what Skyrme suggested.

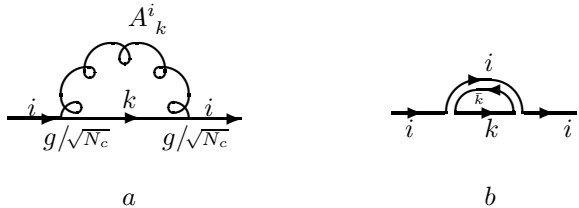
\begin{figure}
\begin{center}
\THICKLINES
\begin{picture}(25000,9000)(-5000,-4000)
\put(18000,-3300){$b$}
\put(4300,-3300){$a$}
\drawline\fermion[\E\REG](0,0)[2000]
\drawloop\gluon[\N 5](\pbackx,\pbacky)
\drawline\fermion[\E\REG](\pbackx,\pbacky)[2000]
\drawarrow[\E\ATTIP](\pbackx,\pbacky)
\drawline\fermion[\W\REG](\pbackx,\pbacky)[7000]
\drawarrow[\E\ATTIP](\pbackx,\pbacky)
\global\advance\pbackx by -550
\global\advance\pbacky by -1000
\put(\pbackx,\pbacky){$\displaystyle g/\!{\scriptstyle \sqrt{N_c}}$}
\global\advance\pbackx by 4500
\put(\pbackx,\pbacky){$\displaystyle g/\!{\scriptstyle \sqrt{N_c}}$}
\put(700,100){$i$}
\put(4300,400){$k$}
\drawarrow[\E\ATTIP](5000,0)
\put(7400,100){$i$}
\put(4500,3500){$A^i_{\phantom{i}k}$}

\put(18000,200){\oval(2900,2500)[t]}
\put(18000,200){\oval(2200,1800)[t]}
\drawarrow[\W\ATTIP](18000,1080)
\drawarrow[\E\ATTIP](18500,200)
\drawline\fermion[\E\REG](19540,200)[2000]
\put(21000,-700){$i$}
\put(18000,-700){$k$}
\put(15000,-700){$i$}
\put(17600,540){$\scriptscriptstyle \bar k$}
\put(18000,1800){$i$}
\drawline\fermion[\W\REG](19200,200)[2300]
\drawarrow[\E\ATTIP](15700,200)
\drawarrow[\E\ATTIP](20700,200)
\drawline\fermion[\W\REG](16520,200)[2000]
\drawarrow[\E\ATTIP](18000,1440)
\end{picture}
\end{center}
\caption{Gluon correction to quark propagator in  standard ($a$) and { 
't~Hooft-Witten} ($b$) notations.}
\label{gluoncorrection}
\end{figure}
Let us now briefly explain why meson-meson coupling constant should scale 
like $\sim 1/N_c$. To this end let us reformulate QCD for an arbitrary 
number of colours $N_c$. In such a theory there are $N_c$ quark degrees of 
freedom and $N_c^2-1\approx N_c^2$ gluonic degrees of freedom (for large 
$N_c$). Consider then the simple gluonic correction to the quark propagator 
depicted in Fig.~\ref{gluoncorrection}${}_a$. 
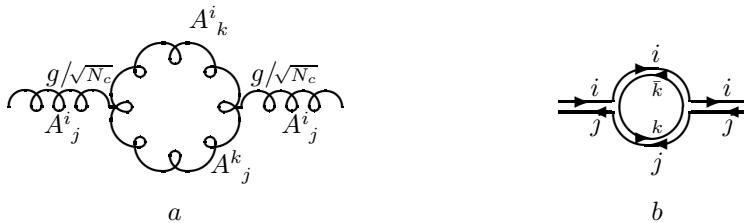
\begin{figure}
\begin{center}
\THICKLINES
\begin{picture}(25000,10500)(-5000,-6500)
\put(18000,-4300){$b$}
\put(-300,-4300){$a$}
\drawloop\gluon[\NE 0](0,0)
\frontstemmed\drawline\gluon[\E\CENTRAL](\loopbackx,\loopbacky)[3]
\global\advance\loopbackx by 600
\global\advance\loopbacky by 1000
\put(\loopbackx,\loopbacky){$\displaystyle g/\!{\scriptstyle \sqrt{N_c}}$}
\put(500,3000){$A^i_{\phantom{i}k}$}
\put(4000,-1000){$A^i_{\phantom{i}j}$}
\put(1300,-2500){$A^k_{\phantom{k}j}$}
\put(-5000,-1000){$A^i_{\phantom{i}j}$}
\frontstemmed\drawline\gluon[\W\FLIPPEDCENTRAL](\loopfrontx,\loopfronty)[3]
\global\advance\loopfrontx by -2700
\global\advance\loopfronty by 1000
\put(\loopfrontx,\loopfronty){$\displaystyle g/\!{\scriptstyle 
\sqrt{N_c}}$}
\put(18000,0){\circle{2500}}
\put(18000,200){\oval(2900,2500)[t]}
\put(18000,-200){\oval(2900,2500)[b]}
\drawline\fermion[\E\REG](19540,200)[2000]
\drawline\fermion[\E\REG](19540,-200)[2000]
\drawarrow[\E\ATTIP](15700,200)
\drawarrow[\E\ATTIP](20700,200)
\drawline\fermion[\W\REG](16520,200)[2000]
\drawline\fermion[\W\REG](16520,-200)[2000]
\drawarrow[\W\ATTIP](20700,-200)
\drawarrow[\W\ATTIP](15700,-200)
\drawarrow[\E\ATTIP](18000,1440)
\drawarrow[\W\ATTIP](18000,-1400)
\drawarrow[\W\ATTIP](18000,1220)
\drawarrow[\E\ATTIP](18000,-1180)
\put(18000,1700){$i$}
\put(18000,400){$\scriptstyle \bar k$}
\put(18000,-1000){$\scriptstyle k$}
\put(18000,-2400){$j$}
\put(15700,400){$i$}
\put(20700,400){$i$}
\put(20700,-900){$j$}
\put(15700,-900){$j$}
\end{picture}
\end{center}
\caption{Gluon loop in standard ($a$) and {'t~Hooft-Witten} 
($b$) notations.}
\label{gluonvertex}
\end{figure}

Even after we specify the colour index of the external quark, this diagram 
receives a combinatorial factor of $N_c$ corresponding to the $N_c$ possible 
values for the index of the internal quark. This is easy to see in 
't~Hooft-Witten notations Fig.~\ref{gluoncorrection}${}_b$, where gluon in 
combinatorial sense (and only in this sense) is equivalent to 
quark-antiquark pair. The resulting loop corresponds to the summation over 
all possible quark index values and is responsible for the combinatorial 
factor $N_c$. 

If we want the theory to have a smooth --- but nontrivial --- limit as 
$N_c\rightarrow\infty$ we must compensate this combinatorial factor. Thus we 
require that the $\bar q qg$ vertex scale like $\sim 1/\sqrt{N_c}$. The same 
result is obtained for the trilinear meson-meson (gluon-gluon) coupling 
constant as can be seen from diagrams Fig.~\ref{gluonvertex}${}_a$ and 
Fig.~\ref{gluonvertex}${}_b$.

{\it G.~'t~Hooft} noted that not all diagrams are of the same 
significance when $N_c\rightarrow\infty$. Simple power counting, similar to 
those just described, implies~\cite{thooft74} that in this limit only {\it 
planar} diagrams become important. Analysis of all planar  
diagrams~\cite{thooft74}, which is out of the scope of this work, together 
with confinement assumption of large $N_c$ QCD leads to the following 
conclusions:
\begin{itemize}
\item Mesons are stable and to leading order non-interacting particles. 
Their number is finite.
\item Amplitudes of elastic meson-meson scattering are of the order of $\sim 
1/N_c$ and are expressed as sum only of tree level diagrams\footnote{Tree 
level diagrams in this case describe only meson (but not gluon 
or quark) exchange.}.
\item 
Baryon-meson  scattering can also be analysed in similar 
fashion~\cite{mattisphd,mattis84,mattis89,mattis89a,mattis90}.
\end{itemize}
In other words, QCD seems to reduce smoothly to an effective theory of mesons
(and glueballs) with the effective coupling constant of the order of 
$\sim 1/N_c$.

\chapter{Classical Skyrme model}
\label{chaptwo}
The chapter deals with classical Skyrme model. Classical model assumes 
that dynamical model variables and its time derivatives 
commute. The model is usually formulated in rotation $R_{ij}$
and Pauli $\Btau$ matrix formalism, illustrated in 2D nonlinear $\sigma$ 
model. Wigner $\BD^j$ matrices and \su\ algebra generators $\BJ$ represented 
in circular basis are, nevertheless, more convenient for model 
formulation in arbitrary reducible representation and, therefore, 
will be followed in this and subsequent chapters.
\section{Formulation}
\label{formulation}
\setcounter{equation}{0}
\setcounter{table}{0}
\setcounter{figure}{0}
The section serves as a formulation of the classical 
SU(2) Skyrme model in arbitrary irreducible representation. 
An emphasis is put on expression dependence on representation.
Physical quantities (mass, coupling constants, etc.) are independent of the 
representation after the proper model parameter renormalization is 
employed.
\subsection{Parametrization of the symmetry group}
\label{introduction}
Chiral group is a group of transformations in the internal (isotopic) space, 
under action of which left and right states transform independently. 
Simple nonabelian (six-parameter) chiral group is obtained by multiplying
two rotation groups directly $\SU_{L_{eft}}\otimes \SU_{R_{ight}}$. There 
is no linear realization of the group in 3D isospace. One can choose either 
to extend the isospace to 4D, where the linear representation exists or to 
construct the nonlinear representation. The natural nonlinear representation 
(which we follow further in the work) is obtained when the group parameters 
space (manifold) is identified with the space where the abstract group 
transformations are realized. For example, \SU\ matrix in well-known 
Euler-Rodrigues parametrization $a_0,\Ba$ takes a form\footnote{
{T.H.R.~Skyrme}
\protect\cite{skyrme61b} formulated his model in terms of 
$\Bsigma$ and pion \Bpi-fields 
$\BU=(\Bsigma+i\Btau\cdot\Bpi)$.}~\cite{bidenkharnbook}\begin{equation}
\BU(a_0,\Ba)=
\begin{pmatrix}
a_0+i a_3&i a_1+a_2\\
i a_1-a_2&a_0-i a_3
\end{pmatrix}
=a_0\cdot\mathbf{1}+i \Ba\cdot\Btau,
\label{rodpar}
\end{equation}
where the group parameters space\footnote{Sometimes the set $a_0,\Ba$ is 
called a 4-isovector. The reader should be aware that because of the 
constraint \protect\eqref{aconstr} the set isn't a vector 
space.} $a_0,\Ba$
itself is restricted by the constraint
\begin{equation}
a_0^2+\Ba^2=1;\qquad a_0,\Ba \in \R.
\label{aconstr}
\end{equation}

The presence of constraint \eqref{aconstr} gives rise to additional problems 
in quantization of the theory. From our point of view  unconstrained 
parameters are more suitable for this purpose, but see~\cite{cebula93}. 
Such an unconstrained parameters are, for example, triple of Euler angles 
$\Balpha\equiv(\alpha^1,\alpha^2,\alpha^3)$~\cite{varshalovichbook}
\begin{equation}
0\leq \alpha^1 < 2\pi,\quad 0\leq \alpha^2 \leq \pi,\quad 0\leq
\alpha^3 <4\pi.\label{defeulang}
\end{equation}
An arbitrary reducible \SU\ matrix in the Euler angles parametrization can 
be expressed as a direct sum of Wigner $\BD^j(\Balpha),\quad 
(j=\frac12,1,\frac32,\dotsc$) functions. 
\subsection{The Lagrangian}
For reasons of simplicity and without 
lose of generality\footnote{Formulation of the model in arbitrary reducible 
representation simply involves summation over representation $j$ and is 
explained in Sec.~\ref{redrep}.} let us formulate the model in the 
arbitrary {\em irreducible}\/ \SU\ representation $j$. Euler angles 
$\Balpha(\mathbf{r},t)$ become the functions of space-time point 
$(\mathbf{r},t)$ and form the dynamical variables of the theory. Model 
is formulated in terms of unitary field
\begin{equation} 
\mathbf{U}(\Br,t)=\BD^j\bigl(\Balpha(\Br,t)\bigr),\label{defu}
\end{equation}
all physical quantities being functions of this field \BU. In the quark 
picture the analogue of $U^{ij}$ is the complex $2\times 2$ matrix 
$\bar q^i\frac{1-\gamma_5}2 q^j$, corresponding to 
pseudoscalar mesons~\cite{zahed86}. Note that this analogue is only valid in 
the fundamental representation of \SU. Unitary field can also be expressed 
in terms of pion fields $\Bpi$ and unphysical $\Bsigma$ field
\begin{equation}
\BU(\Bx)=\frac1{f_\pi}\bigl(\Bsigma +i\Btau\cdot\Bpi\bigr).
\label{interppi}
\end{equation}

The basic Skyrme model is described by chirally symmetric Lagrangian 
density\footnote{We consider chiral transformations in detail in 
Sec.~\ref{symetries} and Sec.~\ref{noethercurr} of Chapter~\ref{chapthree}.}
\begin{equation}
{\CL}=-\frac{f_\pi^2}{4} \Tr\{\BR_\mu \BR^\mu\} 
+\frac1{32e^2}\Tr\{[\BR_\mu,\BR_\nu][\BR^\mu,\BR^\nu]\}
,\label{genlagden}
\end{equation}
where the "right" current\footnote{
The theory as well can be formulated in terms of "left" current
$\mathbf{L}_\mu=\BU^\dag(\partial_\mu \BU )$. Recent political tendency 
rendered "right" current more popular, what we believe defined our 
choice.} $\BR_\mu $, known for mathematicians as {Maurer-Cartan} form, 
is defined as
\begin{equation}
\BR_\mu=(\partial_\mu \BU )\BU^\dag,
\end{equation}
$f_\pi$ (pion decay constant) and $e$ being parameters\footnote{
Note, however, that the parameter $f_\pi$ value cannot be determined 
within the framework of strong  interactions only, because pions are by 
far the lightest strongly interacting particles and, thus, are stable in 
this theory. Experimental value of $f_\pi$ is $93$~MeV.
It is claimed~\cite{adkins86} that parameter $e$ value can be
extracted from the $\pi\pi$ scattering data using formulas given in 
Ref.~\cite{donoghue84}. The result is $e=7.4$. The Skyrme constant $e$ also 
has been roughly estimated by assuming that the Skyrme term arises by 
"integrating out" the effects of a $\rho (770)$ meson; this yields 
$e=m_\rho/(2f_\pi)=5.83$~\cite{jain86}.} of 
the theory. Let us explore the Lagrangian \eqref{genlagden} algebraic 
structure more closely. To this end it is convenient to introduce a
contravariant circular coordinate system. The unit vector $\bar\Bx$ in these 
(contravariant) circular coordinates is defined in respect to Cartesian, 
spherical and circular covariant coordinate systems as\begin{subequations}
\label{cirsys}
\begin{alignat}{4}
&x^{+1}&=&-\frac1{\sqrt{2}}(x_1-ix_2)&\quad 
=&-\frac1{\sqrt{2}}\sin\vartheta \e^{-i\varphi}&\quad =&-x_{-1},\\
&x^0&=&x_3&\quad =&\cos\vartheta &\quad =&x_0,\\
&x^{-1}&=&\frac1{\sqrt{2}}(x_1+ix_2)&\quad =&\frac1{\sqrt{2}}\sin\vartheta 
\e^{i\varphi}&\quad =&-x_{+1},
\end{alignat}
\end{subequations}
respectively. Then the general inner (scalar) product of two algebra 
elements can be defined as
\begin{equation}
\Tr\langle j{\cdot}|J_a J_b|j{\cdot}\rangle 
=(-1)^a\frac16j(j+1)(2j+1)\delta_{a,-b},\label{tratwo}
\end{equation}
where the \su\ generators $\BJ$  satisfy the commutation 
relation 
\begin{equation}
[J_a,\, J_b]=\Bigl[
\begin{matrix}
1 & 1 & 1\\a & b &c
\end{matrix}
\Bigr] J_c;\qquad c=a+b.\label{comrel}
\end{equation}
The factor on the r.h.s. in \eqref{comrel} is the Clebsch-Gordan 
coefficient $(1a\ 1b|1c)$ in a more transparent notation.
Wigner $\BD^j$ function parametrization in the form
\begin{equation}
\langle j{\cdot}|\BD^j(\Balpha)|j{\cdot}\rangle =\langle 
j{\cdot}|\exp\bigl(i\sqrt{2}\,\alpha^1J_0\bigr)\,\exp\bigl(-\alpha^2(J_{+} 
+J_{-})\bigr)\,\exp\bigl(i\sqrt{2}\,\alpha^3 J_0\bigr)|j{\cdot}\rangle,
\end{equation}
makes it easy to obtain the following relations:
\begin{subequations}
\label{pdrel}
\begin{align}
\frac{\partial}{\partial 
\alpha^i}\,D_{mn}^{j}(\Balpha)&=C_i^{(a)}(\Balpha)\,\langle jm| 
J_a|jm'\rangle \,D_{m'n}^{j}(\Balpha),
\\
\frac{\partial}{\partial 
\alpha^i}\,D_{mn}^j(-\Balpha )&=-C_i^{(a)}(\Balpha
)D_{mn^{\prime }}^j(-\Balpha )\left\langle jn^{\prime }\left| J_a\right|
jn\right\rangle ,
\\
\frac{\partial}{\partial 
\alpha^i}\,D_{mn}^j(\Balpha )&=C_i^{\prime (a)}(\Balpha
)D_{mm^{\prime }}^j(\Balpha )\left\langle jm^{\prime }\left| J_a\right|
jn\right\rangle ,
\\
\frac{\partial}{\partial 
\alpha^i}\,D_{mn}^j(-\Balpha )&=-C_i^{\prime (a)}(\Balpha
)\left\langle jm\left| J_a\right| jn^{\prime }\right\rangle D_{n^{\prime
}n}^j(-\Balpha ),
\end{align}
\label{ddifeq}
\end{subequations}
where the coefficients 
\begin{subequations}
\begin{align}
C^{(a)}_i(\Balpha)&=D^1_{a,a^\prime}(\Balpha)\,C^{\prime 
(a^\prime)}_i(\Balpha)&\Balpha&\equiv(\alpha_1,\alpha_2,\alpha_3),\\
C^{\prime 
(a)}_i(\Balpha)&=D^1_{a,a^\prime}(-\Balpha)\,C^{(a^\prime)}_i(\Balpha)
&-\Balpha&\equiv(-\alpha_3,-\alpha_2,-\alpha_1),
\end{align}
\end{subequations}
have the explicit form~\cite{norvaisas94}
\begin{subequations}
\begin{alignat}{6}
&C_1^{(+)}(\Balpha)&=&0&\quad &C_2^{(+)}( 
\Balpha)&=&-\e^{-i\alpha^1}&\quad &C_3^{(+)}(\Balpha)&=&-i \sin \alpha^2\, 
\e^{-i\alpha^1},\\
&C_1^{(0)}(\Balpha)&=&i\sqrt{2}&\quad 
&C_2^{(0)}(\Balpha)&=&0&\quad &C_3^{(0)}(\Balpha)&=&i\sqrt{2}\cos 
\alpha^2,\\
&C_1^{(-)}(\Balpha)&=&0&\quad &C_2^{(-)}( 
\Balpha)&=&-\e^{i\alpha^1}&\quad &C_3^{(-)}(\Balpha)&=&i \sin \alpha^2\, 
\e^{i\alpha^1},
\end{alignat}
\label{defccoef}
\end{subequations}
and satisfy orthogonality relations
\begin{subequations}
\begin{alignat}{3}
&\sum_m C^{(m)}_i(\Balpha) C^j_{(m)}(\Balpha)&=&\sum_m 
C^{\prime(m)}_i(\Balpha) C^{\prime 
j}_{(m)}(\Balpha)&=&\delta_{i,j},\\
&\sum_i C^{(m)}_i(\Balpha) C^i_{(n)}(\Balpha)&=&\sum_i 
C^{\prime(m)}_i(\Balpha) C^{\prime i}_{(n)}(\Balpha)&=&
\delta_{m,n}.
\label{cpriortreltwo}
\end{alignat}
\end{subequations}
Using formulas \eqref{pdrel} the right current $\mathbf{R}_\mu$ can be 
reduced to the form
\begin{equation}
(R_\mu)_{mm'}=\partial_\mu \alpha^i C_i^{(a)}(
\Balpha)\,\langle jm|J_a|jm'\rangle ,
\label{rigcur}
\end{equation}
and clearly have values in \su\ algebra.  Relations 
\eqref{tratwo}, \eqref{comrel} together with formula \eqref{rigcur}
allow us to express the Lagrangian density \eqref{genlagden} in 
terms of the Euler angles~\cite{norvaisas94}
\begin{equation}
\begin{split}
\CL=&\frac13 j(j+1)(2j+1)\bigg( \frac{f_\pi^2}4\Bigl( 
\partial_\mu\alpha^i\partial^\mu\!\alpha^i
+2\cos \alpha^2\, \partial_\mu \alpha^1 \, \partial^\mu\!\alpha^3\Bigr)
\\
&
-\frac1{16e^2}\Bigl( \partial_\mu \alpha^2\partial^\mu
\!\alpha^2(\partial_\nu \alpha^1 \partial^\nu\!\alpha^1+\partial_\nu
\alpha^3\partial^\nu\!\alpha^3)
-(\partial_\mu \alpha^1 \partial^\mu\!\alpha^2)^2\bigskip
\\
&\phantom{-\frac1{16e^2}\Bigl(}
-(\partial_\mu \alpha^2 \partial^\mu\!\alpha^3)^2
+\sin^2\!\alpha^2(\partial_\mu \alpha^1 \partial^\mu\!\alpha^1 \,
\partial_\nu \alpha^3 \partial^\nu\!\alpha^3-(\partial_\mu \alpha^1
\partial^\mu\!\alpha^3)^2)\bigskip
\\
&\phantom{-\frac1{16e^2}\Bigl(}
+2\cos \alpha^2(\partial_\mu \alpha^2 \partial^\mu\!\alpha^2\,
\partial_\nu \alpha^1 \partial^\nu\!\alpha^3-\partial_\mu \alpha^1
\partial^\mu\!\alpha^2 \partial_\nu \alpha^2\partial^\nu
\!\alpha^3) \Bigr) \bigg).
\end{split}\label{parlagden}
\end{equation}
The only dependence on the dimension of the representation
is in the overall factor $j(j+1)(2j+1)$ as it could 
be expected from \eqref{tratwo}. This implies that the equation of motion 
for the dynamical variable \Balpha\ is independent of the dimension of the 
representation. 
\newline
{\bf Note.} We introduce additional normalization factor $1/N$ in the 
definition of quantum Skyrme Lagrangian in Chapter~\ref{chapthree}. The 
motivation comes from considerations below.
\subsection{The topological current}
The following construction of "right" currents is called topological 
current density (cf.~\eqref{topcurrent1} and \eqref{nonlinsigmodcurrent}):
\begin{equation}
\CB^\mu=\frac{1}{3\cdot8N\pi^2} \epsilon^{\mu \nu \beta \gamma} \Tr 
\{\BR_\nu \BR_\beta\BR_\gamma\}.
\label{genbarcur}
\end{equation}
The integral associated with \eqref{genbarcur} is a conserved quantity.
The normalization factor $N$ depends on the dimension of the
representation and has the value $1$ in the fundamental 
($j=\frac12$) representation. The baryon number\footnote{Topological index 
(due to its conservation) is identified with baryon number in the Skyrme 
model. The following expressions are used as synonyms in the physical
literature: topological index, {Chern-Pontryagin} index, winding 
number, soliton number, particle number, baryon number.}
${\goth B}$ is obtained as 
the spatial integral of the time component $\CB^0$. In terms of 
Euler angles $\Balpha$ the baryon current density takes the form
\begin{equation}
\CB^\mu=-\frac{1}{3\cdot6\cdot8N\pi^2}j(j+1)(2j+1)\sin \alpha^2 
\,\epsilon^{\mu \nu \beta\gamma}\  \epsilon_{ikl}\,\partial_\nu 
\alpha^i\partial_\beta\alpha^k\partial_\gamma \alpha^l.
\label{parbarcur}
\end{equation}
As the dimensionality of the representation appears in this expression
in the same overall factor as in the Lagrangian density \eqref{parlagden} 
it follows that all calculated dynamical observables will be independent
of the dimension of the representation at the classical level. The 
same overall factor in \eqref{parbarcur} as in \eqref{parlagden} and 
\eqref{tratwo} also indicates that the topological (or baryon) current 
density $\CB^\mu$ can be expressed in terms of scalar product of algebra 
elements. This is indeed the case~\cite{fadeev76}
\begin{equation}
\CB^\mu\sim\epsilon^{\mu\nu\rho\sigma}\Tr\{\bigl[\BR_\nu, 
\BR_\rho\bigr]\BR_\sigma \}.\label{fadbarcurden}
\end{equation}
The forms \eqref{genbarcur}, \eqref{fadbarcurden} make no difference 
for baryon number $B$ in the quantum case. This can be seen both from 
\eqref{genbarcur} and \eqref{fadbarcurden} as time derivatives $\BR_0$  
are not involved in the expressions. A more symmetric form 
\eqref{genbarcur} is usually used.
\subsection{The hedgehog ansatz}
\label{hedgehog}
\begin{figure}
\begin{center}
\includegraphics*{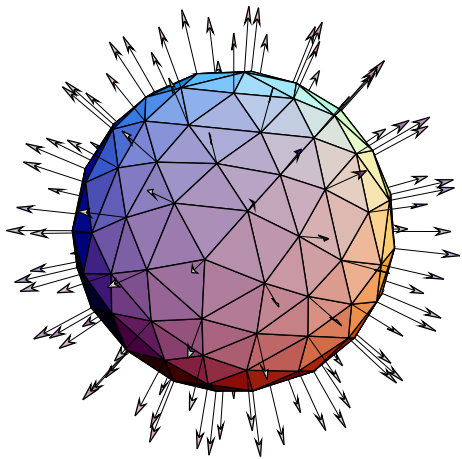}
\end{center}
\caption{Hedgehog configuration. Arrows indicate the directions of the 
isovector field $F(r)$ at different points in coordinate space.}
\label{fig8}
\end{figure}

The general solution of equations of motion, which follows from 
variation of the Lagrangian \eqref{parlagden}, is not found. 
Skyrme suggested the {\it static soliton solution} in 
the fundamental representation of SU(2)
\begin{equation}
\BU_0=\mathrm{e}^{i(\Btau\cdot \bar \Br) F(r)}.\label{skyanz}
\end{equation}
Here $\Btau$ is isovector of Pauli-isospin matrices and  $\bar \Br$ 
denotes unit spatial vector. The object described 
by \eqref{skyanz} has a very peculiar geometric structure 
(see Fig.~\ref{fig8}): at each point $\Bx$ in 3D space the associated 
isovector $F(r)$ points in a radial direction with respect to the spatial 
origin $\Bx=0$, where the centre of the object is located. This radial 
structure has prompted the handy name of "hedgehog" for the configuration 
\eqref{skyanz}.
In order to find its generalizations for representations of higher 
dimension one may compare it to the matrix elements 
$D_{mn}^{1/2}(\Balpha)$, and thus obtain  the explicit expressions 
for the Euler angles $\Balpha$ in terms of the chiral angle $F(r)$. The 
result\footnote{When $\varphi,\vartheta,F(r)$ in \eqref{eultopol} run over
values $0\le\varphi< 2\pi,\quad 0\le\vartheta\le\pi ,\quad 0\le F(r)<\pi$,
the range of $\alpha_1,\alpha_2,\alpha_3$ is
$\quad -\pi\le\alpha_1< 2\pi,\quad -\pi\le\alpha_2\le 0 ,\quad
-2\pi\le\alpha_3 <\pi$ and thus differs from \eqref{defeulang} range.
This, however, can be fixed by dividing the parameters area in  a proper way 
and moving each part by some fraction of $\pi$.} is~\cite{norvaisas94}
\begin{subequations}
\label{eultopol}
\begin{align}
\alpha^1&=\varphi-\arctan (\cos \vartheta \, \tan F(r))-\pi/2,\\
\alpha^2&=-2\arcsin(\sin \vartheta\, \sin F(r)), \\
\alpha^3&=-\varphi-\arctan(\cos \vartheta\, \tan F(r))+\pi/2.
\end{align}
\end{subequations}
Here the angles $\varphi, \vartheta$ are the polar angles that define the
direction of the unit vector $\bar \Br$ in spherical coordinates.

Substitution of the expressions \eqref{eultopol} into the general 
expression \eqref{defu} for the unitary field \BU\ then gives the 
hedgehog field in a representation with arbitrary $j$. As an example, the 
hedgehog field in the representation $j=\frac12$ has the form 
\cite{norvaisas94,varshalovichbook}
\begin{equation}
\BU_c^{\frac12}=
\begin{pmatrix} 
G&i\sin F\sin\vartheta \mathrm{e}^{-i\varphi}\\
i\sin F\sin\vartheta \mathrm{e}^{i\varphi}&G^*
\end{pmatrix},
\end{equation}
where we have used abbreviation $G=\cos F+i\sin F\cos\vartheta $. For $j=1$ 
the same substitution yields
\begin{equation}
\BU_c^{1}=
\begin{pmatrix} 
G^2&
i\sqrt{2} \sin F\sin\vartheta \mathrm{e}^{-i\varphi}G
&-(\sin F\sin \vartheta e^{-i \varphi})^2\\
i\sqrt{2} \sin F\sin\vartheta \mathrm{e}^{i\varphi}G
&1-2\sin^2 F\sin^2\vartheta
&i\sqrt{2} \sin F\sin\vartheta \mathrm{e}^{-i\varphi}G^*\\
-(\sin F\sin \vartheta e^{i \varphi})^2&
i\sqrt{2} \sin F\sin\vartheta \mathrm{e}^{i\varphi}G^*&
G^{*2}
\end{pmatrix}.
\end{equation}

The Lagrangian density \eqref{parlagden} reduces to the following simple 
form, when the hedgehog ansatz \eqref{eultopol} is employed:
\begin{multline}
{\CL}=-\frac43j(j+1)(2j+1)\bigg( \frac{f_\pi^2}4\Bigl( 
F^{\prime 2}+\frac{2}{r^2}\sin^2 \!F\Bigr)
\\
+\frac{1}{4e^2}\frac{\sin^2 \!F}{r^2}\Bigl( 2F^{\prime 2}+\frac{\sin^2 
\!F}{r^2}\Bigr) \bigg).
\end{multline}
For $j=\frac12$ this reduces to the result of Ref.~\cite{adkins83}. The 
corresponding mass density is obtained by reverting the sign of \CL, as the 
hedgehog ansatz is a static solution.

The requirement that the soliton mass be stationary yields the
following equation for the chiral angle $F(r)$~\cite{adkins83}:
\begin{multline}
f_\pi^2\Bigl( F^{\prime\prime}+\frac{2}{r}F^\prime-\frac{\sin 2F}{r^2}\Bigr) 
-\frac{1}{e^2}\Bigl(\frac{1}{r^4}\sin^2\!F\,\sin 2F
\\
-\frac{1}{r^2} (F^{\prime 2}\sin 2F+2F^{\prime\prime} \sin^2\!F)\Bigr) =0.
\label{clachiang}
\end{multline}
It is independent of the dimension of the representation. Note that
the differential equation is nonsingular only if $F(0)=n\pi,\quad n\in \Z$.

For the hedgehog form the baryon density reduces to the expression
\begin{equation}
\CB^0=-\frac{1}{3N\pi^2}j(j+1)(2j+1)\frac{\sin^2 F}{r^2} F'.
\end{equation}
The corresponding baryon number is
\begin{equation}
{B}=\int \di^3 r\CB^0=\frac{2}{3N\pi} 
j(j+1)(2j+1)\bigl(F(0)-\frac12\sin2F(0)\bigr).
\end{equation}
Combining the requirement that $F(0)$ to be an integer multiple of $\pi$
with the requirement that the lowest nonvanishing baryon number to be $1$
gives the general expression for the normalization factor $N$ as
\begin{equation}
N=\frac23 j(j+1)(2j+1).
\label{norfac}
\end{equation}

The equation of motion for chiral angle in the form \eqref{clachiang}
depends on parameter $f_\pi$ and $e$ values. It is convenient to 
introduce a dimensionless variable $\rt =ef_\pi r$ in which \eqref{clachiang}
takes the form
\begin{multline}
F^{\prime \prime}(\rt)\Bigl(1+\frac{2\sin^2\! F(\rt)}{\rt^2}\Bigr)+
F^{\prime 2}(\rt)\frac{\sin 2F(\rt)}{\rt^2}+\frac{2}{\rt}F^\prime (\rt)\\
-\frac{\sin 2F(\rt)}{\rt^2}-\frac{\sin 2F(\rt) \sin^2\! F(\rt)}{\rt^4}=0.
\label{redclachiang}
\end{multline}
\begin{figure}
\begin{center}
\includegraphics*{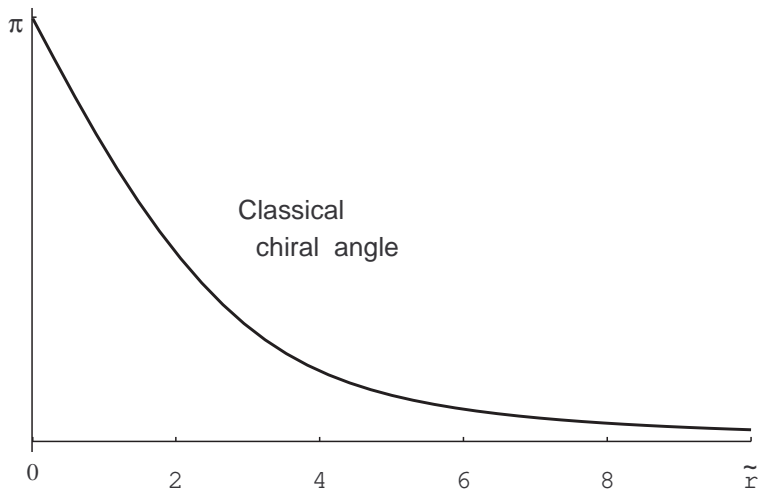}
\end{center}
\caption{Classical chiral angle solution, as taken from 
Ref.~\cite{adkins83} ($e=5.45$, $F_\pi=129$~MeV, $\tilde r=e F_\pi r$).}
\label{fig9}
\end{figure}
Numerical investigation~\cite{skyrme62} of \eqref{redclachiang}
leads to classical chiral angle solution $F(\rt)$ shown in 
Fig.~\ref{fig9}, when boundary conditions $F(0)=\pi,\quad F(\infty)=0$ 
ensuring baryon number $1$ are imposed.
\subsection{Higher order terms}
\label{higher}
There exists an infinite class of alternate stabilizing terms for the
Lagrangian density \eqref{genlagden}, combinations of which can be used
in place of Skyrme's quartic stabilizing term or be added to it
\cite{marleau92}. An alternate term of quartic order (which for $j=1/2$
yields the same result as the Skyrme term) is~\cite{pari91}
\begin{equation}
{\CL}'_4=\frac1{16 e^{\prime 2}}\Bigl(\Tr \{\BR_\mu \BR_\nu\}
\Tr\{\BR^\mu \BR^\nu\}-(\Tr \{\BR_\mu
\BR^\mu\})^2\Bigr).\label{lagpri}
\end{equation}
When this term is expressed in terms of the Euler angles  
\eqref{defeulang}, the resulting Lagrangian density has the 
form \eqref{parlagden}, with the exception that the stabilizing term that 
is proportional to $e^{-2}$ has an additional factor $\frac23j(j+1)(2j+1)$ 
\cite{norvaisas94}. Hence invariance of the physical predictions 
requires that the parameter $1/e^{\prime 2}$ of the stabilizing term 
\eqref{lagpri} be taken to be proportional to 
$\bigl(j(j+1)(2j+1)\bigr)^{-2}$, and the 
parameter $f_\pi$ of the quadratic term to be proportional to 
$\bigl(j(j+1)(2j+1)\bigr)^{-1}$, 
when a representation of dimension $2j+1$ is 
employed. Thus, $\CL_4^\prime$ has different representation dependence from 
a similar term in \eqref{genlagden}.

Consider then the sixth order stabilizing term~\cite{marleau90,marleau92}
\begin{equation}
{\CL}_6=e_6\Tr\{[\BR_\mu,\BR^\nu][\BR_\nu,
\BR^\lambda][\BR_\lambda,\BR_\mu]\}.\label{lagsix}
\end{equation}
In terms of the Euler angles $\Balpha$ this Lagrangian
density takes the form~\cite{norvaisas94}
\begin{multline}
{\CL}_6=-e_6{\frac{j(j+1)(2j+1)}{6}}\epsilon_{i_1i_2i_5}\,\epsilon_{i_3i_4i_6}\, 
\sin^2\!\alpha^2
\\
\times \partial_\mu\alpha^{i_1}\, \partial^\nu\!\alpha^{i_2}\, 
\partial_\nu\alpha^{i_3}\partial^\lambda \alpha^{i_4}\, 
\partial_\lambda\!\alpha^{i_5}\,\partial^\mu\!\alpha^{i_6}.\label{lagsixful}
\end{multline}
This result reveals that the dependence on the dimension of the
representation of this term is contained in the same overall factor
$j(j+1)(2j+1)$ as in the Skyrme model Lagrangian \eqref{parlagden}. Hence
addition of the term ${\CL}_6$ maintains the simple overall 
dimension dependent factor of the original Skyrme model.

As in the case of the quartic term one can construct an alternative
sixth order term, which is equivalent to \eqref{lagsix} in the case of
the fundamental representation, but which differs in its dependence on
$j$
\begin{equation}
{\CL}'_6=e'_6\, \epsilon^{\mu
\nu_1\nu_2\nu_3}\,\epsilon_{\mu\eta_1\eta_2\eta_3}\,
\Tr\{\BR_{\nu_1}\BR_{\nu_2}\BR_{\nu_3}\BR^{\eta_1}\BR^{\eta_2}
\BR^{\eta_3}\}.
\label{lagsixpri}
\end{equation}
In terms of the Euler angles $\Balpha$ this term also reduces
to the expression \eqref{lagsixful}, with the exception of an additional
factor $j(j+1)(2j+1)e'_6/e_6$. Its dependence on $j$ is thus
different from \eqref{lagsix}, although by adjusting the values of the
parameters $e_6$ and $e'_6$ differently in each representation equivalent
dynamical predictions in the classical\footnote{This is not the case in 
quantum Skyrme model.} model can be maintained.
Obviously we cannot express \eqref{lagpri} and \eqref{lagsixpri}
as inner product of group generators, whereas for \eqref{lagsixful} 
this should be possible.
\section{The Lagrangian symmetries}
\setcounter{equation}{0}
\label{symetries}
We start from construction of chirally invariant terms of the lowest order, 
which  satisfy additional physical requirements. Maximal symmetry 
requirement together with $B=1$ condition directly lead to Skyrme hedgehog 
solution in fundamental representation. This solution then is generalised to 
any SU(2) representation.
\subsection{Chiral symmetry breaking}
It is commonly accepted that chiral symmetry is the symmetry of QCD (theory 
of strong interactions) in the zero quark mass limit. There is only one 
second order chirally invariant term 
\begin{equation}
\CL_{(2)}\sim \int\Tr\{\BR_\mu\BR^\mu\}\di^3\!\Bx ,
\end{equation}
whereas there are three\footnote{Here we analyze only terms where all 
currents $\BR_\mu$ enter under {\it single} trace symbol. Generally terms 
\eqref{fourorderterm} and the term
$\int \Tr\{\BR_\mu\BR^\mu\} 
\Tr\{\BR_\nu\BR^\nu\}\di^3 \Bx$ contribute to the same order in chiral 
perturbation theory~\protect\cite{koch97}.}
linearly independent invariants of the order four
\cite{mahankovbook}
\begin{equation}
\CL_{(4)}\sim a\Tr\{\BR_\mu\BR^\mu\, \BR_\nu\BR^\nu\}+b\Tr\{\BR_\mu\BR_\nu
\BR^\mu\BR^\nu\}
+c\Tr\{\partial_\mu\BR_\nu\partial^\mu\BR^\nu\},
\label{fourorderterm}
\end{equation}
$a,b,c$ being some constants. All three of them are Lorenz invariants so 
there are no reasons to prefer any one of them. But if we want to ensure
positive energy density and (as a consequence) to avoid pathology in system 
dynamics we should take care that time components of "right" currents 
entered the Lagrangian only in quadratic form. The requirement is satisfied 
by the only combination of the order four\footnote{$\CL_4$ and 
$\CL^\prime_4$ \protect\eqref{lagpri} coincide up to the overall constant 
factor in the fundamental representation of \SU\ group. In SU(3) case these 
terms are different.}
\begin{equation}
\CL_{(4)}=\Tr\{\BR_\mu\BR^\mu\,\BR_\nu\BR^\nu\}-\Tr\{\BR_\mu\BR_\nu
\BR^\mu\BR^\nu\}
=-\frac12\Tr\{\bigl[\BR_\mu,\BR_\nu\bigr]
\bigl[\BR^\mu,\BR^\nu\bigr]\},
\end{equation}
which is exactly the term suggested by {\it T.H.R.~Skyrme}.
The Lagrangian \eqref{genlagden} is invariant under {\it global}\/ (point 
independent) chiral group $\SU_{L_{eft}}\otimes \SU_{R_{ight}}$ of 
transformations of unitary field \BU
\begin{equation}
\BU^\prime(\Bx)=\BV \BU(\Bx) \BW^{-1};\qquad \BV\in\SU_L;\quad\BW\in\SU_R.
\end{equation}
The group $\SU_L\otimes\SU_R$, however, is not a symmetry group of classical 
vacuum state (the highest symmetry field from $Q_0$ sector which 
takes on the constant value $\mathbf{1}$) 
\begin{equation}
\BU(\infty,t)=\BU_{vac}(\Bx,t)\equiv\mathbf{1},\quad \text{unless}\quad 
\BV=\BW.\end{equation}
As a consequence, maximal global invariant subgroup of configuration space 
of the model is 
\begin{equation}
\diag \bigl(\SU_L\otimes\SU_R\bigr)\approx\SU_{I_{sotopic}},
\end{equation}
where the standard notation $\diag(G_1\otimes G_2)$ denotes the subgroup of
$G_1\otimes G_2$ with parameters of $G_1$ and $G_2$ being identified. 
\subsection{Hedgehog ansatz as a lowest energy solution}
In the baryon number zero sector the field which takes on the constant 
value $\mathbf{1}$ is the field of the highest symmetry. It is fully 
Poincar\'e invariant and provides a classical description of the vacuum 
state. 

We expect that the ground state for $B=1$ would be described by a 
configuration $\BU$ with the maximal possible symmetry. When the winding 
number $B$ is not zero, the field \BU\ cannot possess translational 
invariance. A translational invariant field is a constant and corresponds 
to $B=0$. When $B\ne0$, \BU\ cannot be rotationally invariant either. This 
is because a spherically symmetric unitary field \BU\ depends only on the 
radial distance $r$\begin{equation}
\BU^\dagger\partial_i\BU=\bar\Bx_i\BU^\dagger\frac{\partial\BU}{\partial r},
\end{equation}
where $\bar\Bx$ is a unit vector in Cartesian coordinate system. Then
\begin{equation}
\CB(\BU)\sim \epsilon_{ijk}\Tr\{\bigl(\BU^\dagger\partial_i\BU\bigr)
\bigl(\BU^\dagger\partial_j\BU\bigr)
\bigl(\BU^\dagger\partial_k\BU\bigr)\}=0.
\end{equation}
To obtain one with $B\neq 0$ let us blend isotopic rotations $\SU_I$ with 
space rotations\footnote{SO(3) is homomorphic to SU(2). We keep 
notation SO(3) for spatial rotations (for a while) to make the separation 
more clear.}
$\mathrm{SO(3)}$ to form a group 
$\diag\bigl(\SU_I\otimes\mathrm{SO(3)}\bigr)$. The unitary field \BU\ 
transforms under $\diag\bigl(\SU_I\otimes\mathrm{SO(3)}\bigr)$ as 
follows~\cite{elliottbook1}:
\begin{equation}
\BU_g(\Bx)=T(g)\BU(g^{-1}\Bx)=\BU(\Bx);\quad g\in
\diag\bigl(\SU_I\otimes\mathrm{SO(3)}\bigr).
\end{equation}
Substituting expressions for the explicit rotation generators in the 
fundamental representation of $\SU_I$ yields differential equation
\begin{equation}
-i\bigl(\Bx\times\mathbf{\nabla}\bigr)_i\BU(\Bx)+
\bigl[\frac{\Btau_i}2,\BU(\Bx)\bigr]=0.\label{solsymeq}
\end{equation} 
The solution\footnote{The method of solution of 
equation \protect\eqref{solsymeq} is described,
for example, in Ref.~\protect\cite{mahankovbook}.} of 
\eqref{solsymeq} is 
\begin{equation}
\BU(\Bx)=\BU_c(\Bx)\equiv\cos F(r)\cdot\mathbf{1}+i\Btau\cdot\Bx\sin F(r).
\label{solsymeqsol}
\end{equation}
Pictorially it is illustrated in Fig.~\ref{fig8}, where arrow length $|\sin 
F(r)|$ (for function $F(r)$ itself see Fig.~\ref{fig9}) goes to zero as 
$r\rightarrow\infty$ and $r\rightarrow0$. The solution of 
\eqref{solsymeq} is exactly the Skyrme's hedgehog ansatz \eqref{skyanz}. 
Generalization to the arbitrary representation is straightforward. Instead 
of \eqref{solsymeq} we have
\begin{equation}
i\bigl( \Bx\times \mathbf{\nabla }\bigr)_i\BU(\Bx)+\sqrt{2}\bigl[ 
J_i,\BU(\Bx)\bigr] =0,
\label{gensolsyseq}
\end{equation}
where circular components \eqref{cirsys} are used both for the vector and 
isovector. The solution of \eqref{gensolsyseq} is a generalized hedgehog 
ansatz
\begin{equation}
\BU_c(\Bx)=\exp \bigl(-i\sqrt{2}\,J_a\bar x^a\,F(r)\bigr).
\label{genskyanz}
\end{equation}
Sometimes the hedgehog ansatz \eqref{solsymeqsol}, \eqref{genskyanz} is 
referred to as "spherically symmetric" solutions. These solutions are 
"spherically symmetric" only in the sense that a coordinate rotation is 
equivalent to an isospin rotation of the constant matrix $\mathbf{A}$
\begin{equation}
\BU_c(\Bx)=\mathbf{A}\exp \bigl(-i\sqrt{2}\,J_a\bar x^a\,F(r)\bigr) 
\mathbf{A^\dagger},\quad \mathbf{A}\in \SU .
\end{equation}

To summarize, the highest symmetry solution for $B\neq0$ sector leads to  
certain mixing of indices associated with internal and 
geometric invariance (which are --- a priori --- completely unrelated). 
Similar examples are given by the monopole and instanton configurations 
which occur in \SU\ gauge theories.

"One may wonder whether such a blend of internal and geometric symmetries 
may exist at a more fundamental level as a general feature of field theory 
and not simply in specific field configurations of particular models. This 
feature would be very attractive for the construction of a unified theory 
of all fundamental interactions including gravity. That this is {\it 
not}\/ possible is expressed by the so called no-go theorems, in particular 
the theorem of {Coleman} and {Mandula}, which essentially says the 
following: the most general invariance group of a relativistic quantum field 
theory is a {\it direct}\/ product of the Poincar\'e group and an internal 
symmetry group, i.e. there is no mixture of these symmetry transformations.

However, these no-go theorems do {\it not}\/ claim that such a mixture 
cannot exist if the set of all symmetry transformations represents a more 
general algebraic structure than a Lie group. Indeed, a famous result 
known as the theorem of Haag, Lopusz\`anski and Sohnius~\cite{sohnius85} 
states that the most general {\it super}\/ Lie group
of local field theory is the N-{\it 
extended super Poincar\'e group}\/ in which there is a non-trivial mixing 
of geometric transformations and internal SU(N) transformations. As a 
matter of fact, this result can also be viewed as a good argument in favour 
of the existence of supersymmetry as an invariance of nature since it 
states that {\it supersymmetry is the natural}\/ (only possible) {\it 
symmetry}\/ if one allows for super Lie groups as symmetry structures"  
\cite{gieres98}. 

The existence of nontrivial mixing for certain configurations in the Skyrme 
model as well as in supersymmetric models may serve as a strong argument for 
further investigations of the model which is much more simple to deal 
with than those of supersymmetric theories. 
\subsection{Higher sectors solutions}
It is proved~\cite{mahankovbook} (see, however,~\cite{kaulfuss85})
that "spherically symmetric" hedgehog ansatz  
leads to the absolute energy minimum only when $B=1$. For $B=2$ value of 
baryon number it is hoped~\cite{mahankovbook} that axial symmetric states 
\begin{equation}
F=F(r,\vartheta);\quad  
\Theta=\Theta(r,\vartheta);\quad\Phi=m\varphi;\qquad m\in\NN,
\end{equation}
realise energy minimum configurations. The statement was 
verified numerically~\cite{kopeliovic87,verbaarshot}. The value is
$E_{B=2}/E_{B=1}=1.92$ for the ratio of energies for axial symmetric 
solution of sector $B=2$ and spherically symmetric hedgehog ansatz with 
$B=1$. Energy/baryon densities for $B=2$ configuration possess a 
toroidal symmetry (see Fig.~\ref{fig7} in Appendix~\ref{app3}). Stable 
ansatz for $\BU$ minimizing energy and with baryon numbers $B\ge2$ have 
been numerically found by various groups 
\cite{carson91,sutcliffe97b,sutcliffe97a}. Energy densities for these 
static configurations have been plotted and a remarkable fact has been 
discovered that they are invariant under discrete subgroups $\GG_R$ of the 
spatial rotation group SO(3). Some of them are shown~\cite{sutcliffe97b} in 
Appendix~\ref{app3}. The group $\GG_R$ is the symmetry group of energy 
density. It is not necessarily the invariance group of the static $\BU$ 
field. Published work~\cite{sutcliffe97a} does not report on the symmetry 
group of the latter. 

Configurations with $B>1$ are important in nuclear physics~\cite{carson91} 
since proton and nuclei could be related to quantized 
states of these soliton-like fields. Several recent studies support this 
point of view~\cite{sutcliffe97c} and suggest that the structures of 
heavier nuclei could resemble those of fullerene molecules, at least at the 
classical level.
\subsection{Reducible representations}
\label{redrep}
Generalization of the model to arbitrary reducible representation 
is a bit straightforward. One needs only to sum over all irreducible 
representations involving explicit dependence on representation.
Thus, substitution for \eqref{defu} is 
\begin{equation}
\BU(\Br,t)=\sum_k \oplus \BD^{j_k}\left(\Balpha(\Br,t)\right).
\end{equation}
The general scalar product \eqref{tratwo} then modifies to
\begin{equation}
\Tr\langle j{\cdot}|J_a J_b|j{\cdot}\rangle 
=(-1)^a\frac16\sum_k j_k(j_k+1)(2j_k+1)\delta_{a,-b},\label{redtratwo}
\end{equation}
and the normalization factor \eqref{norfac} takes a form
\begin{equation}
N=\frac23\sum_{k}j_{k}(j_{k}+1)(2j_{k}+1).
\label{rednorfac}
\end{equation}
Other formulas do not involve changes.
%\section[Algebraic structure]{Algebraic structure, symmetries and 
%topological invariants of the Skyrme model}

%\section{Phenomenology of the classical Skyrme model}
%\subsection{Effective model ideology}
%\subsection{Mass and Lagrangian density}
%\subsection{Radius, formfactors, magnetic momenta}
%\subsection{Coupling constants problem}

\chapter{Quantum Skyrme model}
\label{chapthree}
This chapter contains main results. After brief remarks on quantization 
problems in curved space we skip to collective coordinate approach and 
consider the Skyrme Lagrangian quantum mechanically {\it ab initio}.
Assuming noncommutativity of dynamical variables we calculate 
expressions of Noether 
currents, magnetic momenta, axial coupling constant, etc. and 
numerically evaluate physical quantities using the classical chiral angle 
solution in various \SU\ representations. These numerical results then are 
used as starting input for self-consistent quantum chiral angle 
determination procedure. Numerical results of quantum chiral angle 
calculations are presented in Appendices~\ref{app1} and~\ref{app2}.

\section{Quantization in curved space}
\setcounter{equation}{0}
\setcounter{table}{0}
\setcounter{figure}{0}
\label{qincs}
The purpose of the section is to remind readers the Dirac method of
constrained quantization as well as problems of traditional
quantization in curved space. The justification of the actual quantization 
method is considered without going into details in the last subsection.
The section, thus, provides the context for quantization procedure followed 
further but contains no new material.
\subsection{General remarks}
\label{genideas}
Questions may be raised concerning the justification for quantizing the
Skyrme Lagrangian at all, since it is not a fundamental field theory, but
rather a classical model that results from taking the limit of such a
theory, including only some degrees of freedom of the original theory.
Nevertheless, there is a rich experience from the nonrelativistic
many-body problems, for example, from nuclear physics~\cite{walet91},
suggesting the validity of such an approach for the study of collective
properties at low energies.

The second remark concerns general quantization strategy. At the University
level  the construction of quantum theory passes three steps, namely
{\bf Lagrangian (classical) $\rightarrow $ Hamiltonian
(classical) $\rightarrow $ Hamiltonian (quantum) $\equiv$ Quantum theory}.
The quantization in collective coordinate approach~\cite{fujii87}, which
we will follow in the work, slightly modifies this sequence. It starts from
the quantum Lagrangian from the outset. By quantum Lagrangian we mean
that dynamical coordinates $q^i$ and its time derivatives (velocities) 
$\dot q^i$ do not commute. The explicit commutation relations at the 
moment are unknown. These relations are extracted from the standard 
commutation relations $[q^i,p^j]$ after we pass to quantum Hamiltonian 
(and define canonical momenta $p^j$). It can be shown that this modified 
formalism~\cite{sugano71,kimura71,kimura72} leads to consistent quantum 
description, which we will follow in subsequent sections.

The results of modified and usual quantization sequences generally will 
differ. Noncommuting quantum variables will generate additional terms while 
passing from quantum Lagrangian to quantum Hamiltonian. These terms are lost 
when we impose canonical commutation relations after Hamiltonian is 
obtained. The problem is similar to well known operator ordering problem. 
Further in the work we will refer to "canonical quantization method" when 
we start from quantum Lagrangian in the very beginning and to "semiclassical 
quantization method" when canonical commutation relations are imposed after 
Hamiltonian is calculated. It should be noted, however, that these two 
choices do not use up all possibilities. In Ref.~\cite{asano91}, for example, 
the following possibilities are discussed:
\begin{equation}
\renewcommand{\arraystretch}{2}
\begin{array}{ccccc}
L_{\text{Skyrme}}(\mathbf{\Phi},\dot \mathbf{\Phi})&\Longrightarrow 
&L_{\text{hedg.}}(F,\dot F,\BA,\dot \BA)&\Longrightarrow 
&L_{\text{coll.}}(\Bq,\dot \Bq)
\\ \Downarrow \text{case I}& &\Downarrow 
\text{case II} & & \Downarrow \text{case III}\\
\hat H_{\text{Skyrme}}(\mathbf{\Phi},\mathbf{\pi})
&\Longrightarrow &\hat H_{\text{hedg.}}(F,\mathbf{\Pi}_F, 
\BA,\mathbf{\Pi}_A) &\Longrightarrow &\hat H_{\text{coll.}}(\Bq,\dot \Bp)\\
\end{array}.
\end{equation}
In  case~I, the Hamiltonian gets by quantizing directly the Lagrangian in 
the Skyrme model from the beginning. After this quantization one can 
introduce the hedgehog ansatz and the collective coordinates. As another 
method (case~II) we can do the quantization of the classical Lagrangian, 
which is obtained by introducing the hedgehog ansatz and after this we 
introduce the collective coordinates. Case~III starts by getting the 
classical Lagrangian with the collective coordinates, using the hedgehog 
ansatz and getting the quantum Hamiltonian. The case~II is free of the 
ordering problem among the operators. However, in both cases~I and III, 
there are the ordering problems among the operators, and then quantum 
Hamiltonian cannot been determined uniquely. In summary, the problem is in 
which steps we do the quantization. Following~\cite{fujii87} we utilize the 
following detailed quantization sequence:
\begin{enumerate}
\item Introduce quantum collective coordinates 
$\mathbf{A\bigl(\Bq(t)\bigr)}$ \eqref{gencolcoo}. They are quantum in the 
sense that time
differentiation of $\mathbf{A}$ requires the Weyl ordering (see 
Sec.~\ref{weylorderingsection}).
\item Make the Lagrangian quantum. Quantum dynamical variables are $\Bq(t)$.
\item Following the method described in 
\cite{fujii87,sugano71,kimura71,kimura72,kimura71a,sugano72,sugano73a} pass 
to quantum Hamiltonian.
\item Introduce the hedgehog ansatz\footnote{Actually 
we introduce hedgehog ansatz before passing to explicit Hamiltonian. This 
is done only for reasons of simplicity of intermediate expressions and 
cannot affect the quantization itself. The only requirement of the 
quantization sequence in hand is that we do assume existence of solitonic 
solution~\cite{fujii87}.}, and solve integro-differential equation for 
quantum chiral angle.
\end{enumerate}

Another important point is symmetry properties of classical Lagrangian and 
quantum Hamiltonian derived from it. There exist quantization methods (for 
example, general covariant Hamiltonian method~\cite{dewitt57}) preserving 
original classical Lagrangian symmetries. The symmetric (Weyl) ordering of 
operators  $\Bq$ and $\Bp$ (used in the work), however, cannot avoid a risk 
that the quantum Hamiltonian has chiral symmetry breaking  term 
\cite{asano91}. The feature thus can be used to provide (or explain) 
the origin of finite pion mass.

The subsections below are intended to
shed a bit light on the justification of the quantization method we
will follow.
\subsection{Towards quantum theory}
There where attempts~\cite{dyson90} to quantize classical equations of
motion in the form
\begin{equation}
m \frac{\di^2 x^i}{\di t^2}=f^i,
\label{claequ}
\end{equation}
without resort to a Lagrangian (or Hamiltonian). The idea is to start with
classical equations \eqref{claequ} and give commutation relations for the
operators $X^i$ and $\dot X^j$
\begin{subequations}
\begin{align}
\bigl[ X^i, X^j\bigr]&=0\label{komrelone}
\\
\bigl[ X^i, \dot X^j\bigr]&=i\hbar \delta^{i,j}.
\end{align}
\end{subequations}
The result was that the existence of Lagrangian for \eqref{claequ}
essentially comes from \eqref{komrelone}. Thus Feynman's hope to quantize
without a Lagrangian (or Hamiltonian) was doomed when he set the very
reasonable condition that coordinates commute~\cite {hojman91}. Let us
briefly review traditional quantization methods following 
Ref.~\cite{klauder98}.
\subsection{Traditional quantization methods}
We will mention four of them here in the order of historical evolution.
\newline
{\bf Heisenberg quantization} is based on the following postulates
(for the first postulate see below):
\begin{enumerate}
\setcounter{enumi}{1}
\item Introduce matrices $\mathbf{Q}=\{Q_{mn}\}$ and
$\mathbf{P}=\{P_{mn}\}$, here $m,n\in\{1,2,3,\ldots\}$, that satisfy
$[\mathbf{Q},\mathbf{P}]_{mn}\equiv{\displaystyle\sum_p}(
Q_{mp}P_{pn}-P_{mp}Q_{pn})=i\hbar\delta_{m,n}$.
\item Build a Hamiltonian
matrix $\mathbf{H}=\{H_{mn}\}$ as a function (e.g., polynomial) of the
matrices, $H_{mn}=H(\mathbf{P},\mathbf{Q})_{mn}$, that is the same function
as the classical Hamiltonian $H(p,q)$. (In so doing there may be operator
ordering ambiguities which this prescription cannot resolve; choose an
ordering that leads to a Hermitian operator.)
\item Introduce the equation of motion $i\hbar{\dot
X}_{mn}=[\mathbf{X},\mathbf{H}]_{mn}$ for the elements of a general
matrix $\mathbf{X}=\{X_{mn}\}$.
\end{enumerate}
Along with these postulates
comes the implicit task of solving the above equations of motion
subject to suitable operator-valued boundary conditions.
After the principal paper on quantization~\cite{born26}, it subsequently
became clear to {\it W.~Heisenberg} 
that it is {\it necessary}\/ to make this
promotion from $c$-number to $q$-number variables only in {\it
Cartesian}\/ coordinates. Thus here is implicitly another postulate
\cite{dirac58}:
\begin{enumerate}
\item Express the classical kinematical
variables $p$ and $q$ in Cartesian coordinates prior to promoting them to
matrices $\{P_{mn}\}$ and $\{Q_{mn}\}$, respectively.
\end{enumerate}
{\bf Schr\"odinger quantization}
assumes the following postulates~\cite{schrodinger26}:
\begin{enumerate}
\item Express the classical kinematical variables $p$ and $q$ in 
Cartesian coordinates.
\item Promote the classical momentum $p$ to the
differential operator $-i\hbar(\partial/\partial q)$ and the classical
coordinate $q$ to the multiplication operator $\hat q$, a choice that
evidently satisfies the commutation relation $[\hat q,-i\hbar(\partial
/\partial  q)]=i\hbar$.
\item Define the Hamiltonian operator $\hat H$ as the classical
Hamiltonian with the momentum variable $p$ replaced by the operator
$-i\hbar(\partial /\partial q)$ and the coordinate variable $q$ replaced
by the operator $\hat q$. (In so doing there may be operator ordering
ambiguities which this prescription cannot resolve; choose an ordering
that leads to a Hermitian operator.)
\item For $\psi(q)$
a complex, square integrable function of $q$, introduce the dynamical
equation $i\hbar{\dot\psi}=\hat H\psi$.
\end{enumerate}
Implicit with these postulates is the instruction to solve the 
Schr\"odinger equation for a dense set of initial conditions and 
a large class of Hamiltonian operators.
It is interesting to note that 
{\it E.~Schr\"odinger} himself soon became aware
of the fact that his procedure generally works only in Cartesian 
coordinates.
\newline
{\bf Dirac quantization}\footnote{The Dirac quantization although
traditionally used is, actually, Hamilton formulation method, because 
operator ordering problems are not considered here. We give a very brief 
summary of the theory in Sec.~\ref{diractheory}.}
recipe deals with constrained 
dynamical systems. Constrains appear in the Hamilton formulation of all 
gauge theories we know of. Dirac-Bergmann constraint theory puts all  these 
constraints into first or second classes. All second class constraints 
$S_a$ can be eliminated from the theory, whereas Dirac prescription for the 
implementation of first class constraints $F_a$ in quantum theory is that 
they be imposed as conditions on the physically allowed states 
$|{\cdot}\rangle$:
\begin{equation}\hat F_a |{\cdot}\rangle =0.
\end{equation}
Here $\hat F_a$ is the quantum operator corresponding to the classical 
function $F_a$.
\newline
{\bf Feynman quantization} focuses on the solution to the 
Schr\"odinger equation and postulates that the propagator, an integral 
kernel that maps the wave function (generally in the Schr\"odinger 
representation) at one time to the wave function at a later time, may be 
given by means of a path integral expression~\cite{feynman48}.
On the surface, it would seem that the (phase space) path integral, using 
only concepts from classical mechanics, could get around the need 
for Cartesian coordinates. That is not the 
case~\cite{stuckens86}. As postulates for 
a path integral quantization scheme we have:
\begin{enumerate}
\item Express the classical kinematical variables $p$ and $q$ 
in Cartesian coordinates.
\item
\label{posttwo}
Given that $|q,t\rangle$, where $Q(t)|q,t\rangle=q|q,t\rangle$, denote
sharp position eigenstates, write the transition matrix element in the form
of a path integral as
\begin{equation} 
\langle q'',T|q',0\rangle
=\mathcal{M}\int\exp\Bigl( (i/\hbar)\int\bigl(p{\dot q}-
\hat{H}(p,q)\bigr)\,\di
t\Bigr)\,\mathcal{D} p\,\mathcal{D} q.
\label{feyintgen}
\end{equation}
\item Recognize that the formal path integral \protect\eqref{feyintgen}
is {\it effectively undefined}\/ and replace it by a {\it regularized}\/
form of path integral, namely,
\begin{multline}
\langle q'',T|q',0\rangle
=\lim_{N\rightarrow\infty}
M_N\int\exp\Bigl(\frac{i}{\hbar}{\displaystyle
\sum_{l=0}^N}\bigl(p_{l+\scriptscriptstyle 1/2}(q_{l+1}-q_l)\\ 
-\frac{\epsilon}2
\hat{H}(p_{l+\scriptscriptstyle 1/2},q_{l+1}+q_l)\bigr)\Bigr)\,
\Pi_{l=0}^N\di p_{l+\scriptscriptstyle 1/2}\,\Pi_{l=1}^N\di q_l,
\label{regpathint}
\end{multline}
where $q_{N+1}=q''$, $q_0=q'$, $M_N=(2\pi\hbar)^{-(N+1)}$, 
$p_{l+\scriptscriptstyle 1/2}=(p_l+p_{l+1})/2$,
and $\epsilon 
=T/(N+1)$.
\end{enumerate}
Implicit in the latter expression is a {\it Weyl ordering}\/ choice to
resolve any operator ordering ambiguities. Observe that the naive lattice 
formulation of the classical action leads to correct quantum mechanical 
results, generally speaking, only in Cartesian coordinates. Although the 
formal phase space path integral
\eqref{feyintgen} appears
superficially to be covariant under canonical coordinate transformations,
it would be incorrect to conclude that was the case inasmuch as it would
imply that the spectrum of diverse physical systems would be identical. In
contrast, the naive lattice prescription applies only to Cartesian
coordinates, the same family of coordinates singled out in the first
postulate of each of the previous quantization schemes.

It is essential that traditional quantization methods start from
global Cartesian coordinate system\footnote{Needless to say that there are
no ways to introduce global Cartesian coordinates onto arbitrary
configuration (phase) space.}
and Hamiltonian function. But if we want Lorenz {\it covariant}\/ theory,
then it is not a good idea to start formulation from the Hamiltonian as
there are no easy ways to ensure Lorenz {\it covariance}\footnote{
Lorenz covariant theory (Lorenz invariant Lagrangian) places time and space
on the same footing, whereas time plays a special role in Hamiltonian and,
therefore, in quantum theory (quantum mechanics). Also, there are no ways to
make time an operator \protect\cite{crewther95}. Indeed, if time where an
operator $\hat t$ it would be the component of a four-position operator
$\hat \mathbf{X}=(\hat t, \hat \Bx)$ conjugate to the Hamiltonian $\hat H$
in the four momentum $\hat \BP=(\hat H,\hat \Bp)$
\begin{equation}
\bigl[P^\mu,X^\nu\bigr]=i g^{\mu \nu};\quad g^{\mu \nu}=\{1,-1,-1,-1\}.
\end{equation}
Then commutator $[\hat H,\hat t\,]=i$ implies
\begin{equation}
\exp (-i\epsilon \hat t\,) \hat H \exp (i\epsilon \hat t\,)=\hat H-\epsilon,
\end{equation}
for any constant $\epsilon$. Thus the operator $\exp (i\epsilon \hat t)$ 
applied to any eigenstate $|E\rangle$ of $\hat H$ with energy eigenvalue $E$ 
produces another eigenstate $\exp (i\epsilon \hat t\,) |E\rangle$ 
with shifted
eigenvalue $E-\epsilon$. That indicates the presence of a continuous energy
spectrum with range $-\infty <E<\infty$, contrary to the requirement that
$E$ be bounded from below. Also, it contradicts the fact that generally, $E$
is quantized \protect\cite{pauli33}.}. The easiest way to get Lorenz {\it
covariant}\/ theory is to start formulation from Lorenz {\it invariant}
Lagrangian. This way leads us directly both to Hamilton formulation and 
operator ordering problems.
\subsection{Weyl ordering}
\label{weylorderingsection}
A path integral formalism sometimes is referred to as a quantum method
resolving the operator ordering problem~\cite{berezinbook}. This is because 
the Weyl ordered expressions  are used in the {\it regularized}\/ form
\eqref{regpathint} of a path integral in order to resolve the
ambiguities\footnote{Path integral \protect\eqref{feyintgen} value still
strongly  depends on the finite-dimensional approximations. Points
$p_k,q_k$ in \protect\eqref{regpathint} are chosen in such a way (usually 
in the centre of the interval) that \protect\eqref{regpathint} limit 
coincides with operator $\mathrm{e}^{\frac{i}{h}t\hat H}$ matrix element 
only when $\frac12(\hat p\hat q+\hat q\hat p)$ corresponds to classical 
expression $pq$. Interesting, but the exceptions are known, when the middle 
point is not appropriate~\cite{stuckens86}. As far as we know, there is no 
general recipe how to choose these points in the case of arbitrary curved 
space.}.

Let us illustrate the problem for an unconstrained system. Namely, let us
recall the well known harmonic oscillator example. The problem is that
two identical {\it classical} Hamiltonians
\begin{subequations}
\begin{alignat}{2}
H_1 &\sim   x^2+p^2;&\qquad H_2 &\sim  (x-ip)(x+ip),
\label{oscham}\\
\intertext{lead to different energies in quantum theory}
E_1 &\sim \hbar \omega (n+\frac 12);\qquad\text{and}&\qquad E_2 &\sim \hbar 
\omega n,
\end{alignat}
\end{subequations}
respectively, when the {\it same}\/ commutation relations $[\hat p,\hat
q]$ is imposed. The reason is that the two {\it quantum}\/ Hamiltonians
\eqref{oscham} differ exactly by the commutator $[\hat p,\hat q]$, what is
indicated by additional $\frac12\hbar \omega$ 
term in the system energy. The Weyl 
ordered Hamiltonian\footnote{The second Hamiltonian $H_2$ is not Weyl 
ordered. Weyl ordering of it leads to the first one, because the Weyl 
ordering in two operator case is simple symmetrization.} $H_1$ results to 
the true\footnote{Although in this case the reference point can be 
shifted by $-\frac12\hbar\omega$ 
to ensure the same result.} energy $\hbar\omega 
(n+\frac 12)$. How can Weyl ordering can be applied in the general case?

Because of nonvanishing commutator $[\dot q,q]\neq 0$ we need to state
more explicitly what we understand under the symbol $\partial_t G(q)$.
The most natural seems to be the definition
\begin{equation}
\bigl(\partial_t G(q)\bigr)_{W_{\text{eyl ordering}}} =\frac12\Bigl\{\dot
q,\, \frac{\di G(q)}{\di q}\Bigr\},
\label{weylord}
\end{equation}
which is a consequence of application of the Newton-Leibnitz rule to Taylor
series expansion of arbitrary function $G(q)$,
\begin{equation}
G(q)=G(q_0)+G'(q)\Bigl|_{q=q_0}\,q+\frac12
G''(q)\Bigl|_{q=q_0}\,q^2+\dotsb .
\end{equation}
Indeed,
\begin{subequations}
\begin{gather}
(\partial_t q^2)_W=\partial_t (q\,q) =\dot q q+q\dot q=
\frac12 \{\dot q,\,\frac{\di (q^2)}{\di q}\},\\
(\partial_t q^3)_W=\partial_t (q\,q\,q) =\dot q q^2+q\dot q q+
q^2\dot q=\frac32 (\dot q q^2+q^2\dot q)=
\frac12\{\dot q,\,\frac{\di (q^3)}{\di q}\},\\
\hbox to 3cm{\dotfill}\notag \\
(\partial_t q^n)_W=\partial_t \underbrace{(q\dots q)}_n =
\underbrace{\dot q
q+\dotsb+q\dot q}_n=\frac12\Bigl\{\dot
q,\,\frac{\di q^n}{\di q}\Bigr\}.
\end{gather}
\end{subequations}
Here the notation $(\partial_t q^n)_W$ is related to the usual Weyl
ordering notation $(a^n b)_W$ in an obvious way
\begin{equation}
\frac1n \bigl(\partial_t q^n\bigr)_W=\bigl( q^{n-1}\dot q \bigr)_W\ .
\end{equation}
The general Weyl ordered term $(q^n\dot q)_{W_1}$ has a form
\begin{equation}
(q^n\dot q)_{W_1}=\frac1{n+1}\sum_{l=0}^n q^{n-l}\dot q q^l.
\end{equation}
It is straightforward to prove~\cite{tdleebook} that the above form of Weyl
ordering is identical to the definition\footnote{Let us illustrate this for
$(q^2\dot q)_{W_1}$ and $(q^2\dot q)_{W_2}$ terms. Indeed the
sequence $(q^2\dot q)_{W_2}=\frac14
\bigl(q^2\dot q+2q\dot q q+\dot q q^2 \bigr)=\frac14
\bigl(q^2\dot q+(\frac43+\frac23)q\dot q q+\dot q q^2
\bigr)=\frac14
\bigl(\frac43q^2\dot q+\frac43q\dot q q+\frac43\dot q q^2
\bigr)=\frac13 \bigl(q^2\dot q+q\dot q q+\dot q q^2
\bigr)=(q^2\dot q)_{W_1}$ shows the
result.}
\begin{equation}
(q^n\dot q)_{W_2}=(\frac12)^n\sum_{l=0}^n
\frac{n!}{l!(n-l)!}q^{n-l}\dot q q^l.
\end{equation}
The Weyl ordering has a number of interesting features~\cite{berezinbook}
and is widely used. Further in the work we will follow definition
$(q^n\dot q)_{W_1}$, which is identical to \eqref{weylord} form.
\subsection{Hamilton formulation}
Passage from Lagrangian to Hamiltonian sometimes requires additional
assumptions. Important class of theories, where
standard Hamiltonization fails, is, for  example,
nonlinear models of elementary particles, including the Skyrme model.

Let us concentrate on Lagrangian theories only, namely on local Lagrangian
theories of the form
\begin{equation}L=\int \CL (\phi, \partial _\mu \phi
)\di^n \Bx,
\end{equation}
allowing non-ambiguous equations of motion\footnote{
For example, Lagrangian $\CL =q$ leads to ambiguous Euler-Lagrange
equation $1=0$.}. All Lagrangians can be classified into two large groups,
depending the system Hessian
\begin{equation}
M_{ij}=\frac{\partial ^2 \CL }{\partial \dot q^i \partial \dot q^j},
\end{equation}
is singular or not. In the case of nonsingular Hessian the usual method
of Hamilton formulation is valid~\cite{faddeev88}.
\subsection{Singular theories}
When $M_{ij}$ is a singular matrix ($\det M_{ij}=0$), then we cannot
express all velocities $\dot \Bq$ as functions of momenta $\Bp$ and
coordinates $\Bq$. To show this we simply rewrite Lagrange second order
equations
\begin{equation}
\frac{\delta S}{\delta q^i}=\frac{\partial \CL }{\partial q^i}-
\frac{\di}{\di t}\frac{\partial \CL }{\partial \dot q^i}=0;\quad S=\int \CL
\di t ,\label{hamsyst}
\end{equation}
into two first order equations
\begin{equation}
M^v_{ij} \dot v^j=K^v_i, \qquad \dot q^i=v^i,
\end{equation}
where the upper index $v$ means that all $\dot \Bq$ are changed to $\Bv$. 
It is easy to see that the passage from Lagrangian to Hamiltonian is just 
variables exchange $(\Bq,\Bv)\rightarrow (\Bq,\Bp)$. The exchange Jacobian 
is\begin{equation}
\frac{D(\Bq,\Bv)}{D(\Bq,\Bp)}=\frac1{\det \Bigl\|\frac{\partial ^2 \CL 
^v}{\partial v^i \partial v^j}\Bigr\|},
\end{equation}
and, therefore should differ from zero. It is a simple task to check
wheather Skyrme Lagrangian is singular or not.
This depends on parametrization. In the Euler-Rodrigues parametrization
\eqref{rodpar} we can express one differential as a function of others,
because of the constraint \eqref{aconstr} on the fields. Thus we have
singular theory in Euler-Rodrigues parametrization, whereas Hessian is
clearly nonsingular in Euler angles parametrization \eqref{defeulang}.

\subsection{Dirac-Bergmann theory of constraints}{\!\!\!\!\!\!}\footnote{The 
presentation of the subsection follows Ref.~\cite{balachandran92}.}
\label{diractheory}
Let $\GM$ be the space of "coordinates" appropriate to a Lagrangian $L$. We 
denote the points of $\GM$ by $\Bq =(q_1,q_2,\dots)$. Now given any manifold 
$\GM$, it is possible to associate two spaces $TM$ and $T^*M$ to $\GM$. The 
space $TM$ is called the {\it tangent}\/ bundle over $\GM$. The coordinate 
of a point $(\Bq, \dot\Bq)$ of $TM$ can be interpreted as a position and a 
velocity. The Lagrangian is a function on $TM$. The space $T^*M$ is called 
the {\it cotangent\/} bundle over $\GM$. The coordinate of a point 
$(\Bq,\Bp)$ of $T^*M$ can be interpreted as a coordinate and a momentum.
At each point $\Bq$, momenta
$\Bp$ belongs to the vector space dual to the vector space of 
velocities. Now given a Lagrangian $L$, there exists a map from $TM$ to 
$T^*M$ defined by
\begin{equation}
(\Bq,\dot\Bq)\rightarrow \biggl(\Bq,\frac{\partial 
L(\Bq,\dot\Bq)}{\partial\dot\Bq}\biggr).
\label{tmmap}
\end{equation}
If this map is globally one to one and onto, the image of $TM$ is $T^*M$ and 
we can express velocity as a function of position and momentum (see also 
previous section). This is the case in elementary mechanics and leads to the 
familiar rules for the passage from Lagrangian to Hamiltonian mechanics. It 
may happen, however, that the image of $TM$ under the map \eqref{tmmap}  is 
not all of\, $T^*M$. Suppose, for instance, that it is a submanifold of 
$T^*M$ defined by the equations
\begin{equation}
P_a(\Bq,\Bp)=0;\quad a=1,2,\dots .
\label{psurf}
\end{equation}
Then we are dealing with a theory with constraints. The constraints $P_a$ 
are said to be primary.

The functions $P_a$ do not identically vanish on $T^*M$: their zeros define 
a submanifold of $T^*M$. A reflection of the fact that $P_a$ are not zero 
functions on $T^*M$ is that there exist functions $G$ on $T^*M$ such that 
their Poisson brackets\footnote{Recall that in quantum theory Poisson 
brackets are turned into commutators.} $\{G,P_a\}$ do not vanish on the 
surface $P_a=0$. These functions $G$ generate canonical transformations 
which take a point of the surface  $P_a=0$ out of this surface. It follows 
that it is incorrect to take Poisson brackets of arbitrary functions with 
both sides of the equations $P_a=0$ and equate them. This fact is 
emphasized by rewriting \eqref{psurf}, replacing the "strong" equality 
signs $=$ of these equations by "weak" equality signs $\approx$: 
$P_a\approx 0$. When $P_a(\Bq,\Bp)$ are weakly zero, we can in general set 
$P_a(\Bq,\Bp)$ equal to zero only after evaluating all Poisson brackets.

In the presence of constraints, the Hamiltonian can be shown to be 
\cite{diracbook,gitmanbook}
\begin{subequations}
\begin{align}
H&=\dot q_a \frac{\partial L(\Bq,\dot\Bq)}{\partial \dot q_a} -L(\Bq,\dot\Bq)
+V_aP_a(\Bq,\Bp)\\
&=H_0+V_aP_a(\Bq,\Bp).
\end{align}
\end{subequations}
In obtaining $H_0$ from the first two terms of the first line, one can 
freely use the primary constrains. The functions $V_a$ are as yet 
undetermined Lagrange multipliers. Some of them may get determined later in 
the analysis while the remaining ones will continue to be unknown with even 
their time dependence arbitrary.

Consistency of dynamic requires that the primary constrains are preserved in 
time. Thus we require that
\begin{equation}
\bigl\{P_a,H\bigr\}\approx 0.
\label{secondarydef}
\end{equation}
These equations may determine some of the $V_a$ or they may hold identically 
when the constraints $P_a\approx 0$ are imposed. Yet another possibility 
is that they lead to the "secondary constraints" 
$P^\prime_a(\Bq,\Bp)\approx 0$. The requirement 
$\bigl\{P^\prime_a,H\bigr\}\approx 0$ may determine more of the Lagrange 
multipliers, lead to tertiary constraints or be identically satisfied when
\eqref{secondarydef} and $P^\prime_a\approx 0$ are imposed. We proceed in 
this fashion until no more new constraints are generated.

Let us denote all the constraints one obtains in this way by $C_b\approx 
0$. Dirac divides these constraints into the {\it first\/} and the {\it 
second\/} class constraints. First class constraints $F_a\approx 0$ are 
those for which $\bigl\{ F_a,C_b\bigr\}\approx 0,\quad \forall b$. In 
other words, the Poisson brackets of $F_a$ with $C_b$ vanish on the  surface 
defined by $C_b\approx 0$. The remaining constraints $S_a$ are defined to 
be second class. It can be shown that 
\begin{equation}
\bigl\{F_a,F_b\bigr\}=C^c_{ab}F_c,
\end{equation}
where $C^c_{ab}=-C^c_{ba}$ are functions on $T^*M$. The proof can be found 
in~\cite{balachandran92,diracbook}. 

Let $\GC$ be the submanifold of $T^*M$ defined by the constraints:
\begin{equation}
\GC =\bigl\{(\Bq,\Bp) | C_b(\Bq,Bp)=0\bigr\}.
\end{equation}
Then since the canonical transformations generated by $F_a$ preserve the 
constraints, a point of $\GC$ is mapped onto another point of $\GC$ under 
the canonical transformations generated by $F_a$. Since the canonical 
transformations generated by $S_a$ do not preserve the constraints, such is 
not the case for $S_a$. Second class constraints can be eliminated by 
introducing the so-called Dirac brackets. They have the basic property that 
the Dirac bracket of $S_a$ with any function on $T^*M$ is weakly zero.
Let $\GF$ be the set of all functions which have zero Poisson brackets with 
$S_a$. So long as we work with only such functions, we can use the 
constraints $S_a\approx 0$ as strong constraints $S_a=0$. Assuming that 
there are no first class constraints, the number $n$ of functionally 
independent functions $\GF$ is $\dim (T^*M)-s$, $s$ being number of second 
class constraints. Thus $s$ second class constraints eliminate $s$ 
variables. Since matrix $(\bigl\{S_a,S_b\bigr\})$ is nonsingular and 
antisymmetric, $s$ is even. Since $\dim (T^*M)$ is even as well, $n$ is even.

Let us apply this theory to the Skyrme model~\cite{cebula93}. In 
{Euler-Rodrigues} parametrization~\eqref{rodpar} we have primary 
constraint~\eqref{aconstr} 
\begin{subequations} 
\label{sonestwo} 
\begin{equation}
S_1=\Phi_a \Phi_a-1=0;\quad a=1,2,3,4.\label{priconstr}
\end{equation}
The further requirement that the condition \eqref{priconstr} not vary in 
time can be satisfied by imposing a secondary constraint
\begin{equation}
S_2=\Phi_a M_{ab}\dot\Phi_b-1=0;\quad a,b=1,2,3,4\quad,
\end{equation}
\end{subequations}
where $M_{ab}$ is inertia density matrix in {Euler-Rodrigues}
parametrization explicit form of which is not important for further 
consideration (see \cite{cebula93} for details). It can be shown 
(using the canonical equations of motion) that this secondary 
constraint is independent of time~\cite{cebula93}. These two constrains
\eqref{sonestwo} are of the second class, because Poisson bracket of the 
fields $S_1$ and $S_2$ is non-vanishing. This is also consistent with the 
fact that second class constraints come in pairs.

If one carries out a canonical quantization by the usual commutators, 
trouble ensues from this noncommutativity $\bigl[ S_1,S_2 
\bigr]_{\text{PB}}\neq 0$. The quantum expression of the constraints is that 
every vector in Hilbert space must be annihilated by the constraint 
operators $\hat S_1$ and $\hat S_2$. It follows trivially that every vector 
must also be annihilated by their commutator and this conclusion is 
inconsistent since the commutator in question is itself nonvanishing in the 
canonical quantization. One resolution of this difficulty is to introduce 
modified classical brackets, the Dirac brackets, which share with the 
Poisson brackets all its basic algebraic properties, but are designed so 
that the Dirac bracket of any pair of second class constraints vanishes, in 
our case, $\bigl[ S_1,S_2 \bigr]_{\text{DB}} =0$. The Dirac brackets replace 
the Poisson brackets for determining the time evolution of relevant 
quantities. This replacement also eliminates the need to introduce a 
Lagrange multiplier field.

It can be shown~\cite{gitmanbook} that for second class constraints there 
exists a canonical transformation leading to complete elimination of 
dependent dynamical variables. By canonical transformation here we mean 
non-singular transformation of dynamical variables 
$\BBeta\rightarrow\BBeta^\prime,\quad \BBeta\equiv(\Bq,\Bp); \quad 
\BBeta^\prime\equiv(\Bq^\prime,\Bp^\prime)$ if for arbitrary functions 
$G_1(\BBeta)$ and $G_2(\BBeta)$ Poisson brackets are invariant in the 
following sense
\begin{equation}
\bigl\{G_1(\BBeta),G_2(\BBeta)\bigr\}=\bigl\{G^\prime_1(\BBeta^\prime), 
G^\prime_2(\BBeta^\prime)\bigr\},\quad 
G^\prime_i(\BBeta^\prime)=G_i(\BBeta),\quad i=1,2.
\label{cantr}
\end{equation}
Note, however, that finding this eliminating
transformation is nontrivial task, general 
solution of which is unknown~\cite{gitmanbook}. Dirac brackets in these new 
variables $\BBeta^\prime$  coincide with Poisson brackets.
In the \SU\ Skyrme model this elimination fortunately  can easy be done by 
introducing Euler angles parametrization. 

In SU(3) Skyrme model the 
problem cannot be completely solved by appropriate parametrization, 
because of the first class constraints appearance in the theory. In 
this case Dirac prescription for dealing with first class constraints should 
be followed. Quantization of SU(3) Skyrme model is also complicated by 
Wess-Zumino term, which eliminates an extra discrete symmetry that is not 
a symmetry of QCD~\cite{witten83,rabinovici84,witten83a}. Wess-Zumino term,
however, vanishes in the \SU\ case even in quantum model~\cite{norvaisas94}.

To summarize, despite Dirac quantization\footnote{More precisely 
Hamiltonian formulation scheme.} (with its constraints classification into 
primary/secondary/tertiary/$\dots$, first/second class) is convenient and 
traditionally often followed, it is, in fact not 
mandatory~\cite{faddeev88}, due  to existence~\cite{gitmanbook} of 
canonical transformations \eqref{cantr}, which eliminate dependent 
dynamical variables. If this elimination is technically formidable task, 
then Dirac procedure provides us with consistent quantization method. Euler 
angles \eqref{defeulang} parametrization automatically eliminates dependent 
dynamical variables, and thus is consistent with Dirac prescription.

\section{Quantization of skyrmion in collective coordinate approach}
\setcounter{equation}{0}
The section deals with quantization of the "zero frequency modes" or
"collective coordinates" of the classical skyrmion. We proceed here
(as in the classical case) with arbitrary irreducible representation. The 
case of reducible representation is investigated in the last subsection.
\subsection{Collective coordinate approach}
An approximation of "zero modes" or "collective coordinates", which retains
just a few modes out of a possible infinite number of modes, requires
justification. It has been the subject of some criticism. We shall, however,
proceed with our calculations using this approximation.

Following {\it G.S.~Adkins} et~al.~\cite{adkins83} we shall employ collective
rotational coordinates\footnote{The method of collective coordinates
originally was introduced by {\it N.N~Bogolyubov} in~\cite{bogolubov50}.}
to separate the variables which depend on the time and spatial coordinates
\begin{equation}
\BU\bigl(\Bx,\Bq(t)\bigr)=\BA\bigl(\Bq(t)\bigr) \BU_c(\Bx)\BA^\dagger
\bigl(\Bq(t)\bigr),\qquad \BA\bigl(\Bq(t)\bigr)\in \SU_I.
\label{gencolcoo}
\end{equation}
Three real independent parameters $\Bq(t)=\bigl(q^1(t),q^2(t),
q^3(t)\bigr)$ are dynamical quantum variables --- skyrmion rotation 
(Euler) angles in the internal (isotopic) space $\SU_I=\diag\bigl( 
\SU_L\otimes\SU_R\bigr)$ but not in geometric space $\mathrm{SO(3)}$. The 
skyrmion remains static in geometric space.

Quantum fluctuations near the classical solution\footnote{Recall that the
most general collective rotational variables $\BU(\Bx,t)=\BV(t)\BU_c(\Bx)
\BW^{-1}(t)$ corresponding to the full chiral invariance group
$\SU_L\otimes\SU_R$ of the Lagrangian \eqref{genlagden} are not appropriate,
because we are interested only in fluctuations preserving
$\BU(\Bx,t)=\mathbf{1}$, when $|\Bx|\rightarrow \infty$.}
can be put into two different classes. Namely, fluctuation modes which
are generated by action or Hamiltonian symmetries and modes orthogonal to
the symmetric one. Symmetric fluctuation modes are of primary importance in
quantum description because the infinitely small energy perturbation can
lead to reasonable deviations from classical solution.
As a consequence, collective rotation matrices 
$\mathbf{A}(t),\mathbf{A^\dagger}(t)$ in \eqref{gencolcoo} are not
required to be small (i.e.~close to the identity matrix).
\subsection{Commutation relations}
We shall consider the Skyrme Lagrangian \eqref{genlagden} quantum
mechanically {\it ab initio}. The generalized coordinates $\Bq(t)$ and
velocities $\dot \Bq (t)$ then satisfy the commutation
relations\footnote{We quantize only internal (isotopic) rotational degrees
of freedom of the static soliton. Note that this does not imply that the
quantization cannot affect the shape of the solution in geometric space.
Conversely, because internal and geometric indices are mixed in the
solution \protect\eqref{genskyanz}
the shape of quantum hedgehog ansatz is significantly modified
(see Fig.~\ref{fig10}).}
\begin{equation}
\bigl[\dot q^r,\,q^k\bigr]=-ig^{rk}(\Bq).
\label{gencomrel}
\end{equation}
Here the tensor $g^{rk}(\Bq)$ is a function of generalized
coordinates \Bq\ only, the explicit form of which is determined after
the quantization condition has been imposed\footnote{Assumption
\protect\eqref{gencomrel} actually is a consequence
\protect\cite{sugano71,lin70}
of canonical commutation relation \protect\eqref{cancomrel} and canonical
momentum definition \protect\eqref{canmomdef}.}. The tensor $g^{rk}$ is
symmetric with respect to interchange of the indices $r$ and $k$ as a
consequence of the commutation relation $[q^r,\,q^k]=0$. Indeed, 
differentiation of the relation gives $[\dot q^r,\,q^k]= [\dot
q^k,\,q^r]$, from what it follows that $g^{rk}$ is symmetric. The commutation
relation between a generalized velocity component $\dot q^k$ and arbitrary
function $G(\Bq)$ is given by
\begin{gather}
\bigl[\dot q^k,\,G(\Bq)\bigr]=-i\sum_r g^{kr}(\Bq)\frac\partial {\partial
q^r}G(\Bq).
\label{velcom}\\
\intertext{We shall employ the Weyl ordering for the noncommuting
operators $\dot\Bq, G(\Bq)$ throughout}
\partial_t G(\Bq)=\frac12 \Bigl\{\dot q^r,\,\frac{\partial 
G(\Bq)}{\partial q^r}\Bigr\},
\label{weylordnew}
\end{gather}
where symbol $\partial_t G(\Bq)$ further in the work is understood as
$\bigl(\partial_t G(\Bq)\bigr)_W$. The {\it operator} ordering is 
fixed by the form of the classical Lagrangian \eqref{genlagden} and no
further ambiguity associated with it appears at the level of the
Hamiltonian. In order to find the explicit form of $g^{rk}(\Bq)$ one can
substitute \eqref{gencolcoo} into \eqref{genlagden} and keep only terms
quadratic in velocities\footnote{Lagrangian formulation and absence of the
first order constrains suggest that terms linear in $\dot\Bq$ should not
appear. This is indeed the case \protect\cite{fujii87}. Terms which are
independent of $\dot\Bq$ do not contribute to momenta and thus to $g^{rk}$
also.}
\begin{align}
\hat{L}(\dot \Bq,\Bq,F)=&\frac1N \int \ 
\hat{\!\!\!\CL}\bigl(\Bx,\dot\Bq (t),\Bq(t),F(r)\bigr)
r^2\sin\vartheta \di r\di\vartheta \di\varphi
\notag \\=& \frac 12\dot q^r g_{rk}(\Bq)\dot q^k
+\text{$(\dot \Bq)^0$-order  term},
\label{notallqlag}\\
\intertext{where}
N=&\frac23 j(j+1)(2j+1),
\end{align}
and we introduce here from the very beginning a normalization factor in the
Lagrangian \eqref{genlagden} in order to ensure baryon number $1$. The
$3\times 3$ metric tensor $g_{rk}(\Bq)$ is defined~\cite{norvaisas94} as 
the scalar product of a set of functions~\eqref{defccoef} 
$C_r^{(m)}(\Bq)$ or 
$C_r^{\prime (m)}(\Bq)$ 
\begin{align}
g_{rk}(\Bq)&=-\frac12 a(F)\sum_m(-)^m C_r ^{(m)}(\Bq)C_k^{(-m)}(\Bq)=
-\frac12 a(F)\sum_m(-)^m C_r^{\prime (m)}(\Bq)C_k^{\prime(-m)}(\Bq)\notag
\\
&=a(F)\delta _{r,k}+a(F)(\delta _{r,1}\delta _{k,3}+\delta_{r,3}
\delta_{k,1})\cos q^2,
\label{defmetr}
\end{align}
where $a(F)$ (soliton inertia moment) is the following integral:
\begin{align}
a(F)&=\int \di^3\Bx \CA\bigl(F(r)\bigr)=
\frac 1{e^3f_\pi }\tilde a(F)\notag \\
&=\frac 1{e^3f_\pi }\frac{8\pi }3\int \di
\tilde r\tilde r^2\sin ^2F\Bigl( 1+F^{\prime 2}+\frac{\sin
^2F}{\tilde r^2}\Bigr).
\label{defatil}
\end{align}
The appropriate definition for the canonical momentum $p_r$ (which is
conjugate to $q^r$) is
\begin{align}
p_r(\dot \Bq,\Bq,F)=\frac{\partial \hat{L}(\dot
\Bq,\Bq,F)}{\partial
\dot q^r}&=\frac 12\bigl\{\dot q^k ,g_{rk}(\Bq)\bigr\}.
\label{canmomdef}
\\
\intertext{The canonical commutation relation}
\left[ p_r (\dot \Bq,\Bq,F),q^k \right]&=-i\delta _{r}^{\phantom{r}k}  ,
\label{cancomrel}\\
\intertext{
then yields the following explicit form for functions $g^{rk}(\Bq)$:
}
g^{rk}(\Bq)&=g_{rk}^{-1}(\Bq).
\label{invdef}
\end{align}
Note that for the time being we do not require $\left[p_r,p_k\right]=0$.
\subsection{Angular momentum operators and remark on
$\left[p_r,p_k\right]$}
Because of the model spherical symmetry\footnote{Important here is the
spherical symmetry of dynamical degrees of freedom $\Bq:\quad
\SU_I\sim\text{SO(3)}_I$.} it is convenient to introduce operators
$\hat\BJ,\hat\BJ^\prime$ instead of canonical momentum operators $\Bp$
\begin{subequations}
\begin{equation}
\hat J_a^{\prime }=-\frac i2\left\{ p_r,C_{\left(a\right) }^{\prime
r}(\Bq)\right\} =(-1)^a\frac{ia(F)}4\left\{ \dot q^r,C_r^{\prime
\left(-a\right) }(\Bq)\right\},
\label{defjstrix}
\end{equation}
\begin{equation}
\hat J_a=-\frac i2\left\{ p_r,C_{\left(a\right) }^r(
\Bq)\right\} =(-1)^a\frac{ia(F)}4\left\{ \dot q^r, C_r^{\left( -a\right)
}(\Bq)\right\}.
\end{equation}
\end{subequations}
Straightforward but lenghty calculations then show that the following
relations hold~\cite{fujii87}:
\begin{multline}
\Bigl[\bigl\{p_l,C^{\prime l}_{(a)}(\Bq)\bigr\},
\bigl\{p_k,C^{\prime k}_{(b)}(\Bq)\bigr\}\Bigr]=\Bigl\{
C^{\prime l}_{(a)}(\Bq),\bigl\{C^{\prime k}_{(b)}(\Bq),
\bigl[p_l,p_k\bigr]\bigr\}\Bigr\}\\
+2i
\Bigl[
\begin{matrix}
1 & 1 & 1\\a & b &a+b
\end{matrix}
\Bigr] \bigl\{p_k,C^{\prime k}_{(a+b)}(\Bq)\bigr\}.
\label{komanticomrel}
\end{multline}
In obtaining \eqref{komanticomrel} the below equalities  are useful:
\begin{multline}
\Bigl[\bigl\{a,b\bigr\},\bigl\{c,d\bigr\}\Bigr]=
\Bigl\{a,\bigl\{c,[b,d]\bigr\}\Bigr\}+
\Bigl\{b,\bigl\{c,[a,d]\bigr\}\Bigr\}
+\Bigl\{a,\bigl\{d,[b,c]\bigr\}\Bigr\}\\
+\Bigl\{b,\bigl\{d,[a,c]\bigr\}\Bigr\},
\label{komoftwoanticom}
\end{multline}
\begin{align}
\Bigl[\bigl\{a,b\bigr\},c\Bigl]=&
\Bigl\{a,\bigl[b,c\bigr]\Bigr\}+
\Bigl\{\bigl[a,c\bigr],b\Bigr\},
\label{komofanticom}
\\
\partial_k C^{\prime(a+b)}_l(\Bq)
-\partial_l C^{\prime(a+b)}_k(\Bq)=&
\Bigl[
\begin{matrix}
1 & 1 & 1\\a & b &a+b
\end{matrix}
\Bigr] C^{\prime(a)}_k(\Bq) C^{\prime(b)}_l(\Bq),
\label{rulecpri1}
\\
\bigr[p_i,G(\Bq)\bigr]=&-i\partial_i G(\Bq),
\label{prule}
\end{align}
where $G(\Bq)$ is an arbitrary function of $\Bq$. Thus from
\eqref{komanticomrel} we have
\begin{equation}
\bigl[\hat J^\prime _a,\hat J^\prime _b\bigr]=-\frac14\Bigl(\Bigl\{
C^{\prime l}_{(a)} (\Bq) ,\bigl\{ C^{\prime k}_{(b)}(\Bq),\bigl[p_l,p_k\bigr]
\bigr\}\Bigr\}\Bigr)+
\Bigl[
\begin{matrix}
1 & 1 & 1\\a & b &a+b
\end{matrix}
\Bigr] \hat J^\prime _{a+b}.
\label{jjcomwithp}
\end{equation}
Let us examine commutator $\bigl[p_r,p_k\bigr]$. From \eqref{defjstrix} we
have
\begin{equation}
p_r=\frac{i}2\Bigl\{C^{\prime (a)}_{r}(\Bq),\hat J_a^\prime \Bigr\}.
\label{defp1}
\end{equation}
Introducing vielbeins $h_r^{(a)}(\Bq)$ and dual vielbeins $h^r_{(a)}(\Bq)$
\begin{subequations}
\begin{alignat}{2}
&
\sum_a (-1)^a h_r^{(a)}(\Bq) h_k^{(-a)}(\Bq)=g_{rk}(\Bq),
&\quad &
\sum_a (-1)^a h^r_{(a)}(\Bq) h^k_{(-a)}(\Bq)=g^{rk}(\Bq),
\\
&
h_r^{(a)}(\Bq) =i\sqrt{\frac{a(F)}2}C^{\prime (a)}_{r}(\Bq),
&\quad &
h^r_{(a)}(\Bq) =-i\sqrt{\frac2{a\smash{(F)}}}C^{\prime r}_{(a)}(\Bq),
\end{alignat}
\label{defvielbeins}
\end{subequations}
we can rewrite \eqref{defp1} in a geometrically more suitable form
\begin{equation}
p_r=\frac{1}{\sqrt{2 a\smash{\left(F\right)}}}\Bigl\{h_r^{(a)}(\Bq),\hat 
J_a^\prime\Bigr\}.
\label{defp2}
\end{equation}
Utilizing equations
\begin{subequations}
\begin{align}
\Bigl[\bigl[\hat J_c^\prime,h_k^{(a)}(\Bq)\bigr],h_s^{(b)}(\Bq)\Bigr]&=0,
\\
\Bigl[\bigl[\hat J_c^\prime,h_k^{(a)}(\Bq)\bigr],
\bigl[\hat J_d^\prime,h_s^{(b)}(\Bq)\bigr]\Bigr]&=0 ,
\end{align}
\end{subequations}
which are obtained without recourse to $\bigl[p_r,p_k\bigr]$ we obtain with
the help of \eqref{defp2}, \eqref{komoftwoanticom}, \eqref{rulecpri1},
\eqref{komofanticom}, \eqref{cpriortreltwo} and \eqref{prule}
\begin{equation}
\begin{split}
\bigl[p_r,p_k\bigr]=&\frac{1}{2a(F)}
\Bigl[\bigl\{h_r^{(a)}(\Bq),\hat J_a^\prime\bigr\},\bigl\{h_k^{(b)}(\Bq) ,
\hat J_b^\prime\bigr\}\Bigr]
\\=&\frac{1}{2a(F)}\biggl(\Bigl\{ h_{r}^{(a)} (\Bq)
,\bigl\{ h_{k}^{(b)}(\Bq),\bigl[\hat
J_a^\prime,\hat J_b^\prime\bigr]\bigr\}\Bigr\}
+2\Bigl\{\hat J_b^\prime,\bigl[\hat J_a^\prime
,h_r^{(b)}(\Bq)\bigr] h_k^{(a)}(\Bq)\\&\phantom{
\frac{1}{2a(F)}\Bigl(\Bigl\{ h^{r}_{(a)} (\Bq) ,\bigl\{
h^{k}_{(b)}(\Bq),\bigl[\hat J_r^\prime,\hat J_k^\prime\bigr]\bigr\}\Bigr\}
+2\Bigl\{\hat J_r^\prime,
}
-\bigl[\hat J_b^\prime ,h_r^{(a)}(\Bq)\bigr] h_k^{(b)}(\Bq)\Bigr\}\biggr)\\
=&\frac{1}{2a(F)}\Bigl\{ h_{r}^{(a)} (\Bq) ,\bigl\{
h_{k}^{(b)}(\Bq),\bigl[\hat J_a^\prime,\hat J_b^\prime\bigr]\bigr\}\Bigr\}\\
&+\frac14\Bigl\{\hat J_{m+n}^\prime,\Bigl[
\begin{matrix}
1 & 1 & 1\\m & n &m+n
\end{matrix}
\Bigr]
C^{\prime (m)}_{r}(\Bq) C^{\prime (n)}_{k}(\Bq)\Bigr\}.
\end{split}
\label{ppcomrel}
\end{equation}
It is easy to see that metric \eqref{defmetr} is invariant under local
rotations of vielbeins \eqref{defvielbeins}
\begin{equation}
h^{(a)}_r(\Bq)\rightarrow h^{\prime (a)}_r(\Bq) =D^1_{a,b}\bigl(\Balpha
(\Bq)\bigr) h^{(b)}_r(\Bq).
\label{rotvielbeins}
\end{equation}
This is related to the fact that space defined by metric \eqref{defmetr}
is of constant curvature
\begin{subequations}
\begin{gather}
R=R^{rk}_{\phantom{rk}kr}=-\frac3{2a(F)},
\\
R^{rk}_{\phantom{rk}sl}=g^{kp} \Bigl(\partial_s\Gamma^r_{\phantom{r}pl}
-\partial_l\Gamma^r_{\phantom{r}ps}+
\Gamma^r_{\phantom{r}sh}\Gamma^h_{\phantom{h}pl}
-\Gamma^r_{\phantom{r}lh}\Gamma^h_{\phantom{h}ps}\Bigr),
\\
\intertext{$\Gamma^r_{\phantom{r}pl}$ being Christoffel symbols}
\Gamma^r_{\phantom{r}ls}=\frac12 g^{rp}\bigl( \partial_l g_{ps}
+\partial_s g_{pl}-\partial_p g_{ls}\bigr).
\label{defchristoffel}
\end{gather}
\label{defconstcurv}
\end{subequations}
Equations \eqref{gencomrel} and \eqref{invdef} then imply that one can
locally rotate vielbeins \eqref{rotvielbeins} without affecting commutation
relation \eqref{cancomrel}.

Defining spin connection $A^a_{\phantom{a}b;r}(\Bq)$ in usual way
\cite{fujii87,kimura71a,flandersbook}
\begin{subequations}
\begin{align}
\nabla_r C^{\prime (a)}_k(\Bq)=&\partial_r C^{\prime (a)}_k(\Bq) -
\Gamma^s_{\phantom{s}rk} C^{\prime (a)}_s(\Bq),
\\
\nabla_r C^{\prime (a)}_k(\Bq)=&-A^a_{\phantom{a}b;r}(\Bq) C^{\prime
(b)}_k(\Bq),
\end{align}
\end{subequations}
and employing \eqref{defchristoffel} and \eqref{rulecpri1} we get 
explicit form for connection $A^a_{\phantom{a}b;r}(\Bq)$
\begin{equation}
A^a_{\phantom{a}a-b;r}(\Bq)=-\frac12 \Bigl[
\begin{matrix}
1 & 1 & 1\\b & a-b &a
\end{matrix}
\Bigr] C^{\prime (b)}_r(\Bq),
\end{equation}
which transformes under local rotation of vielbeins \eqref{rotvielbeins}
in well known way
\begin{equation}
A^{\prime a}_{\phantom{\prime a}b;r}(\Bq)= 
D^1_{a,c}\bigl(\Balpha(\Bq)\bigr)A^c_{\phantom{c}d;r}(\Bq) 
D^{1\dagger}_{d,b}\bigl(\Balpha(\Bq)\bigr)-\partial_r 
D^1_{a,c}\bigl(\Balpha(\Bq)\bigr) 
D^{1\dagger}_{c,b}\bigl(\Balpha(\Bq)\bigr). 
\label{contransf}
\end{equation}
Rewriting equation \eqref{ppcomrel} in terms of spin connection 
$A^a_{\phantom{a}b;r}(\Bq)$
\begin{align}
\bigl[p_r,p_k\bigr]=&
\frac{1}{2a(F)}\Bigl\{ h_{r}^{(a)} (\Bq)
,\bigl\{ h_{k}^{(b)}(\Bq),\bigl[\hat J_a^\prime,\hat
J_b^\prime\bigr]\bigr\}\Bigr\}+\frac{i}{\sqrt{2a(F)}}\Bigl\{\hat J_m^\prime
,A^m_{\phantom{m}n;r}(\Bq) h_{k}^{(n)}(\Bq)\Bigr\},\notag\\
\intertext{or in a form reflecting the expression symmetry in respect of  
interchange of indices $r,k$}=&
\frac{1}{2a(F)}\Bigl\{ h_{r}^{(a)} (\Bq)
,\bigl\{ h_{k}^{(b)}(\Bq),\bigl[\hat J_a^\prime,\hat
J_b^\prime\bigr]\bigr\}\Bigr\}\notag
\\ \phantom{=}&
+\frac{i}{2\sqrt{2a(F)}}\Bigl\{\hat J_m^\prime ,A^m_{\phantom{m}n;r}(\Bq)
h_{k}^{(n)}(\Bq)-A^m_{\phantom{m}n;k}(\Bq) h_{r}^{(n)}(\Bq)\Bigr\},
\label{ppsymform}
\end{align}
we are ready to show that one can perform a suitable local rotation
\eqref{rotvielbeins} of vielbeins around a point $P(\Bq)$ to ensure
$\bigl[p_r,p_k\bigr]=0$ at that point. From Jacobi identity
\begin{gather}
\bigl[\xi,\bigl[\eta,\zeta\bigr]\bigr]+
\bigl[\eta,\bigl[\zeta,\xi\bigr]\bigr]+
\bigl[\zeta,\bigl[\xi,\eta\bigr]\bigr]=0,\\
\intertext{we have}
\bigl[\bigl[p_r,p_k\bigr],G(\Bq)\bigr]=0,
\label{jacconseq}
\end{gather}
for an arbitrary function $G(\Bq)$ satisfying $\quad\partial_r \partial_k 
G(\Bq)=\partial_k \partial_r G(\Bq)$. Assume that such a rotation has 
been found. Then \eqref{jjcomwithp} reduces to
\begin{gather}
\bigl[\hat J^\prime _a,\hat J^\prime _b\bigr]=
\Bigl[
\begin{matrix}
1 & 1 & 1\\a & b &a+b
\end{matrix}
\Bigr] \hat J^\prime _{a+b},
\label{jjcom}\\
\intertext{and \eqref{ppcomrel} to}
\bigl[p_r,p_k\bigr]=
\frac{i}{2\sqrt{2a(F)}}\Bigl\{\hat J_m^\prime ,
A^m_{\phantom{m}n;r}(\Bq) h_{k}^{(n)}(\Bq)
-A^m_{\phantom{m}n;k}(\Bq) h_{r}^{(n)}(\Bq)\Bigr\},
\label{ppcomzero}
\end{gather}
where we have used symmetry reflecting form \eqref{ppsymform}.
Substitution of \eqref{ppcomzero}
into \eqref{jacconseq} leads to 
\begin{equation}
\Bigl(
A^m_{\phantom{m}n;r}(\Bq) h_{k}^{(n)}(\Bq)
-A^m_{\phantom{m}n;k}(\Bq) h_{r}^{(n)}(\Bq)
\Bigr)h^s_{(m)} \partial_s G(\Bq)=0.
\end{equation}
This implies that $A^m_{\phantom{m}n;r}(\Bq)\simeq 0$ in  the very vicinity 
of the point $P(\Bq)$. Thus, equation \eqref{contransf} with
$A^{\prime a}_{\phantom{\prime a}b;r}(\Bq)=0$ gives us explicit
partial differential equations for local rotation angles $\Balpha (\Bq)$.

When $\bigl[p_r,p_k\bigr]=0$, operators $\hat \BJ^\prime,\hat\BJ$ become
angular momentum operators with usual commutation relations \eqref{jjcom}
and 
\begin{equation}
\bigl[\hat J_a,\hat J_b\bigr]=
\Bigl[
\begin{matrix}
1 & 1 & 1\\a & b &a+b
\end{matrix}
\Bigr] \hat J_{a+b},
\label{jjc}
\end{equation}
respectively. The operator $\hat \BJ^{\prime }$ is then a ''right 
rotation'' generating matrix $\BD^\ell(\Bq)$
\begin{align}
\bigl[ \hat J_a^{\prime },D_{m,m^{\prime }}^\ell (\Bq)\Bigr]
=&-\Bigl\langle
\ell ,m^{\prime }+a\Bigl| \hat J_a\Bigr| \ell ,m^{\prime }\Bigr\rangle
D_{m,m^{\prime }+a}^\ell (\Bq),
\\
\intertext{and $\hat \BJ$
is a ''left rotation'' generating matrix $\BD^\ell (\Bq)$}
\Bigl[ \hat J_a,D_{m,m^{\prime }}^\ell (\Bq)\Bigr]=&\Bigl\langle
\ell ,m\Bigl|
\hat J_a\Bigr| \ell ,m-a\Bigr\rangle D_{m-a,m^{\prime }}^\ell ( \Bq).
\end{align}
\subsection{The Hamiltonian and state vectors}
In order to calculate all terms in quantum Lagrangian expression 
\eqref{notallqlag}, we substitute collective rotational coordinates 
\eqref{gencolcoo} into properly normalized Lagrangian density 
\eqref{genlagden}. Utilizing commutation rule \eqref{velcom} and relations 
\eqref{pdrel} for Wigner $\BD^j$~functions we can pull out all velocities to 
one or another side in symmetric fashion.
Formulas\footnote{Formulas \eqref{norform} are derived by 
{\it E.~Norvai\v sas} (private communication).}
\begin{subequations}
\label{norform}

\begin{align}
\Tr\{\bigl\langle jm\bigr|J^\prime_a  
J^\prime_b\bigl|jm\bigr\rangle\} &=(-1)^a\frac16j(j+1)(2j+1)\delta_{a,-b},\\
\Tr\{\bigl\langle jm\bigr| J^\prime_a J^\prime_b J^\prime_c
\bigl| jm\bigr\rangle\}&=
-(-1)^a \frac{j(j+1)(2j+1)}{3\cdot4}\Bigl[
\begin{matrix}
1 & 1 & 1 \\
c & b & -a
\end{matrix}\Bigr],\\
\Tr\{\bigl\langle jm\bigr| J^\prime_a J^\prime_b J^\prime_c
J^\prime_d \bigl| jm\bigr\rangle\}&=
\sum_k (-1)^{a+b}\frac14 j^2(j+1)^2(2j+1)^2
\Bigl\{
\begin{matrix}
j & j & k \\
1 & 1 & j
\end{matrix}
\Bigr\}^2\notag\\
&\phantom{=\sum_k}\times
\Bigl[
\begin{matrix}
1 & 1 & k \\
a & b & a+b
\end{matrix}\Bigr]
\Bigl[
\begin{matrix}
1 & 1 & k \\
c & d & c+d
\end{matrix}\Bigr],
\end{align}
\label{fullqlag}
\end{subequations}
then allow us to take trace explicitly. The resulting expression still 
contains a lot (up to $6$ for Lagrangian density and up to $8$ for Noether 
current densities) of sums over repeated group indices. We have used computer 
algebra system to make the explicit summation\footnote{Recently we have made 
some progress in completing the summation manually, but intermediate 
results are still very large. Computer algebra system also was used to 
check many of symbolic manipulations mentioned above.}. The result of all 
computations is the following explicit form for the quantum Lagrangian
\cite{acus96a}:
\begin{subequations}
\begin{align}
\hat L(\dot \Bq,\Bq,F)&=-M(F)-\Delta M_j(F)+\frac 1{a(F)}\hat \BJ^{\prime 
2},\\&=-M(F)-\Delta M_j(F)+\frac 1{a(F)}\hat \BJ^2,
\end{align}
\label{exqlag}
\end{subequations}
and its density
\begin{equation}
\hat{\!\!\!\CL}(\Br,\dot\Bq,\Bq)=\frac{3\CA(F)}{2\,a^{2}(F)}\Bigl(
\hat \BJ^{\prime 2}-(\hat \BJ^{\prime }\cdot \bar \Bx)(\hat \BJ
^{\prime }\cdot \bar \Bx)\Bigr) -\Delta \CM_j(F)-\CM(F).
\label{defqlagden}
\end{equation}
Here $M(F)$ is classical soliton mass
\begin{align}
M(F)=&\int \di^3\Bx\CM(F(r))=\frac{f_{\pi }}{e}\tilde{M}(F)
\notag \\
=&2\pi \frac{f_{\pi }}{e}\int \di\tilde{r}\tilde{r}^{2}\biggl( F^{\prime
2}+\frac{\sin ^{2}\!F}{\tilde r^2}\Bigl( 2+2F^{\prime 2}+\frac{\sin
^2\!F}{\tilde r^2}\Bigr) \biggr),
\label{defclamas}\\
\intertext{and}
\Delta M_j(F)=&\int \di^3\Bx\Delta\CM_j(F(r))=
e^3f_\pi \Delta \tilde{M_j}(F)\notag \\
=&-\frac{2\pi e^3f_\pi
}{5\tilde a^2}\!\!\int \!\di\tilde r\tilde r^2\sin ^2\!F
\Bigl(5+2(2j-1)(2j+3)\sin ^2\!F+\bigl(8j(j+1)-1\bigr)F^{\prime 2}\notag\\
&+\bigl(2j(j+1)+1\bigr)\frac{\sin ^2F}{\tilde r^2}
-2(2j-1)(2j+3)F^{\prime 2}\sin
^2\!F\Bigr),\label{defdeltam}
\end{align}
is quantum mass correction.
The angular momentum operator on the r.h.s. of \eqref{defqlagden} can be
separated into scalar and tensor terms in the usual way
\begin{align}
\hat \BJ^{\prime 2}-(\hat \BJ^{\prime }\cdot \bar \Bx)(\hat \BJ
^{\prime }\cdot \bar \Bx) &=\frac{2}{3}\hat \BJ^{\prime
2}-\frac{4\pi }{3}Y_{2,m+m^{\prime }}^*(\vartheta ,\varphi
)
\Bigl[
\begin{matrix}
1 & 1 & 2 \\
m & m^{\prime } & m+m^{\prime }
\end{matrix}
\Bigr] \hat J_m^{\prime }\hat J_{m^\prime }^\prime ,
\label{spherharm}
\end{align}
where $Y_{l,m}(\vartheta ,\varphi )$~\cite{varshalovichbook} is a 
spherical functions.

The volume integral of the Lagrangian density \eqref{defqlagden}
gives the Lagrangian \eqref{exqlag}. In the fundamental representation
($j=1/2$), the second rank tensor part of \eqref{spherharm} vanishes. This
implies that the quadrupole moment of the $\Delta_{33}$ resonance cannot be
described within the fundamental representation.

It is known~\cite{lin70} that in the quantum mechanics, the Hamilton
formalism is inconsistent (see also Sec.~\ref{qincs}) with Lagrange 
one for velocity dependent potentials if the Hamiltonian is defined by the 
ordinary method 
\begin{equation}
\hat{K}=\frac12\bigl\{p_r,\dot q^r \bigr\}-\hat{L},
\end{equation}
and that this $\hat K$ does not satisfy the canonical equations of motion
\begin{equation}
\frac{\partial \hat{H}(\Bq,\Bp)}{\partial p_i}=\dot q^i;\qquad
\frac{\partial \hat{H}(\Bq,\Bp)}{\partial q^i}=\dot p_i,
\label{caneq}
\end{equation}
with $\hat H$ replaced by $\hat K$. In a number of works 
\cite{
sugano71,kimura71,kimura72,sugano72,sugano73a,lin70,sugano73,kimura73,kiang69} 
it has been shown that
consistent Lagrange and Hamilton formalism exists\footnote{Moreover,
consistent variation in connection with Noether theorem exists for the case
of constant curvature \protect\cite{sugano72}.} for constant curvature
\eqref{defconstcurv} spaces. The correct Hamiltonian is given
\cite{fujii87} by
\begin{equation}
\hat{H}=\frac12\bigl\{p_r,\dot q^r \bigr\}-\hat{L}(\dot \Bq,\Bq)-Z(\Bq),
\end{equation}
satisfying the canonical equations \eqref{caneq} of motion. The extra term
$Z(\Bq)$ arises from noncommutativity of operators and can be expressed in
terms of $g^{rk}(\Bq)$ and $g_{rk}(\Bq)$, so that it does not involve $\dot
\Bq$ or $\Bp$. There exist a few explicit forms of $Z(\Bq)$.
Calculations become shorter if one uses the following expression of 
$Z(\Bq)$, given by {\it R.~Sugano}~\cite{sugano71}:
\begin{equation}
Z(\Bq)=-\frac1{16} g^{rk}\bigl(\partial_r g^{sp}\bigr)\bigl(
\partial_k g_{sp}\bigr)
+\frac18 g^{rk}\bigl(\partial_r g^{sp}\bigr)\bigl(
\partial_s g_{kp}\bigr).
\end{equation}
By direct calculations we see the following identity to hold:
\begin{equation}
\frac12 \bigl\{p_r,\dot q^r \bigr\}-Z(\Bq)=\frac2{a(F)} \hat \BJ^{\prime 2}.
\end{equation}
Thus the true Hamiltonian is written as
\begin{subequations}
\begin{align}
\hat H_j(F)=&M(F)+\Delta M_j(F)+\frac 1{a(F)}\hat \BJ^{\prime 2},\\
=&M(F)+\Delta M_j(F)+\frac 1{a(F)}\hat \BJ^2.
\end{align}
\label{hamdef}
\end{subequations}
Note, that {\it both} the second and the third terms on the right-hand side 
of~\eqref{hamdef} are of the order of $\hbar^2$, the magnitude being
characteristic of operator ordering contribution~\cite{hayashi92,dewitt57}.
The Hamiltonian \eqref{hamdef} yields~\cite{fujii87} canonical equations 
of motion \eqref{caneq} under the condition $\hat{\dot \BJ}=\smash{\hat{\dot 
\BJ}}^\prime =0$. This condition is consistent with $\hat 
H(\Bq,\Bp)$ to be a Hamiltonian, because we have
\begin{equation}
\bigl[\hat\BJ^\prime ,\hat H(\Bq,\Bp)\bigr]=0;
\qquad \bigl[\hat\BJ,\hat  H(\Bq,\Bp)\bigr]=0,
\end{equation}
from commutation relations \eqref{jjcom} and \eqref{jjc}.
The most important feature of this result is that the quantum
correction $\Delta M_j(F)$ is negative definite and that it depends
explicitly on the dimension of the representation of the
$SU(2)$ group. This term is lost in the usual semiclassical treatment
\cite{adkins83} of the Skyrme model even in the fundamental representation
\cite{fujii87} of $SU(2)$, because that ignores the commutation relations.
For the Hamiltonian \eqref{hamdef} the normalized {\it state vectors}
with fixed spin and isospin $\ell $ are
\begin{equation}
\genfrac{|}{\rangle}{0pt}{}{\ell }{m_{\text{isospin}},m^\prime_{\text{spin}}
} =\frac{\sqrt{2\ell +1}}{4\pi } D_{m,m^\prime }^\ell (\Bq)
\bigl| 0\bigr\rangle .
\label{hamnotoper}
\end{equation}
These have the eigenvalues
\begin{equation}
H(j,\ell ,F)=M(F)+\Delta M_j(F)+\frac{\ell (\ell +1)}{2a(F)}.
\label{defqmass}
\end{equation}
This expression is the quantum version of the mass formula of the Skyrme
model.
The Hamiltonian density corresponding to the
Hamiltonian \eqref{hamdef} has the following matrix elements for baryon
states with spin and isospin $\ell >1/2$:
\begin{align}
\genfrac{\langle}{|}{0pt}{}{\ell }{m_i,m_s}\ \,
\hat{\!\!\!\CH}(\Br,\Bq)
\genfrac{|}{\rangle}{0pt}{}{\ell }{m_i,m_s}
=&\CM(F(r))+\Delta \CM_j(F(r))+\frac{\CA(F(r))}{2\,a^2(F)}
\biggl( \ell (\ell
+1)\notag \\
&-\sqrt{\frac23}\pi \bigl(3m_{s}^{2}-\ell (\ell +1)\bigr)
Y_{2,0}(\vartheta ,\varphi )\biggr).
\label{defhamdens}
\end{align}
For nucleons $\ell =1/2$ the dependence on angles is absent and the quantum
skyrmion is, therefore, spherically symmetric as required. The states
with larger spin than $1/2$ are, thus, described as deformed.
\subsection{Reducible representations}
Here we provide modifications which are necessary when passing
to the general reducible representation
\begin{equation}
\BD^j\rightarrow \sum_k \oplus \BD^{j_k}.
\end{equation}
Thus $\Delta M_j$ \eqref{defdeltam} in quantum Lagrangian
\eqref{exqlag} (and its density \eqref{defqlagden}), Hamiltonian
\eqref{hamdef} (and its density \eqref{defhamdens}) and
quantum mass formula \eqref{defqmass} is modified to
\begin{align}
\Delta M_{\Sigma_j}(F)=&\int \di^3\Bx\Delta \CM_{\Sigma_
j}(F(r))=e^{3}f_\pi \Delta \tilde M_{\Sigma_j}(F)\notag\\
=&-\frac{2\pi e^3 f_\pi}{15\tilde a^2}  
\!\!\!\int\!\di\tilde r\tilde r^2\sin ^{2}\!F\Bigl(15+4d_{2}\sin
^{2}\!F(1-F^{\prime 2})+2d_{3}\frac{\sin
^{2}\!F}{\tilde r^2}+2d_{1}F^{\prime2}\Bigr).
\label{newdefdeltam}
\end{align}
The coefficients $d_1,d_2,d_3$ in these expressions are given as
\begin{subequations}
\begin{align}
d_1 &=\frac{1}{N}\sum_{k}j_{k}(j_{k}+1)(2j_{k}+1)\bigl(
8j_{k}(j_{k}+1)-1\bigr),\label{d1} \\
d_2 &=\frac{1}{N}\sum_{k}j_{k}(j_{k}+1)(2j_{k}+1)(2j_{k}-1)(2j_{k}+3),
\label{d2} \\
d_3 &=\frac{1}{N}\sum_{k}j_{k}(j_{k}+1)(2j_{k}+1)\bigl(
2j_{k}(j_{k}+1)+1\bigr),
\label{d3}
\end{align}
\end{subequations}
with the same generalized  normalization factor as in classical model
\eqref{rednorfac}
\begin{equation}
N=\frac23\sum_{k}j_{k}(j_{k}+1)(2j_{k}+1).
\label{qrednorfac}
\end{equation}
Further in the work we proceed with the general reducible representation.
\section{The Noether currents}
\setcounter{equation}{0}
\label{noethercurr}
In the section we introduce vector and axial-vector
transformations of unitary field $\BU$ and calculate vector and axial-vector
currents assuming non-commutativity of dynamical variables in the
Lagrangian from the outset. Obtained expressions then are used to define
nucleon and $\Delta_{33}$-resonance 
magnetic momenta and axial coupling constant
$g_A$. We show that in this approach axial symmetry becomes broken, whereas
vector symmetry is still conserved. In the end we provide definitions of
various experimentally measurable radii following Ref.~\cite{adkins83}.

\subsection{Vector and axial transformations}\!\!\footnote{The 
presentation of the subsection follows Ref.~\cite{koch97}.}
Long before QCD was believed to be the theory of strong 
interactions, the phenomenological indications for the existence of chiral 
symmetry came from the study of the nuclear beta decay: $n\rightarrow 
p+e+\bar\nu$. There one finds that the weak coupling constants for the vector 
and axial-vector hadronic currents, $V$ and $A$, did not (in the case of $V$) 
or only slightly(25\% in the case of $A$) differ from those for the leptonic 
counterparts. Consequently, strong interaction "radiative" corrections to 
the weak vector and axial vector "charge" are absent. The same is true for 
the more familiar case of the electric charge and there we know that it is 
its conservation, which protects it from radiative corrections. Analogously, 
we expect the weak vector and axial vector charge, or more generally, 
currents, to be conserved due to some symmetry of the strong interaction. In 
case of the vector current, the underlying symmetry is the well known 
isospin symmetry of the strong interactions and thus the hadronic vector 
current is identified with the isospin current. Vector-transformation, 
therefore, is the isospin rotation. In terms of pions this can be written as
\begin{equation}
\Bpi\rightarrow\Bpi + \boldsymbol{\Theta}\times\Bpi,
\end{equation}
which states that the isospin direction of the pion is rotated by angle
$\boldsymbol{\Theta}$. The isospin, thus, is a constant of motion
associated with vector transformation. Also note that vector
transformation leaves the vacuum $\BU =\mathbf{1}$ invariant, due to
rotation in the isospace only.

The identification of the axial current, on the other
hand, is not so straightforward. This is due to another, very important and
interesting feature of the strong interaction, namely that the symmetry
associated with the conserved axial-vector current is "spontaneously
broken". By that, one means that while the Hamiltonian possesses the
symmetry, its ground state does not. An important consequence of
spontaneous breakdown of a symmetry is the existence of a massless mode, the
so-called Goldstone boson. In our case, the  Goldstone boson is the pion. If
chiral symmetry were a perfect symmetry of QCD, the pion should be
massless. Since chiral symmetry is only approximate, we expect the pion to
have a finite but small (compared to all other hadrons) mass. This is indeed
the case.
Interpretation of axial-transformation is also not so straightforward. It
can be shown~\cite{koch97} that this transformation mixes pion and
$\sigma$ meson states
\begin{align*}
&\Bpi\rightarrow\Bpi + \boldsymbol{\Theta}\sigma ,\\
&\sigma\rightarrow\sigma-\boldsymbol{\Theta}\cdot\Bpi .
\end{align*}
The pion thus is "rotated" into sigma meson under the axial
transformation and vice versa.

Let us consider the matrix element of the axial current between the vacuum
and the pion: $\bigl\langle 0\bigr|\ \
\hat{\!\!\!\!\CA}_\mu\bigl|\Bpi\bigr\rangle$. Because of
parity, the matrix element describes the weak decay of the pion and must be
proportional to the pion momentum (this is the only vector around)
\begin{equation}
\bigl\langle 0\bigr|\ \
\hat{\!\!\!\!\CA}^a_\mu(\Bx)\bigl|\pi^b(\Bp)\bigr\rangle=
-i f_\pi p_\mu \delta^{a,b}\mathrm{e}^{-i\Bp\cdot\Bx},
\label{pionaxialcurrent}
\end{equation}
where $\Bp$ is pion momentum, indices $a,b$ refer to isospin, $\mu$
indicates the Lorenz vector character of axial current, and $f_\pi$ is 
the pion decay constant determined from the experiment\footnote{Instead of 
$\mathrm{e}^{-i\Bp\cdot\Bx}$ generally one should use $f_p(\Bx)$ --- the 
asymptotic pion wave function as a solution of the Klein-Gordon 
equation~\cite{hayashi92}.}. Taking divergence of
\eqref{pionaxialcurrent} we obtain the relation which is often in the
literature referred to as the PCAC relation (partial conservation of axial
current)
\begin{equation}
\bigl\langle
0\bigr|\partial^\mu\ \ 
\hat{\!\!\!\!\CA}^a_\mu(\Bx)\bigl|\pi^b(\Bp)\bigr\rangle=
-i f_\pi m_\pi^2 \delta^{a,b}\mathrm{e}^{-i\Bp\cdot\Bx}=
-i f_\pi m_\pi^2\pi^a \delta^{a,b}.
\label{pcacusual}
\end{equation}
From this equation we see that to the extent that the pion mass is small
compared to hadronic scales, the axial current is approximately conserved.
Or in other words, the smallness of the pion mass is directly related to the
partial conservation of the axial current, i.e. to the fact that the axial
transformation is an approximate symmetry of QCD.

\subsection{Vector and axial currents in the Skyrme model}
The Lagrangian density of the Skyrme model is invariant under left
and right transformations of the unitary field $\BU(\Bq)$
\begin{subequations}
\begin{align}
&\BU\rightarrow\text{(left)}\BU\equiv\bigl(1-i2\sqrt{2}\omega^a
J^\prime_a\bigr)\BU ,
\\
&\BU\rightarrow\text{(right)}\BU\equiv\BU\bigl(1+i2\sqrt{2}\omega^a
J^\prime_a\bigr) .
\end{align}
\end{subequations}
The vector and axial Noether currents are nevertheless simpler and directly
related to physical observables. They are associated with the 
transformations~\cite{acus96a}
\begin{equation}
\BU(\Bx)\xrightarrow[\text{Axial transf.}]{\text{Vector transf.}} \left(
1-i2\sqrt{2}\omega^a J_a^\prime\right) \BU(\Bx)\left( 1{\displaystyle\pm}
i2\sqrt{2}\omega ^a J_a^\prime\right) ,
\label{defvecaxtrans}
\end{equation}
respectively. The factor $-2\sqrt{2}$ before the generators is introduced
so that the transformation \eqref{defvecaxtrans} for $j=1/2$ matches
the infinitesimal transformation in Ref.~\cite{adkins83}.
The corresponding Noether currents can be expressed in terms of the
collective coordinates~\eqref{gencolcoo}. After this substitution
Noether currents become operators in terms of the generalized collective
coordinates $\Bq$ and the generalized angular momentum operators $\hat
\BJ^{\prime }$ \eqref{defjstrix}. Long manipulations similar to those,
described in obtaining quantum Hamiltonian, lead to the
explicit expression for the vector current density
\begin{equation}
\begin{split}
\hat{\!\!\!\mathcal{V}}_b^a=&
\frac{\partial \ \,\hat{\!\!\!\CL}_V}{\partial \left(
 \nabla ^b\omega_a\right) }=(\text{left})
\frac{\partial \ \,\hat{\!\!\!\CL}}{\partial 
\left( \nabla^b\omega_a\right) }+(\text{right})
\frac{\partial \ \,\hat{\!\!\!\CL}}{\partial \left( \nabla^b\omega_a\right)}
\\
=&\frac{2\sqrt{2}\sin ^2\!F}r\Biggl(i
\biggl( f_\pi ^2
+\frac 1{e^2}\Bigl(
F^{\prime 2}
+\frac{\sin ^2\!F}{r^2}-\frac{2d_2+5}{4\cdot 5\cdot
a^2}\sin ^2\!F
\Bigr) \biggr)
\Bigl[
\begin{matrix}
1 & 1 & 1 \\
u & s & b
\end{matrix}
\Bigr] D_{a,s}^1(\Bq)\bar x_u\\
&
-\frac{\sin ^2\!F}{\sqrt{2} e^2 a^2}(-1)^s\biggl(\bigl[
\hat
\BJ^{\prime }\times \bar {\Bx}\bigr]_{-s}D_{a,s}^1(\Bq)\Bigl[
\bigl[
\hat \BJ^{\prime }\times \bar {\Bx}\bigr] \times \bar {
\Bx}\Bigr]_b \\
&
\phantom{
-\frac{\sin ^2F}{\sqrt{2} e^2 a^2(F)}(-)^s\Bigl(+
}
+\Bigl[\bigl[ \hat \BJ^{\prime }\times \bar {\Bx}\bigr] \times
\bar {
\Bx}\Bigr] _b D_{a,s}^1(\Bq)\bigl[ \hat \BJ^{\prime }\times
\bar { \Bx
}\bigr]_{-s}\biggr)\Biggr).
\end{split}
\label{isovcurr}
\end{equation}
Here $\nabla ^b$ is a circular component of the gradient operator. The
indices $a$ and $b$ denote isospin and spin components respectively. The
time (charge) component of the vector current density becomes~\cite{acus96a}
\begin{equation}
\begin{split}
\hat{\!\!\!\mathcal{V}}_t^a=&\frac{\partial 
\ \,\hat{\!\!\!\CL}_V}{\partial \left( \partial
_0\omega_a\right) }=
(\text{left})\frac{\partial 
\ \,\hat{\!\!\!\CL}}{\partial \left( \nabla
_0\omega_a\right) }+(\text{right})
\frac{\partial \ \,\hat{\!\!\!\CL}}{\partial \left( \nabla
_0\omega_a\right) }
\\
=&\frac{2\sqrt{2}(-1)^s}{a}\sin ^2\!F\biggl(
f^2_\pi
+\frac 1{e^2}\Bigl( F^{\prime 2}+\frac{\sin ^2\!F}{r^2}\Bigr) \biggr)
\Bigl( D_{a,-s}^1(\Bq)\hat J_s^{\prime }-D_{a,-s}^1({\bf
q})\bar
x_s(\hat \BJ^{\prime }\cdot \bar {\Bx})\Bigr).
\end{split}
\label{isovcurrtime}
\end{equation}
The explicit expression for the axial current density takes the
form~\cite{acus96a}
\begin{equation}
\begin{split}
\hat{\!\!\!\mathcal{A}}_b^a=&\frac{\partial 
\ \,\hat{\!\!\!\CL}_A}{\partial \left(
 \nabla ^b\omega_a\right) }=
(\text{left})\frac{\partial \ \,\hat{\!\!\!\CL}}{\partial \left( \nabla
^b\omega_a\right) }-(\text{right})
\frac{\partial \ \,\hat{\!\!\!\CL}}{\partial \left( \nabla
^b\omega_a\right) }
\\
=&\biggl( f_\pi ^2\frac{\sin2F}r+\frac 1{e^2}\frac{\sin 2F}r\Bigl(
F^{\prime 2}\frac{\sin^2\!F}{r^2}
-\frac{\sin ^2\!F}{4 a^2}\Bigr) \biggr)
D_{a,b}^1(\Bq)+\biggl( f_\pi ^2\Bigl(2F^{\prime
}-\frac{\sin2F}r\Bigr)
\\
&
-\frac 1{e^2}\Bigl(\frac{F^{\prime 2}\sin 2F}r+
\frac{\sin ^2\!F\sin 2F}{r^3}
-\frac{4F^{\prime}\sin ^2\!F}{r^2}-\frac{\sin ^2\!F\sin 2F}{4 a^2r}
\Bigr)\biggr) (-1)^s D_{a,s}^1(\Bq)\bar x_{-s}\bar x_b
\\
&
-
\frac{2F^{\prime}\sin ^2\!F(-1)^s}{e^2 a^2}
\biggl(D_{a,s}^1(\Bq)\bar x_{-s}\hat \BJ^{\prime 2}+\hat
\BJ^{\prime 2}D_{a,s}^1(\Bq)\bar x_{-s}-2 D_{a,s}^1(\Bq)\bar
x_{-s}(\hat \BJ^{\prime }\cdot \bar {\Bx})(\hat \BJ^{\prime }\cdot
\bar {\Bx})\biggr) \bar x_b
\\
&
-\frac{\sin ^2\!F\sin 2F}{e^2 a^2 r}(-1)^s\biggl( \Bigl[
\bigl[
\hat \BJ^{\prime }\times \bar {\Bx}\bigr] \times \bar {
\Bx}\Bigr]
_{-s}D_{a,s}^1(\Bq)\Bigl[ \bigl[ \hat \BJ^{\prime }\times \bar
{\Bx
}\bigr] \times \bar {\Bx}\Bigr] _b
\\
&
\phantom{
-\frac{\sin ^2\!F\sin 2F}{e^2 a^2 r}(-1)^s\biggl(
}
+\Bigl[ \bigl[ \hat \BJ^{\prime }\times \bar {\Bx}\bigr]
\times
\bar {\Bx}\Bigr] _b D_{a,s}^1(\Bq)\Bigl[ \bigl[ \hat
\BJ^{\prime
}\times \bar {\Bx}\bigr] \times \bar {\Bx}\Bigr]
_{-s}\biggr) .
\end{split}
\label{axialisovcurr}
\end{equation}
The operators \eqref{isovcurr}, \eqref{isovcurrtime} and
\eqref{axialisovcurr} are well defined for all representations $j$ of the
classical soliton and for fixed spin and isospin $\ell$ of the quantum
skyrmion. The new terms which are absent in the semiclassical case are those
that have the factor $a^2(F)$ in the denominator.

The matrix element of the divergence of the vector current density
\eqref{isovcurr} vanishes\footnote{The vector-transformation 
\eqref{defvecaxtrans} (finite) $\BU\rightarrow \mathbf{B}\BU\mathbf{B}^{-1}$ 
and collective coordinate rotations \eqref{gencolcoo} possesses the same 
symmetry, whereas axial-vector transformation (finite)
$\BU\rightarrow \mathbf{B}\BU\mathbf{B}$  does not.}.
\begin{equation}
\genfrac{\langle}{|}{0pt}{}{\ell}{m_i,m_s}
\nabla ^{b}\ \,\hat{\!\!\!\mathcal{V}}_{b}^{a}
\genfrac{|}{\rangle}{0pt}{}{\ell}{m_i,m_s}=0 .
\end{equation}
The result just confirms validity of the variation procedure on constant
curvature space~\cite{sugano73}.

\subsection{Baryon current density and magnetic momenta operators}
\begin{figure}
\begin{center}
\includegraphics*[width=12cm]{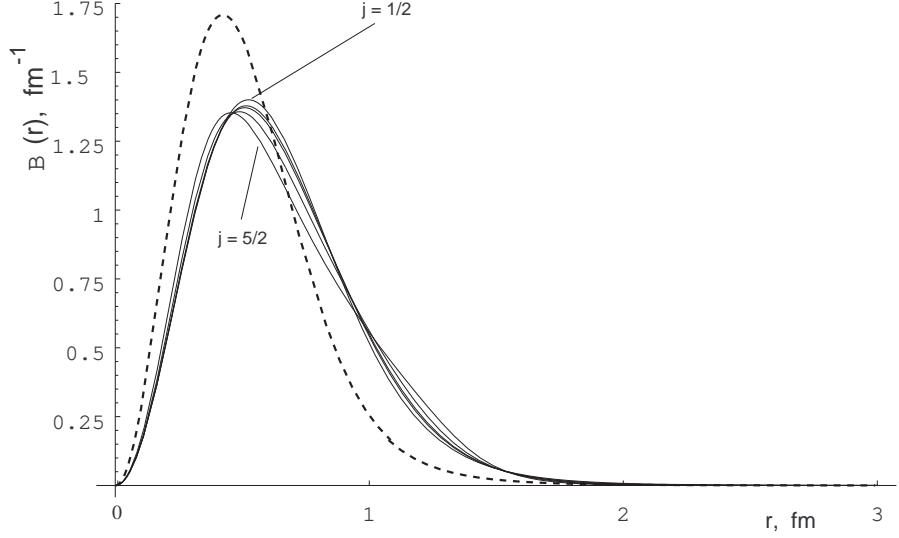}
\end{center}
\caption{Baryon charge density distribution. Dashed line denotes
classical chiral angle distribution, with parameters $e=5.45;\quad
f_\pi=64.5$~MeV taken from Ref.~\cite{adkins83}. Quantum
chiral angle distributions for various $j$ values are plotted with solid
lines.}
\label{fig10}
\end{figure}

The conserved topological current density in the Skyrme model is the baryon
current density. For the hedgehog solution its components take
the form~\cite{acus96a}
\begin{equation}
\hat{\!\!\! \CB}_a\bigl(\Bx,F(r)\bigr)=\frac 1{\sqrt{2}\pi ^2 r\,a(F)}
F^\prime \sin^2 \!F
\bigl[ \hat \BJ^\prime \times \bar \Bx \bigr]_a.
\label{qbarcurden}
\end{equation}
It is sketched in Fig.~\ref{fig10} for classical and quantum chiral angles.

Rotating soliton generates isoscalar magnetic moment associated with current
density \eqref{qbarcurden}.
The matrix elements of the third component of the
isoscalar magnetic momentum operator have the form~\cite{acus96a}
\begin{equation}
\begin{split}
\genfrac{\langle}{|}{0pt}{}{\ell }{m_i,m_s}
\bigl[ \hat{\mu}_{I=0}\bigr]_3
\genfrac{|}{\rangle}{0pt}{}{\ell }{m_i,m_s}
=&\genfrac{\langle}{|}{0pt}{}{\ell }{m_i,m_s}
\frac 12\int \di^3 \Bx r\bigl[ \bar \Bx \times \
\hat{\!\!\! \CB}\bigr]_0
\genfrac{|}{\rangle}{0pt}{}{\ell }{m_i,m_s}
\\
=&\frac{e}{f_\pi}\frac{\bigl( \ell(\ell+1)\bigr)^{1/2}}{3\tilde a}
\bigl\langle
\tilde r_{I=0}^2\bigr\rangle
\Bigl[
\begin{matrix}
\ell & 1 & \ell \\
m_s & 0 & m_s
\end{matrix}
\Bigr],
\end{split}
\label{isoscalarpart}
\end{equation}
where the isoscalar electric mean square radius is given as
\begin{equation}
\bigl\langle r_{E,I=0}^2\bigr\rangle=\frac1{e^2f^2_\pi}
\bigl\langle \tilde r_{E,I=0}^2\bigr\rangle;\qquad
\bigl\langle \tilde r_{E,I=0}^2\bigr\rangle =-\frac 2\pi \int \tilde
r^2 F^\prime(\tilde r) \sin^2F(\tilde r) \di \tilde r.
\label{esqrad}
\end{equation}
The matrix elements of the third component of the isovector part of
magnetic momentum operator that is obtained from the isovector current
\eqref{isovcurrtime} have the form~\cite{acus96a}
\begin{equation}
\begin{split}
\genfrac{\langle}{|}{0pt}{}{\ell }{m_i,m_s}
\bigl[ \hat \mu _{I=1}\bigr]_3
\genfrac{|}{\rangle}{0pt}{}{\ell }{m_i,m_s}
=&\genfrac{\langle}{|}{0pt}{}{\ell }{m_i,m_s}
\frac 12\int \di^3 \Bx r\bigl[ \bar \Bx \times \ 
\hat{\!\!\!\mathcal{V}}^{a=3}\bigr]_0
\genfrac{|}{\rangle}{0pt}{}{\ell }{m_i,m_s}
\\
=&
\biggl( \frac{\tilde a}{e^3\, f_\pi}+\frac{e}{f_\pi}\frac{8\pi }{3
\tilde a^2}\int \di \tilde r \tilde r^2\sin ^4F
\Bigl( 1-\frac{\ell(\ell+1)}3 -\frac{d_2}{2\cdot 5}\\
&
+\frac{(-1)^{2\ell}}2{\Bigl({\textstyle\frac{
5\ell(\ell+1)(2\ell-1)(2\ell+1)(2\ell+3)}{2\cdot 3}\Bigr)}^{\frac12}}
\Bigl\{
\begin{matrix}
1 & 2 & 1 \\
\ell & \ell & \ell
\end{matrix}
\Bigr\} \Bigl) \biggr)
\\
&\times
\Bigl[
\begin{matrix}
\ell & 1 & \ell \\
m_s & 0 & m_s
\end{matrix}
\Bigr] \Bigl[
\begin{matrix}
\ell & 1 & \ell \\
m_i & 0 & m_i
\end{matrix}
\Bigr]
,
\end{split}
\label{isovectorpart}
\end{equation}
where the symbol in the curly brackets is a $6j$
coefficient~\cite{varshalovichbook}.

From \eqref{isoscalarpart} and \eqref{isovectorpart} proton and neutron
magnetic moments measured in nuclear magnetons can be extracted using
relations
\begin{subequations}
\label{pnmagneticmom}
\begin{align}
\mu_p=&\frac12\Bigl({\scriptstyle
\genfrac{\langle}{|}{0pt}{}{1/2}{1/2,1/2}
[\hat \mu _{I=0}]_3
\genfrac{|}{\rangle}{0pt}{}{1/2}{1/2,1/2}
+
\genfrac{\langle}{|}{0pt}{}{1/2}{1/2,1/2}
[\hat \mu _{I=1}]_3
\genfrac{|}{\rangle}{0pt}{}{1/2}{1/2,1/2}}\Bigr),\\
\mu_n=&\frac12\Bigl({\scriptstyle
\genfrac{\langle}{|}{0pt}{}{1/2}{-1/2,1/2}
[ \hat \mu _{I=0}]_3
\genfrac{|}{\rangle}{0pt}{}{1/2}{-1/2,1/2}
-
\genfrac{\langle}{|}{0pt}{}{1/2}{-1/2,1/2}
[ \hat \mu _{I=1}]_3
\genfrac{|}{\rangle}{0pt}{}{1/2}{-1/2,1/2}}\Bigr) .
\end{align}
\end{subequations}
Similar formulas exist for $\Delta_{33}$-resonance magnetic moments
\begin{subequations}
\begin{align}
\Delta_{33}^{++}=&\frac12\Bigl({\scriptstyle
\genfrac{\langle}{|}{0pt}{}{3/2}{3/2,3/2}
[\hat \mu _{I=0}]_3
\genfrac{|}{\rangle}{0pt}{}{3/2}{3/2,3/2}
+
\genfrac{\langle}{|}{0pt}{}{3/2}{3/2,3/2}
[\hat \mu _{I=1}]_3
\genfrac{|}{\rangle}{0pt}{}{3/2}{3/2,3/2}}\Bigr),\\
\Delta_{33}^{+\phantom{+}}=&\frac12\Bigl({\scriptstyle
\genfrac{\langle}{|}{0pt}{}{3/2}{1/2,3/2}
[\hat \mu _{I=0}]_3
\genfrac{|}{\rangle}{0pt}{}{3/2}{1/2,3/2}
+
\genfrac{\langle}{|}{0pt}{}{3/2}{1/2,3/2}
[\hat \mu _{I=1}]_3
\genfrac{|}{\rangle}{0pt}{}{3/2}{1/2,3/2}}\Bigr),\\
\Delta_{33}^{0\phantom{+}}=&\frac12\Bigl({\scriptstyle
\genfrac{\langle}{|}{0pt}{}{3/2}{-1/2,3/2}
[\hat \mu _{I=0}]_3
\genfrac{|}{\rangle}{0pt}{}{3/2}{-1/2,3/2}
+
\genfrac{\langle}{|}{0pt}{}{3/2}{-1/2,3/2}
[\hat \mu _{I=1}]_3
\genfrac{|}{\rangle}{0pt}{}{3/2}{-1/2,3/2}}\Bigr),\\
\Delta_{33}^{-\phantom{+}}=&\frac12\Bigl({\scriptstyle
\genfrac{\langle}{|}{0pt}{}{3/2}{-3/2,3/2}
[\hat \mu _{I=0}]_3
\genfrac{|}{\rangle}{0pt}{}{3/2}{-3/2,3/2}
+
\genfrac{\langle}{|}{0pt}{}{3/2}{-3/2,3/2}
[\hat \mu _{I=1}]_3
\genfrac{|}{\rangle}{0pt}{}{3/2}{-3/2,3/2}}\Bigr) .
\label{deltamagneticmom}
\end{align}
\end{subequations}

\begin{figure}
\begin{center}
\includegraphics*[width=12cm]{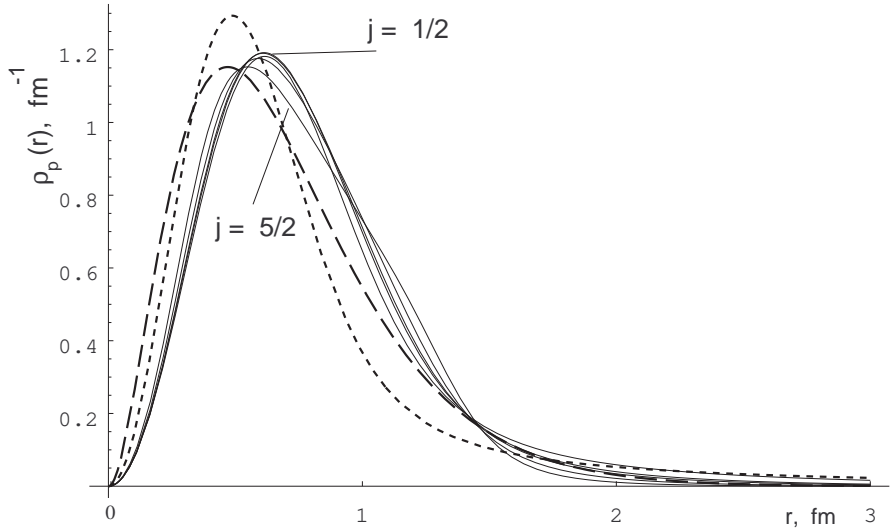}
\end{center}
\caption{Proton charge density distribution. Short-dashed line denotes
classical chiral angle distribution, with parameters $e=5.45;\quad
f_\pi=64.5$~MeV from Ref.~\cite{adkins83}. Quantum
chiral angle distributions for various $j$ values are plotted with solid
lines. Long-dashed line denotes semi-empirical charge density
distribution~\cite{jain90a}.}
\label{fig11}
\end{figure}

\begin{figure}
\begin{center}
\includegraphics*[width=12cm]{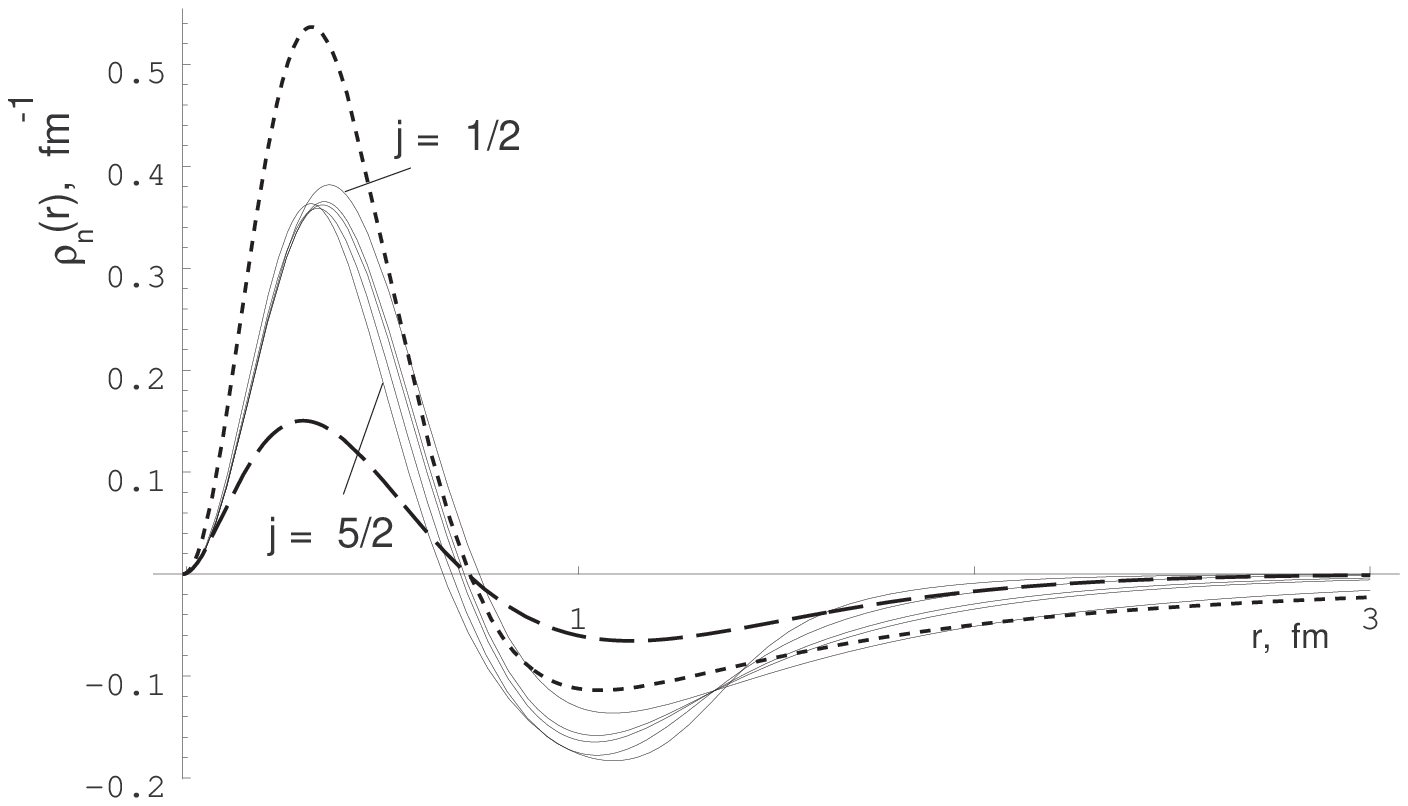}
\end{center}
\caption{Neutron charge density distribution. Short-dashed line denotes
classical chiral angle distribution, with parameters $e=5.45;\quad
f_\pi=64.5$~MeV from Ref.~\cite{adkins83}. Quantum
chiral angle distributions for various $j$ values are plotted with solid
lines. Long-dashed line denotes semi-empirical charge density
distribution~\cite{jain90a}.}
\label{fig12}
\end{figure}

Relations \eqref{pnmagneticmom} also allow us to obtain proton and neutron
charge distributions, when isoscalar and isovector current
densities\footnote{These densities are integrated over angular variables
$\varphi, \vartheta$. The factor $r^2$ (which comes from the Jacobian) is
also included to ensure usual dimensions.}
are used instead of integrated expressions
\eqref{isoscalarpart} and \eqref{isovectorpart}. These distributions are
measurable quantities for which semi-empirical formulas exist~\cite{jain90a}
\begin{subequations}
\begin{align}
\rho_p&=\frac{M^3_D r^2}{2}\exp\bigl(-M_D r\bigr);\qquad
M_D=0.84~\mathrm{GeV},\label{pemdistrib}
\\
\rho_n&=\frac{-\mu_n M_D^4 r}{2(5.6 M_D^2-4 m_N^2)}
\biggl(\frac{8m_N^2\bigl(\exp(-M_D r)-\exp(-2m_N r/\sqrt{5.6})\bigr)}{5.6
M_D^2-4 m_N^2}\notag
\\
&\phantom{=\frac{-\mu_n M_D^4 r}{2(5.6 M_D^2-4 m_N^2)}
\biggl(}
+r M_D\exp(-M_D r)\biggl).
\label{nemdistrib}
\end{align}
\end{subequations}
Distributions for classical (short-dashed) and quantum chiral angles in
various representations (solid lines) are plotted versus semi-empirical
distribution~\eqref{nemdistrib} and~\eqref{pemdistrib} (long-dashed line) in
Fig.~\ref{fig11} for proton and in Fig.~\ref{fig12} for neutron.
The neutron distribution peak value ($\sim 0.5$) in classical Skyrme model
is known to be too large when compared with the empirical value ($0.2-0.3$).
Our results for quantum chiral angle show significant improvement
in neutron charge density distribution, which only weakly depends on the
representation used. Note also that measurement indicates much faster
distribution fall in the asymptotic region even when compared with
semi-empirical formula~\cite{jain90a}. Speed up of quantum chiral soliton
distribution fall thus is also a step in right direction. For
experimental distributions we refer to~\cite{jain90a} and references therein.
\enlargethispage*{2.9cm}
The axial coupling $g_A$ measures the spin-isospin correlation in the
nucleon and is defined as the expectation value of the axial current
$A^a_\mu$ in a nucleon state at zero momentum
transfer~\cite{zahed86}. Comparison of the Fourier transform of the axial
current density~\eqref{axialisovcurr} with the usual phenomenological
expression for the axial-vector current leads to the axial-vector
coupling constant expression\footnote{In the limit of vanishing pion mass,
the resulting expression for $g_A$ should be multiplied by a factor of
$\frac32$. There are subtle points in calculating this quantity for
solutions with $\sim 1/r^2$ asymptotic~\protect\cite{adkins83}. 
This factor should not be used in calculating $g_A$ in finite pion mass
model~\cite{adkins83,hayashi92,jain90a}, and thus in quantum case either.}. 
From the axial current density~\eqref{axialisovcurr} we obtain the axial 
coupling constant $g_A$ of the nucleon as~\cite{acus96a}
\begin{align}
g_A=&
-3\genfrac{\langle}{|}{0pt}{}{1/2}{1/2,1/2}
\int \di^3\Bx \ \ \hat{\!\!\!\!\CA}_0^1
\genfrac{|}{\rangle}{0pt}{}{1/2}{-1/2,1/2}\notag\\
=&-\frac 1{e^2}\tilde{g}_1(F)-\frac{\pi
^2e^2}{3\tilde{a}^2(F)}\bigl\langle
\tilde{r}_{E,I=0}^2\bigr\rangle,
\label{defga}
\\
\intertext{where}
\tilde{g}_1(F)=&\frac{4\pi }3\int \di\tilde{r}\Bigl(
\tilde{r}^2F^\prime +
\tilde{r}\sin 2F+\tilde{r}F^{\prime 2} \sin 2F
+2F^\prime \sin ^2F+\frac{\sin^2\!F}{\tilde{r}}\sin 2F\Bigr) .
\end{align}
All versions of the topological soliton model lead to underprediction of
$g_A$ compared to the empirical value of $1.26$. In the simple Skyrme model
with a pion mass term \cite{adkins84}  the predicted value
for $g_A$ was only $0.65$. By readjusting the parameters so as to fit
empirical values for $f_\pi$ and $\pi NN$ coupling constant the value is
somewhat increased (to $0.82$). In the vector-meson-stabilised model 
\cite{meissner87} the values range between
$0.88$ and $0.99$, depending on the details of the model, but remains below
unity~\cite{riska90}. We shall see (see Tables of numerical results) that
axial coupling constant $g_A$ strongly depends (always grows with increasing
dimension of representation) on the representation employed. In
self-consistent quantum formalism in higher representations ($j>1/2$) $g_A$
is always above unity even for small ($\sim 0.5$~MeV) parameter $f_\pi$
values.
\subsection{Radii}
In addition to electric mean square radius
\eqref{esqrad} the following three mean square radii are used: electric 
isovector, isoscalar magnetic, isovector magnetic. Below we give definitions 
for all of them following Ref.~\cite{adkins83}.

For nucleons the isovector (electric) charge mean square radius becomes
\begin{align}
\left\langle r_{E,I=1}^2\right\rangle &=\frac 1{e^2f_\pi ^2}\left\langle
\tilde r_{E,I=1}^2\right\rangle \notag\\
&=\frac 1{e^2f_\pi ^2}\frac{\int \di \tilde
r\tilde r^4\sin ^2\!F\Bigl( 1+F^{\prime 2}+\frac{\sin ^2\!F}{\tilde r} \Bigr)
}{\int \di\tilde r\tilde r^2\sin ^2\!F\Bigr( 1+F^{\prime 2}+ \frac{\sin 
^2\!F}{\tilde r}\Bigr) }.
\end{align}
The isoscalar magnetic mean square radius is expressed as
\begin{align}
\left\langle r_{M,I=0}^2\right\rangle &=\frac 1{e^2f_\pi ^2}\left\langle
\tilde r_{M,I=0}^2\right\rangle \notag\\
&=-\frac 1{e^2f_\pi ^2}\frac 2\pi \frac{\int
\di\tilde r\tilde r^4F^{\prime }\sin ^2\!F}{\int \di\tilde r\tilde r^2
F^{\prime }\sin^2\!F},
\end{align}
and the isovector magnetic mean square radius takes the form
\begin{align}
\left\langle r_{M,I=1}^2\right\rangle &=\frac 1{e^2f_\pi ^2}\left\langle
\tilde r_{M,I=1}^2\right\rangle \nonumber \\
&=\frac 1{e^2f_\pi ^2}\frac{\int \di\tilde r\tilde r^4\sin ^2\!F\Bigl(
1+F^{\prime 2}+\frac{\sin ^2\!F}{\tilde r} +\frac{e^2\sin ^2\!F}{\tilde 
a^2(F)}\bigl( \frac 34-\frac{d_2}{10}\bigr) \Bigr) }{\int \di\tilde r\tilde 
r^2\sin^2\!F\Bigl( 1+F^{\prime 2}+\frac{\sin ^2\!F}{\tilde r}+\frac{ e^2\sin 
^2\!F}{\tilde a^2(F)}\bigl( \frac 34-\frac{d_2}{10}\bigr) \Bigr) }.
\end{align}
All these radii can be measured experimentally.

\section[The static properties of the nucleon and $\Delta_{33}$-resonance
 \dots]{The
static properties of the nucleon and $\Delta_{33}$-resonance in classical
chiral angle approach}
\setcounter{equation}{0}
\label{subsecstpr}
The normalized state vector with equal fixed spin and isospin $\ell$ are
to be identified with nucleons ($\ell =\frac12$) and $\Delta_{33}$
resonances ($\ell =\frac32$). Before we proceed with the self-consistent
quantum case it is worth to examine quantum formulae with classical chiral
angle solution (see Fig.~\ref{fig9}). We restrict this numerical analysis
to irreducible representations only. Clearly, numerical values for
reducible representations lie in the range between minimal and maximal $j$
values of which reducible representation is constructed.

\begin{table}
\caption{The predicted static baryon observables as obtained with
the quantum Skyrme model for representations of different dimension.
The first column (ANW) are the predictions
for the classical Skyrme model given in Ref.~\protect\cite{adkins83}.
Classical chiral angle function (taken from ANW column) has been used for
evaluation of integrals. The empirical results~\protect\cite{pdg94} are
listed in the last column.}
\label{clatabone}
\begin{center}
\begin{tabular}{|l|l|l|l|l|l|l|l|} \hline \hline
  &  {\bf ANW} &  $\mathbf{j=1/2}$ &
$\mathbf{j=1}$ & $\mathbf{j=3/2}$ & $\mathbf{j=2}$ & $\mathbf{j=5/2}$ &
{\bf Expt.}\\ \hline
  $m_N$       &  Input &  Input &  Input      &  Input
&  Input      &  Input &$\hphantom{0}939$ MeV \\
  $m_\Delta $ &  Input &  Input &  Input      &  Input
&  Input      &  Input &$1232$ MeV \\
  $f_\pi $    &$\hphantom{0}64.5$  &$\hphantom{0}72.1$ &
$\hphantom{0}76.4$ &$\hphantom{0}82.2$
&$\hphantom{0}89.4$ &$\hphantom{0}98.0$  &
$\hphantom{00}93$ MeV \\
 $e $    &$\hphantom{00}5.45$    &$\hphantom{00}5.23$&
$\hphantom{00}5.15$ &$\hphantom{00}5.03$
&$\hphantom{00}4.89$ &$\hphantom{00}4.74$ & \\
 $r_0 $    &  $\hphantom{00}0.59$   &$\hphantom{00}0.55$ &
$\hphantom{00}0.53$ &$\hphantom{00}0.51$
&$\hphantom{00}0.48$      & $\hphantom{00}0.45$  &
$\hphantom{000}0.72$ fm   \\
 $\mu_p $    & $\hphantom{00}1.87 $   &$\hphantom{00}1.90$ &
$\hphantom{00}1.84$ &$\hphantom{00}1.78$
&$\hphantom{00}1.71$ & $\hphantom{00}1.64$ &$\hphantom{000}2.79$   \\
 $\mu_n $    & $ -1.31 $ &$-1.42$ &$-1.40$&$-1.37$
&$-1.35 $     &$-1.33$ &$\hphantom{0}-1.91$    \\
 $g_A $    &$\hphantom{00}0.61 $   &$\hphantom{00}0.62$ &
$\hphantom{00}0.65$ &$\hphantom{00}0.68$
&$\hphantom{00}0.73$  & $\hphantom{00}0.78$  &$\hphantom{000}1.26$
\\ 
$\mu_{\Delta ^{++}}$    &   &$\hphantom{00}3.70$ &
 $\hphantom{m}3.58$ &$ \hphantom{m}3.44$
&$ \hphantom{m}3.29$  & $\hphantom{m}3.15$  
&$3.7-7.5$\protect\footnote{Recent measurements~\cite{bosshard91} obtain 
value $\mu_{\Delta^{++}}=4.52.$}
 \\
$\mu_{\Delta ^{+}}$    &  &$\hphantom{m}1.71$ &
 $\hphantom{m}1.64$ &$ \hphantom{m}1.55$
&$ \hphantom{m}1.46$  & $\hphantom{m}1.37$  &$\hphantom{000}? $
\\
$\mu_{\Delta ^{0}}$    &  &$-0.28$ &
 $-0.31$ &$ -0.34$
&$-0.38$  &$-0.42$  &$\hphantom{000}?$    \\
$\mu_{\Delta ^{-}}$    &  &$-2.27$ &
 $-2.25$ &$ -2.23$
&$-2.21$  &$-2.20$  &$\hphantom{000}?$    \\ \hline
\hline
\end{tabular}
\end{center}
\end{table}
\footnotetext{Recent measurements~\cite{bosshard91} obtain 
value $\mu_{\Delta^{++}}=4.52.$}

Classical chiral angle is obtained by solving the classical equation of
motion \eqref{redclachiang} (with appropriate boundary conditions) that is
given by the requirement that the classical mass \eqref{defclamas} be
stationary. Asymptotic behaviour when $\tilde r \rightarrow\infty$ of
classical solution can be easily found from asymptotic equation
\begin{equation}
F^{\prime\prime}(\tilde r)+\frac2{\tilde r}F^\prime (\tilde r)-
\frac2{\tilde r^2} F(\tilde r)=0.
\label{claasympeq}
\end{equation}
Physical solution (satisfying $F(\tilde r \rightarrow \infty)=0$) of
\eqref{claasympeq} is
\begin{equation}
F(\tilde r \rightarrow \infty)=\frac{k}{\tilde r^2};\qquad
k=\text{const,}
\label{claasymp}
\end{equation}
where $k$ is determined by  derivative continuity requirement and equation
\eqref{redclachiang} solution value at $\tilde r =0$: $F(0)=\pi$.
Classical equations of motion can be solved, for example, in the following
way.
\begin{enumerate}
\item Fix a merge point $\tilde r_0$ (it is assumed that asymptotic
\eqref{claasymp} gives a good approximation to the solution at that point)
and from \eqref{claasymp} calculate function $F(\tilde r_0)$ and its
derivative $F^\prime(\tilde r_0)$ values at that point.
\item Choose arbitrary $k$ and start standard differential equation
iteration procedure until point\footnote{Actually at $\tilde r=0$ equation
\eqref{redclachiang} becomes indefinite due to spherical coordinate system
singularity at the origin. But we can solve the equation until some small
value $\epsilon$. The function value at $\tilde r=0$ then can be
calculated using Taylor-series expansion near the $\tilde r=\epsilon$
point. From practical point of view it is convenient to solve a system of
two first-order differential
equations instead of one of the second order.} $\tilde r=0$ 
is reached.
\item If  $F(0)\neq\pi$ adjust $k$ value, recalculate $F(\tilde
r_0)$ and $F^\prime(\tilde r_0)$ values and start differential equation
iteration procedure again until $F(0)=\pi$ to the required precision
is achieved\footnote{There are, of course, much more efficient algorithms
for boundary value problems, for example, free and well known collocation
software package~\protect\cite{ascher81} (FORTRAN).}.
\end{enumerate}

\begin{table}
\caption{
The predicted static baryon observables for representations of
different dimension with fixed empirical values for the isoscalar radius
and the axial coupling constant.}
\label{clatabtwo}
\begin{center}
\begin{tabular}{|l|l|l|l|l|l|l|l|} \hline \hline
  &$\mathbf{j=1/2}$ &  $\mathbf{j=1}$ &
$\mathbf{j=2}$ & $\mathbf{j=3}$ & $\mathbf{j=4}$ & $\mathbf{j=9/2}$ &
$\mathbf{j=7}$\\ \hline
  $m_N$       &$1434$&$1402$&$1300$ &$1147$
&$\hphantom{0}942$&$\hphantom{0}821$&$\hphantom{00}23$ \\
  $m_\Delta $ &$1552$&$1520$&$1418$&$1265$
&$1060$&$\hphantom{0}939$ &$\hphantom{0}141$ \\
  $f_\pi $&$\hphantom{0}76.2$&$\hphantom{0}76.2$&
$\hphantom{0}76.2$&$\hphantom{0}76.2$
&$\hphantom{0}76.2$&$\hphantom{0}76.2$&
$\hphantom{0}76.2$ \\
 $e $&$\hphantom{00}3.81$&$\hphantom{00}3.81$&
$\hphantom{00}3.81$&$\hphantom{00}3.81$
&$\hphantom{00}3.81$&$\hphantom{00}3.81$&$\hphantom{00}3.81$ \\
$r_0 $&Input&Input&
Input&Input
&Input&Input&
Input \\
 $\mu_p$&$\hphantom{00}4.18$&$\hphantom{00}4.15$&
$\hphantom{00}4.07$&$\hphantom{00}3.96$
&$\hphantom{00}3.80$ & $\hphantom{00}3.70$ &$\hphantom{00}3.08$ \\
 $\mu_n $&$-3.85$ &$-3.83$&$-3.75$&$-3.63$
&$-3.47$&$-3.38$&$-2.76$ \\
 $g_A$&Input&Input&
Input&Input
&Input&Input&Input
\\
$\mu_{\Delta ^{++}}$&$\hphantom{00}7.72$&$\hphantom{00}7.67$&
$\hphantom{00}7.53$&$\hphantom{00}7.32$
&$\hphantom{00}7.03$&$\hphantom{00}6.86$&$\hphantom{00}5.75$
 \\
$\mu_{\Delta ^{+}}$&$\hphantom{00}2.90$&$\hphantom{00}2.89$&
$\hphantom{00}2.84$&$\hphantom{00}2.77$
&$\hphantom{00}2.67$&$\hphantom{00}2.61$&$\hphantom{00}2.24$
\\
$\mu_{\Delta ^{0}}$&$-1.92$ &$-1.90$&
$-1.85$&$-1.78$
&$-1.69$&$-1.63$&$-1.26$ \\
$\mu_{\Delta ^{-}}$&$-6.73$&$-6.69$&
$-6.55$&$-6.33$
&$-6.05$&$-5.88$&$-4.77$ \\ \hline
\hline
\end{tabular}
\end{center}
\end{table}
When classical solution is obtained, the corresponding
values for the Lagrangian parameters can be extracted from
equations \eqref{nucrezmasses} or \eqref{defga} and \eqref{esqrad}, or from
any of two combinations of them. As in Ref.~\cite{adkins83}, we determine the
two parameters in the Lagrangian \eqref{exqlag} so that nucleon and
$\Delta_{33}$-resonance masses take their empirical values. The expressions 
for these masses are then given by \eqref{defqmass}
\begin{subequations}
\begin{align}
m_N&=\frac{f_\pi }e\tilde{M}(F)+e^3f_\pi \Delta
\tilde{M}_j(F)+\frac{3
e^3f_\pi }{2\cdot 4\tilde{a}(F)},
\label{nuclmas}
\\
m_\Delta &=\frac{f_\pi }e\tilde{M}(F)+e^3f_\pi \Delta
\tilde{M}_j(F)+
\frac{15e^3f_\pi }{2\cdot4 \tilde{a}(F)}.
\end{align}
\label{nucrezmasses}
\end{subequations}
In the evaluation of these two masses numerically we employ the classical
chiral angle $F(r)$, the shape of which is shown in Fig.~\ref{fig9}.

In Table~\ref{clatabone} we include the predicted values for static
nucleon properties~\cite{acus96a}, as well as the original predictions 
obtained in Ref.~\cite{adkins83} for the classical Skyrme model. For larger 
values of $j$ the quantum corrections become increasingly important. The key
qualitative feature is that the quantum mass correction $\Delta
M_j(F)$ is negative, with a magnitude that grows with the dimension
of the representation. Therefore it becomes possible to reproduce the
empirical nucleon and $\Delta_{33}$-resonance mass values with increasingly
realistic values of the pion decay constant $f_\pi$ as the dimension
increases. This reaches its empirical value for a representation $j=5/2$ of
dimension 6. There is an accompanying --- if less significant ---
improvement of the numerical value for the axial coupling constant $g_A$.

In the case of the isoscalar radius $r_0$ (Tables~\ref{clatabone} and
\ref{clatabtwo}) of the baryon, there is,
however, no reduction of the difference between the predicted
and the empirical value with increasing dimension of the representation. The
same is true for the magnetic moments. The predicted value for the ratio of
the proton and neutron magnetic moments actually deteriorates slowly with
increasing dimension of the representation.
In Table~\ref{clatabtwo} we show the representation dependence of the
observables for the case where the parameters are determined by matching
the empirical values of isoscalar radius $r_{E,I=0}$ and the axial coupling
constant $g_A$. Because the expressions \eqref{defga} and \eqref{esqrad} are
independent of the dimension of the representation, parameters $f_\pi$
and $e$ remain constant.  The best agreement with the empirical values of
the static properties of baryons in this case are obtained with
the representation $j=4$.

We see that when the Skyrme model is treated consistently quantum
mechanically {\it ab initio} the dimension of the representation becomes
a significant additional model parameter~\cite{acus97,acus97a}. By choosing 
two parameters of the model so as to match the empirical nucleon and 
$\Delta_{33}$-resonance masses it becomes possible to obtain a value for the
pion decay constant, which is very close to the empirical value 
($89.4$~MeV~vs.~$93$~MeV). There was, however, no comparable gain in quality 
of the predictions for the baryon magnetic moments, which deteriorated 
slowly with increasing dimension of the representation. The value of axial 
coupling constant does on the other hand improve with increasing dimension, 
but stay below unity for representations of reasonably low dimension.

When parameters of the model are chosen to match isoscalar radius and axial
coupling constant, the parameters $f_\pi$ and $e$ are {\it constants} and
nucleon and $\Delta_{33}$-resonance observables depend only on the dimension 
of the representation. In contrast to the first match, the magnetic moments
improved with increasing dimension of the representation. Note that the
treatment used here for the quantum skyrmion breaks down when the
dimension of the representation grows so large that the negative quantum
mass correction $\Delta M_j$ becomes of the same order of magnitude as or
larger than the classical skyrmion mass. As shown in 
Table~\ref{clatabtwo} the numerical value of the quantum mass correction
$\Delta M_j(F)$ \eqref{defdeltam} is of the order of $55$~MeV in the
fundamental representation, but it rapidly increases in magnitude as the
dimension of the representation grows. For a representation of dimension
$9$ $(j=4)$ it is large enough to cancel the $\sim 500$~MeV overprediction
of the nucleon mass that is obtained when the empirical value for the pion
decay constant is employed in the classical Skyrme model. For $j=15/2$ the
quantum mass correction exceeds the skyrmion mass and baryon masses
become negative. After these comments we skip directly to quantum
self-consistent treatment.

\section{Self-consistent quantum formalism}
\label{scqf}
\setcounter{equation}{0}
Minimization of the expression for the classical mass
$M(F)$ \eqref{defclamas} leads to the conventional differential equation for
the chiral angle \eqref{redclachiang} according to which $F(\tilde r)$ falls
as $1/{\tilde r^2}$ at large distances. The behaviour is typical
for long-range interaction and, therefore, implies zero pion mass. This is
inconsistent with strong interaction properties, which is known to be
short-range and according to Yukawa~\cite{xenlibook} imply
\begin{equation}
\Bpi(r\rightarrow\infty)=-\frac{{\scriptscriptstyle\text{int. const}}}{r}
\mathrm{e}^{-\tilde m r};\qquad \tilde m
=\frac{mc}{\hbar},
\label{yukawafall}
\end{equation}
pion field $\Bpi(r)$ fall and, therefore, finite pion mass. 
{\it T.H.R.~Skyrme} in his
1962 work~\cite{skyrme62} wrote
"These mesons have zero mass, ultimately on account of the full
rotational symmetry of the Lagrangian \eqref{genlagden}. This symmetry is,
however, destroyed by the boundary condition $\BU(\infty)=\mathbf{1}$, and
we believe that the mass may arise as a self consistent quantal effect."
Thus, {\it T.H.R.~Skyrme} noted that chiral group $\SU_L\otimes\SU_R$ is
spontaneously broken by choosing vacuum state $\BU(\infty)=\mathbf{1}$ to the
subgroup $\diag \bigl(\SU_L\otimes\SU_R\bigr)\sim \SU_I$.
According to the recent point of view just the spontaneous breakdown of
chiral symmetry implies massless pions. In other words, pion is massless
provided that the axial current is perfectly conserved. Because a pion is
massive, we expect that axial symmetry should be an approximate symmetry
and that the axial current should be only approximately (partially)
conserved.

In the semiclassical approach the quantum mass term $\Delta
{M}_{\Sigma_j}$ is absent from the mass expression \eqref{defqmass}. The
absence of negative $\Delta{M}_{\Sigma_j}$ correction
has the consequence that variation of the truncated quantum mass
expression yields no stable solution~\cite{braaten85,bander84}. The
semiclassical approach describes the skyrmion as a ``rotating'' rigid body
with fixed $F(r)$~\cite{adkins83}. In contrast the variation of the full
energy expression \eqref{defqmass} that is obtained in the consistent
canonical quantization procedure~\cite{fujii87} in collective coordinates
approach gives stable solutions.

Minimization of the quantum mass expression \eqref{defqmass} leads to the
following integro-differential equation for the chiral angle $F(\tilde
r)$ in the dimensionless variable $\tilde r$:
\begin{equation}
\begin{split}
\ &F^{\prime \prime }\biggl(-2\tilde r^2-4\sin ^2\!F+\frac{e^4\tilde r
^2\sin ^2\!F}{15\tilde a^2}\Bigl( 80\tilde a\Delta
\tilde M_{\Sigma_j}+20\ell (\ell +1) +4d_1-8d_2\sin ^2\!F\Bigr) \biggr)
\\
+&F^{\prime 2}\biggl(-2\sin
2F+\frac{e^4\tilde r^2\sin 2F}{15\tilde a^2}\Bigl( 40\tilde a\Delta \tilde
M_{\Sigma_j}+10\ell (\ell +1)+2d_1-8d_2\sin ^2\!F\Bigr) \biggr)
\\
+&F^\prime \biggl(-4\tilde r+\frac{e^4\tilde r\sin^2\!F}{15\tilde a
^2} \Bigl( 160\tilde a\Delta \tilde M_{\Sigma_j} +40\ell (\ell
+1)+8d_1-16d_2\sin^2\!F\Bigr)\biggr)
\\
+&\sin2F\biggl(2+2\frac{\sin ^2\!F}{\tilde r^2} -\frac{e^4}{15\tilde
a^2}\Bigl(\bigl( 40\tilde a\Delta \tilde M_{\Sigma_j} +10\ell
(\ell+1)\bigr) \bigl( \tilde r^2+2\sin ^2\!F\bigr)
+15\tilde r^2
\\ \ &\phantom{
\sin2F\biggl(2+2\frac{\sin ^2\!F}{\tilde r^2} -\frac{e^4}{15\tilde
a^2}\Bigl(
}
+4d_3\sin^2\!F +8d_2\tilde r^2\sin^2\!F\Bigr)
\biggr)=0,
\end{split}
\label{intdifeq}
\end{equation}
with the usual boundary conditions $F(0)=\pi $ and $F(\infty )=0$. The
state dependence of this equation is a direct consequence of the
fact that quantization preceded variation (cf. Ref.~\cite{li87}).
Contrary to the classical equation \eqref{redclachiang} quantum
chiral angle equation \eqref{intdifeq} depends on the parameter $e$. This,
however, is not an unusual result: quantization of breathing
modes~\cite{kostyuk95,sawada91} also leads to parameter $e$ dependent
equations.

At large distances this equation reduces to the asymptotic form
\begin{equation}
\tilde r^2F^{\prime \prime }+2\tilde r F^\prime -\bigl(2+\tilde m_\pi
^2\tilde r^2\bigr)F=0,
\label{asympintdif}
\end{equation}
where the quantity $\tilde m_\pi^2$ is defined as
\begin{gather}
\tilde m_\pi^2=-\frac{e^4}{3\tilde a(F)}\Bigl(8\Delta
\tilde M_{\Sigma_j}(F)+\frac{2\ell (\ell +1)+3}{\tilde a(F)}\Bigr),
\label{defmpi}\\
\intertext{
and the corresponding asymptotic solution takes the form }
F(\tilde r)=k\Bigl( \frac{\tilde m_\pi
}{\tilde r}+\frac1{\tilde r^2}\Bigr) \mathrm{e}^{-\tilde m_\pi
\tilde r};\qquad k=\text{const}.
\label{qfasymp}
\end{gather}
The requirement of stability of the quantum skyrmion is that the integrals
\eqref{defatil}, \eqref{defclamas} and \eqref{defdeltam} converge. This
requirement is satisfied only if $\tilde m_\pi^2>0$. For that the
presence of the negative quantum correction $\Delta M_{\Sigma_j}(F)$ is
necessary. The absence of this term leads to the instability
of the skyrmion in the semiclassical approach~\cite{braaten85}. Note that
in the quantum treatment the chiral angle possesses~\cite{fujii87} the
asymptotic Yukawa behaviour \eqref{qfasymp}. The positive quantity $m_\pi
=ef_\pi \tilde m_\pi $, therefore, can be interpreted as an
effective mass for the pion field.

It is known that in classical Skyrme model 3-divergence of axial current
density component $\mathcal{A}^a(\Bx )$ gives the differential
equation \eqref{redclachiang} for chiral angle~\cite{hayashi92}.
"Therefore, in this chiral-symmetric theory the conservation of axial-vector
current is, in fact, the equation of motion, i.e. the Euler-Lagrange
equation for the rotated pion field $\pi_i$"~\cite{li87}. 
Let's assume that in quantum theory axial symmetry is broken and the 
matrix element of the divergence of axial current is
proportional to pion mass (PCAC relation\footnote{
To make explicit comparison with~\eqref{pcacusual} just account
that $F(r)\bar x^a=f_\pi \pi^a$ (see~\eqref{interppi}).
Generally the comparison of \eqref{pcacusual} and \eqref{axcurdivinasymp}
is complicated by the fact that in \eqref{axcurdivinasymp} axial vector 
current is sandwiched between two hadronic states. We will not consider this 
question here, but refer to~\cite{adkins83,hayashi92} and references 
therein.})
\begin{equation}
\genfrac{\langle}{|}{0pt}{}{\ell}{m_i,m_s}
\nabla ^{b}\ \ \hat{\!\!\!\!\CA}_{b}^{a}
\genfrac{|}{\rangle}{0pt}{}{\ell}{m_i,m_s}=
f_{\pi }^{2} m_{\pi }^{2}F(r)
\label{axcurdivinasymp}.
\end{equation}
As a consequence we get the {\it same asymptotic equation} 
\eqref{asympintdif}, when $r\rightarrow \infty$. This result supports the 
interpretation of $m_\pi$ as the effective pion mass.

Note also that the finite pion mass in the Skyrme model is usually 
introduced by adding an explicit term~\cite{kostyuk95,fujii87,adkins84}
\begin{equation}
\CL_{m_\pi}=\frac14 m_{\pi{\scriptscriptstyle (exp)}}^2 f_\pi^2 \Tr
\{\BU+\BU^\dagger -2\},
\label{explmass}
\end{equation}
in the Lagrangian density \eqref{genlagden}, and, therefore, the Lagrangian
\eqref{genlagden} chiral symmetry becomes explicitly broken even in 
classical limit.

Positivity of the pion mass \eqref{defmpi} can obviously only be achieved
for states with sufficiently small values of spin $\ell $. This implies
that the spectrum of states with equal spin and isospin will necessarily
terminate at some finite value of the spin quantum number
$\ell$. The termination
point depends on parameters $e$ and $j$ values. When spin
$\ell$ value increases, the upper $e$ value for which stable soliton solution
exist always decreases (for the same representation $j$).
As the negative quantum mass correction $\Delta M_{\Sigma_j}$ in the
expression \eqref{defdeltam} grows in magnitude with the dimension of the
representation, it is always possible to find a representation in which
the nucleon and the $\Delta _{33}$-resonance are the only stable
particles, as required by experiment.

\begin{figure}
\begin{center}
\includegraphics*[width=12cm]{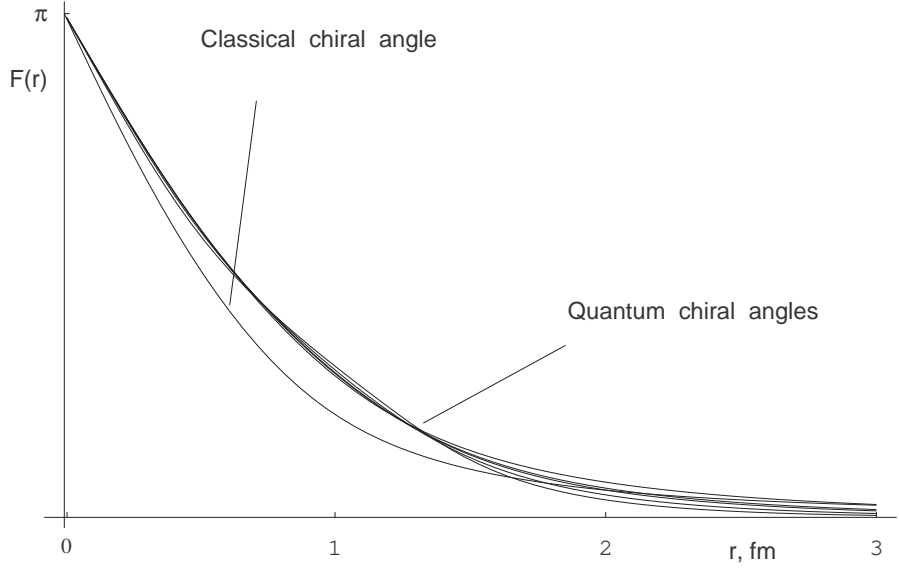}
\end{center}
\caption{Quantum chiral angle solution family for nucleon ($\ell =\frac12$)
in representations
$j=\frac12,1\oplus\frac12\oplus\frac12, 1, \frac32, \frac52$ and the
classical chiral angle solution as taken from Ref.~\cite{adkins83}.}
\label{fig13}
\end{figure}

Integro-differential equation \eqref{intdifeq} can be attacked in the
following way.
\begin{enumerate}
\item Using chiral classical angle Fig.~\ref{fig9} and any pair of
empirical baryon observables, for example, nucleon mass \eqref{nuclmas} and
isoscalar radius \eqref{esqrad} we fit two model parameters $f_\pi$ and
$e$ and calculate all required integrals in the quantum equation
\eqref{intdifeq}, namely, \eqref{defclamas}, \eqref{defatil},
\eqref{defdeltam}, \eqref{defmpi}.
\item Using known asymptotic solution \eqref{qfasymp} (and its derivative)
one can adopt simple differential equation solution procedure described in
Sec.~\ref{subsecstpr} and find the first quantum solution $F_1(\tilde r)$
and the constant $k_1$.
\label{item}
\item  This quantum solution $F_1(\tilde r)$ now can be subsequently used
to recalculate $f_\pi, e$ and integrals \eqref{defclamas}, \eqref{defatil},
\eqref{defdeltam}, \eqref{defmpi}. Then again procedure described in
item~\ref{item} can be used to get the solution $F_2(\tilde r)$ and constant
$k_2$.
\item This procedure can be iterated until convergent solution
and parameters $f_\pi, e$ as well as stable values of $M(F), 
\Delta M_{\Sigma_j}(F), a(F),m_\pi(F)$  are obtained. The 
self-consistent set then can be used to calculate numerous 
phenomenologically interesting quantities.
\end{enumerate}
Quantum chiral angle solutions with model parameters determined from
nucleon mass \eqref{nuclmas} and isoscalar radius \eqref{esqrad}
are shown in Fig.~\ref{fig13}.
If one succeeds in initial guess\footnote{
Solution of classical equation 
\eqref{redclachiang} exist or not for all $e,f_\pi$ 
values. This is not the case for
integro-differential equation \eqref{intdifeq}.},
then it usually takes only $10$--$15$
iterations to get $5$--$6$ fixed digits in all integrals and $f_\pi, e$
values.
For nucleon solutions with spin $\ell=1/2$ exist (in the fundamental
representation $j=1/2$) when $0\le e<7.5$. The largest value of $e$ for
which stable solutions are obtained decreases with increasing
dimensionality of the representation, and there are no restrictions to the
existence of solution from representation $j$ employed. This can be seen
from equation \eqref{intdifeq} itself. Indeed, the formal substitution
$e\rightarrow 0$ into \eqref{intdifeq} yields the classical equation
\eqref{redclachiang}, which does have a solution\footnote{Note that
$e\rightarrow 0$ does not lead to any meaningful limit due to
dimensionless variable $\tilde r=ef_\pi r$ being used. This, however, does
not affect the solution existence argument for arbitrary small $e$, because
only formal similarity of equations is important.}.
\begin{figure}
\begin{center}
\includegraphics*[width=12cm]{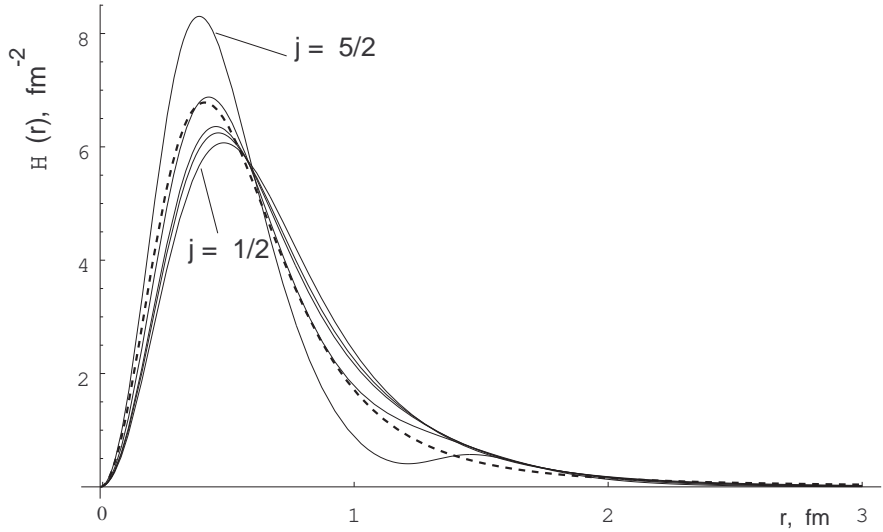}
\end{center}
\caption{Family of nucleon mass densities 
in representations
$j=\frac12,1\oplus\frac12\oplus\frac12,1,\frac32,\frac52$ (model
parameters being calculated from empirical nucleon mass \eqref{nuclmas} and
isoscalar radius \eqref{esqrad}).}
\label{fig14}
\end{figure}
For illustration (see Fig.~\ref{fig14}) we also provide nucleon mass
densities (Hamiltonian densities \eqref{defqmass})
for quantum chiral angles in Fig.~\ref{fig13} and detailed contributions of
classical soliton masses, rotation energies and quantum mass corrections
for two of them ($j=\frac12$ and $j=\frac52$) in Fig.~\ref{fig15}
\begin{figure}
\begin{center}
\includegraphics*[width=12cm]{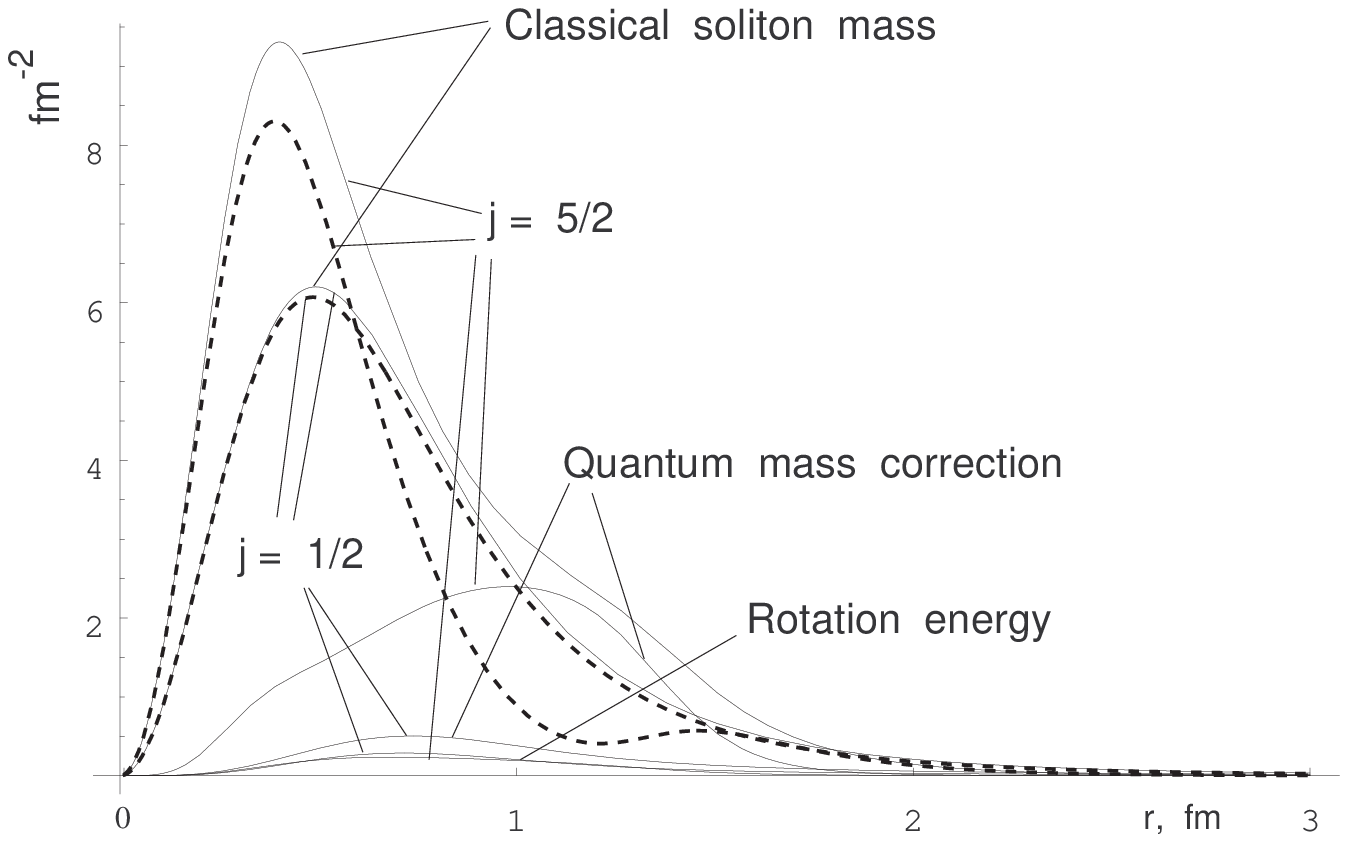}
\end{center}
\caption{
Detailed contributions of
classical soliton masses, rotation energies and quantum mass corrections
for nucleon in representations $j=\frac12$ and $j=\frac52$
(model
parameters being calculated from empirical nucleon mass \eqref{nuclmas} and
isoscalar radius \eqref{esqrad}).}
\label{fig15}
\end{figure}

A remark is given to the $\sigma$ model without the Skyrme term (this
corresponds to the limit $e\rightarrow \infty$ in \eqref{genlagden}).
Although the quantum correction $\Delta M_{\Sigma_j}$ plays a role in
stabilization of soliton, one cannot obtain "quantum" rotating chiral
soliton without the Skyrme term as stabilizer~\cite{fujii92}. Applying higher
representations also cannot help in obtaining this
stability~\cite{krupovnickas97}. This is consistent with Derrick
theorem~\cite{derrick64}.
\enlargethispage*{1cm}
For $\Delta_{33}$-resonances $(\ell=3/2)$ there are no stable solutions in
the fundamental ($j=1/2$) representation, nor in the representation $j=1$.
In the representations $j=3/2$ and $j=2$ there are only stable soliton
solutions for baryons with spin $\ell=1/2$ and $\ell=3/2$. A dimension
with $j=5/2$ allows solitons with $\ell=1/2,\ 3/2$ and $5/2$, and,
therefore, appears to be empirically contraindicated. The numerical results
for nucleons with fixed empirical values for isoscalar radius and mass are
shown in Table~\ref{nuc1} in Appendix~\ref{app1}. For the irreducible
representation $j=1$ the proton magnetic moment calculated in this way is
within 10\% of the empirical value. The calculated values of both the
neutron magnetic moment and the axial coupling constant agree with the
corresponding empirical values to within 1\%. Numerical results for
fixed isoscalar radius and axial coupling constant are represented in
Table~\ref{nuc2} (Appendix~\ref{app1}). For representation $j=1$ nucleon
mass and proton magnetic moment agree with experimental values within 10\%,
neutron magnetic moment within 2\%. In Table~\ref{nuc4} and Table~\ref{nuc6}
we represent numerical results obtained by fixing $m_\pi=138$~MeV and
nucleon mass (Table~\ref{nuc4}) or isoscalar radius (Table~\ref{nuc6}). In
these approaches agreement with experimental data is worse. We note,
however, that pion field mass used here ($138$~MeV) approximately equals
to particles $\pi_0,\pi_{-},\pi_{+}$ mass.
The effective strong interaction field fall can be indirectly
estimated\footnote{Private communications of 
{\it E.~Norvai\v sas} and
{\it D.O.~Riska.}} from
experiments. These measurements correspond to about $150$~MeV effective
boson mass, when extracted from Yukawa formula~\eqref{yukawafall}.

For parameters $e,f_\pi$ fixed from nucleon data, empirical values for
$\Delta_{33}$-resonance observables for different representations can be 
obtained. Results for $e=4.15,\ f_\pi=58.5$~MeV (determined by fit to nucleon
$m_N=939$~MeV and 
$\langle r^2_{E,I=0}\rangle ^{1/2}=0.72$~fm and representation $j=1$)
are demonstrated in Table~\ref{deltaone} in Appendix~\ref{app2}. Results
for the same fit of experimental quantities, but in different
representations are shown in Table~\ref{deltaseven} ($j=\frac12$) and
Table~\ref{deltaeight} ($j=1\oplus\frac12\oplus\frac12$).
Tables~\ref{deltafour}, \ref{deltafive}, \ref{deltasix} represent
numerical results with fixed values for $e$ and $f_\pi$, which are
calculated from nucleon observables $g_A$ and $\langle r^2_{E,I=0}\rangle
^{1/2}$ in representations $j=\frac12;\quad j=1$ and
$j=1\oplus\frac12\oplus\frac12$, respectively. Tables~\ref{deltatwo} and
\ref{deltathree} show similar results for fixed $e,f_\pi$ values, extracted
from nucleon observables $m_\pi$ and $\langle r^2_{E,I=0}\rangle ^{1/2}$ in
representations $j=1$ (Table~\ref{deltatwo}) and
$j=1\oplus\frac12\oplus\frac12$ (Table~\ref{deltathree}).

It is also worth to mention the numerical results for reducible
representations $1\otimes\frac12\otimes\frac12$ for nucleon and
$\frac32\otimes1\otimes\frac12$ for $\Delta_{33}$-resonance. 
The representation
$\frac32\otimes1\otimes\frac12$ is the first one for which the stable
soliton solution is obtained for $\ell=3/2$ state. The numerical results
in these reducible representations are comparable --- if not better --- to
those of representations $j=1$ (for $\ell =1/2$) and $j=3/2$ (for $\ell
=3/2$), respectively. Noting that the SU(3) isomultiplets --- octet $(1,1)$
and decuplet $(2,0)$ --- (to which nucleon and $\Delta_{33}$-resonance 
belong, respectively) split, when restricted to \SU\ subgroup, into
$1\otimes\frac12\otimes\frac12\otimes0$ and
$\frac32\otimes1\otimes\frac12\otimes0$, respectively\footnote{
Representation $j=0$ does not influence the results. Indeed, 
$j=0$ representation of SU(2) is trivial and, thus, the 
Lagrangian~\eqref{genlagden} is identical zero: $\mathcal{L}\equiv 0$.}, the
following explanation is not improbable: it just can indicate that \SU\
Skyrme model in some way (via representations employed) reveals relevant
SU(3) symmetry of strong interactions. If this is indeed the case, then in
SU(3) Skyrme model using higher representations 
($(1,1)$ and $(2,0)$) one can hope to describe
the entire isomultiplets of baryons more successfully. Investigation of
SU(3) Skyrme model is required to confirm or reject these considerations.
\section{Remarks on persisting problems}
\setcounter{equation}{0}
Despite quite successful generalization of quantum Skyrme model to general
representation of \SU\ group a few problems still persist. First of all we
see rather big gap between calculated ($f_\pi\sim 60$ MeV, $e\sim 4$) and
extracted from experiment ($f_\pi=93$ MeV, $e\sim 7.4$) numerical values
of model parameters. The problem is inherited from semiclassical
calculations~\cite{adkins83}. The difference, however, becomes increasingly
important, because all other measurable quantities fit to their
experimental values much better ($\sim 10\%$, see Appendix~\ref{app1}).
Thus the question again arises to what extent do these errors of the
Skyrme model reflect use of the approximation, and to what extent do they
reflect the fact that the Skyrme model is crude approximation to meson
physics. The question has been considered by 
{\it E.~Witten}~\cite{witten84}, who
suspected that in the case of semiclassical approximation "the error mainly
has the latter origin$\dots$ If the proper meson theory were known, the
semiclassical approximation would be equivalent to the $1/N_c$ expansion,
and I personally suspect the error would be much less than $30$\%". Our
calculations, if we exclude $f_\pi ,e$ values, which can be considered just
as model parameters without referring to its physical content, are in much
better agreement with experimental data. However, it should be noted that we
obtain these values due to an additional implicit discrete parameter --- 
the group representation. The quantization procedure also differs from the
semiclassical approach~\cite{adkins83} considered by 
{\it E.~Witten}~\cite{witten84}.
\enlargethispage*{1.9cm}
A problem is also present with the interpretation of representation
dependence itself. Despite the quantization method used here is free of
ambiguity and is in agreement with principles of modern  quantization of
constrained systems, it undoubtedly shows representation dependence of
measurable quantities\footnote{The problem we suspect is not the problem of 
the Skyrme model only but rather of the quantization procedure itself.
As a consequence, we expect representation dependence in other models 
defined on group manifold.}. Thus physical interpretation is welcome and we 
hope that quantum investigation of SU(3) Skyrme model can enlighten this 
problem\footnote{We hope that the situation is somehow similar to that of 
{\it W. Heisenberg} and {\it M.~Gell-Mann} about assumptions of symmetry of 
strong interaction, briefly mentioned in the Preface.}.
%\subsection{Other quantization methods}
%\subsection{Some quantization aspects of the Skyrme model}
%\section{Phenomenology of the quantum Skyrme model}
\chapter*{Concluding statements}
The following statements represent the main results of the work in 
the order of decreasing importance.
\begin{itemize}
\item 
Each of SU(2) representation $j$ yields the different quantum Lagrangian
density. As a consequence, theoretical observables depend on  representation
$j$ which can be treated as a new phenomenological parameter.

\item Quantum chiral solitons exist and possess asymptotic behaviour 
consistent with the massive Yukawa field fall \eqref{yukawafall}. The 
asymptotic shape and PCAC relation leads to the correct asymptotic equation 
\eqref{asympintdif} coinciding with contribution of explicitly broken 
term \eqref{explmass}. This encourages us to suggest that the integral 
$m_\pi$ should be interpreted as an effective pion mass.

\item A nucleon and $\Delta_{33}$-resonance are the only stable states
for irreducible representations $j=\frac32$ and $j=2$.  
Unphysical tower of states $\ell_{\text{spin}} =\ell_{\text{isospin}}$, 
which is artifact of the classical and semiclassical Skyrme model is, 
therefore, terminated by choosing the appropriate \SU\ representations.

\item Higher spin ($\ell > 1/2$) quantum states are not "spherically 
symmetric". The Hamiltonian~\eqref{defhamdens} 
(Lagrangian~\eqref{defqlagden}) density function depends on the polar
angle $\vartheta$. Nucleon states are "spherically symmetric" in various 
representations of $j$, as required. 

\item Each of the spin-isospin state yields the different range of
realizable values of the parameter $e$. A stable quantum 
self-consistent solution exist:
\begin{itemize}
\item  for spin $\ell =\frac12$ states in all \SU\ 
representations,

\item for $\ell =\frac32$ spin states starting at least from reducible 
representation $\scriptscriptstyle \mathbf{\frac 32 \oplus 
{\scriptstyle 1}\oplus \frac12}$. 
\end{itemize}

\item A very good agreement with experimental data is obtained for 
axial coupling constant $g_A$ in higher representations of SU(2), the 
problem being previously 
unsolved by using various extensions of the model in the fundamental
representation of \SU.
\item The basic quantum Skyrme model provides considerable improvements in 
nucleon magnetic momenta and, especially, in neutron charge density 
distribution.
\end{itemize}
The text of PhD thesis can be found at the Website of Institute
of Theoretical Physics and Astronomy\newline
\quad
{\bf http://www.itpa.lt/baryon}

\backmatter
\setcounter{equation}{0}
\setcounter{table}{0}
\setcounter{figure}{0}
\renewcommand{\theequation}{\thechapter.\arabic{equation}}
\appendix
%\chapter[SU(2) Wigner $\BD^j$ matrices]{SU(2) 
%Wigner $\BD^j$ matrices for $j=\frac12,1,\frac32$}
%\label{exwdm}
%\documentclass[10pt]{book}
%\begin{document}
\chapter{Nucleon observables in different representations}
\label{app1}.
\begin{table}
\caption{The predicted static baryon observables for different
representations with fixed empirical values for the effective pion
mass $m_\pi =138$~MeV and nucleon mass $939$~MeV.}
\label{nuc4}
\begin{center}
\begin{tabular}{|c|c|c|c|c|c|}
\hline \hline
$\mathbf{j}$ & $\mathbf{1/2}$ & $\mathbf{1}$ & $\mathbf{3/2}$ &
$\scriptscriptstyle \mathbf{1\oplus \frac 12 \oplus \frac 
12}$&\textbf{Expt.} \\ \hline$m_N$
&Input&Input&Input&Input&$939$~MeV \\ \hline
$f_\pi$& 68.4 & 54.9 & 49.6 & 57.2 & $93$~MeV\\ \hline
$e$ & 4.97 & 3.96 & 3.52 & 4.15 & \\
\hline
$\mu_p$ & 1.63 & 3.20 & 4.44 & 2.80 & $2.79$\\
\hline
$\mu_n$ &$-$1.06 &$-$2.61 &$-$3.85 &$-$2.21 & $-1.91$ \\
\hline
$g_A$ &0.89&1.43&1.77&1.30&$1.26$\\
\hline
$m_\pi$&Input&Input&Input&Input&$138$~MeV\\
\hline
$\sqrt{\langle r^2_{E,I=0}\rangle }$
&0.54&0.83&1.00&0.77&$0.72$~fm \\
\hline\hline
\end{tabular}
\end{center}
\end{table}
\setlength{\floatsep}{1.5cm}

\begin{table}
\caption{The predicted static nucleon observables in different
representations with fixed empirical values for the isoscalar
radius $\langle r^2_{E,I=0}\rangle ^{1/2}=0.72$~fm and nucleon mass 
$939$~MeV.}
\label{nuc1}
\begin{center}
\begin{tabular}{|c|c|c|c|c|c|c|}
\hline \hline
$\mathbf{j}$&$\mathbf{1/2}$&$\mathbf{1}$&$\mathbf{3/2}$ 
&$\mathbf{5/2}$&$\scriptscriptstyle \mathbf{1\oplus \frac{1}{2} \oplus \frac 
{1}{2}}$&{\bf Expt.} \\ \hline
$m_N$ &Input&Input&Input&Input&Input&$939$~MeV \\ \hline
$f_\pi $& 59.8 & 58.5 & 57.7 & 56.6 & 58.8 &$93$~MeV \\
\hline
$e$ & 4.46 & 4.15 & 3.86 &  3.41 & 4.24 &  \\ 
\hline
$\mu_p$ & 2.60 & 2.52 & 2.51 & 2.52 & 2.53 &$2.79$\\
\hline 
$\mu_n$ & $-$2.01 & $-$1.93 & $-$1.97 & $-$2.05 & 
$-$1.93 & $-1.91$ \\
\hline
$g_A$ &1.20&1.25&1.33&1.52&1.23&$1.26$\\
\hline
$m_\pi $& 79.5 & 180. & 248. & 336. & 155.&$138$~MeV 
\\ \hline
$\sqrt{\langle r^2_{E,I=0}\rangle }$ 
&Input&Input&Input&Input&Input&$0.72$~fm \\
\hline
$\sqrt{\langle r^2_{E,I=1}\rangle }$ 
&1.33&1.03&0.97&0.93&1.07&$0.88$~fm\\ 
\hline
$\sqrt{\langle r^2_{M,I=0}\rangle }$ 
&1.05&1.01&1.00&1.00&1.01&$0.81$~fm \\ 
\hline
$\sqrt{\langle r^2_{M,I=1}\rangle }$ 
&1.32&1.03&0.97&0.93&1.07&$0.80$~fm \\ \hline
\hline
\end{tabular}
\end{center}
\end{table}
\begin{table}
\caption{The predicted static baryon observables for different
representations with fixed empirical values for the isoscalar
radius $\langle r^2_{E,I=0}\rangle ^{1/2}=0.72$~fm and axial coupling
constant $g_A=1.26$\, .}
\label{nuc2}
\begin{center}
\begin{tabular}{|c|c|c|c|c|c|c|}
\hline \hline
$\mathbf{j}$ & $\mathbf{1/2}$ & $\mathbf{1}$ & $\mathbf{3/2}$ & $\mathbf{5/2}$ &
$\scriptscriptstyle \mathbf{1\oplus\frac 12 \oplus \frac 12}$ 
&\textbf{Expt.} \\ \hline
$m_N$ & 986. & 948. & 882. & 694. &  
963.  &$939$~MeV \\
\hline
$f_\pi $& 61.4 & 58.9 & 55.7 & 48.1 & 59.7 & $93$~MeV \\
\hline
$e$ & 4.37 & 4.13 & 3.92 &  3.56 & 4.20 & \\
\hline
$\mu_p$ & 2.70 & 2.53 & 2.42 & 2.05 & 2.57 & $2.79$\\
\hline
$\mu_n$ & $-$2.14 & $-$1.95 & $-$1.84 & $-$1.51 & 
$-$1.99 &$-1.91$ \\
\hline
$g_A$ &Input&Input&Input&Input&Input&$1.26$\\
\hline
$m_\pi $& 75.8 & 179. & 259. & 386. & 152. &$138$~MeV \\
\hline
$\sqrt{\langle r^2_{E,I=0}\rangle }$
&Input&Input&Input&Input&Input&$0.72$~fm \\
\hline
$\sqrt{\langle r^2_{E,I=1}\rangle }$
&1.36&1.03&0.96&0.92&1.08&$0.88$~fm\\
\hline
$\sqrt{\langle r^2_{M,I=0}\rangle }$
&1.05&1.01&1.00&1.01&1.01&$0.81$~fm \\
\hline
$\sqrt{\langle r^2_{M,I=1}\rangle }$
&1.35&1.03&0.96&0.92&1.08&$0.80$~fm \\ \hline
\hline
\end{tabular}
\end{center}
\end{table}
\begin{table}
\caption{The predicted static baryon observables for different
representations with fixed empirical values for the isoscalar
radius $\langle r^2_{E,I=0}\rangle ^{1/2}=0.72$~fm and effective pion 
mass  $m_\pi=138$~MeV.}
\label{nuc6}
\begin{center}
\begin{tabular}{|c|c|c|c|c|c|c|}
\hline \hline
$\mathbf{j}$ & $\mathbf{1/2}$ & $\mathbf{1}$ & $\mathbf{3/2}$&  
$\mathbf{5/2}$ &$\scriptscriptstyle \mathbf{1\oplus \frac 12 \oplus \frac 
12}$ &\textbf{Expt.} \\ \hline
$m_N$ & 486. & 1264. & 1872. & 2992. & 1069. &$939$~MeV \\ \hline
$f_\pi$& 41.7 & 68.8 & 84.0 & 106.3 & 63.1 & $93$~MeV \\ \hline
$e$ & 5.56 & 3.76 & 3.12 &  2.48 &4.05 &  \\ \hline
$\mu_p$ & 1.93 & 3.13 & 4.35 & 6.72 &2.77 & $2.79$\\ \hline 
$\mu_n$ & $-$0.86 & $-$2.68 & $-$4.04 & $-$6.52 &
$-$2.24 &$-1.91$ \\ \hline
$g_A$ &0.60&1.65&2.46&3.94&1.39&$1.26$\\ \hline
$m_\pi $&Input&Input&Input&Input&Input&$138$~MeV\\ \hline
$\sqrt{\langle r^2_{E,I=0}\rangle }$ 
&Input&Input&Input&Input&Input&$0.72$~fm\\ \hline
$\sqrt{\langle r^2_{E,I=1}\rangle }$ 
&1.07&1.12&1.12&1.12&1.11&$0.88$~fm\\ \hline
$\sqrt{\langle r^2_{M,I=0}\rangle }$ 
&1.01&1.02&1.02&1.02&1.02&$0.81$~fm\\ \hline
$\sqrt{\langle r^2_{M,I=1}\rangle }$ 
&1.05&1.12&1.12&1.13&1.11&$0.80$~fm\\ \hline
\hline
\end{tabular}
\end{center}
\end{table}

\chapter{$\Delta_{33}$-resonance observables in different representations}
\label{app2}
%nuo cia prasideda delta rezonansu duomenys

\begin{table}
\caption{The predicted static $\Delta _{33}$-resonance observables in 
different representations with fixed values for the 
parameters $e=4.15$ and $f_\pi =58.5$~MeV (from 
nucleon observables $m_N =939$~MeV, $\langle r^2_{E,I=0}\rangle 
^{1/2}=0.72$~fm, representation $j=1$ in Table~\ref{nuc1}).}
\label{deltaone}
\begin{center}
\begin{tabular}{|c|c|c|c|c|}
\hline \hline
$\mathbf{j}$ &$\scriptscriptstyle \mathbf{\frac {3}{2}\oplus 1\oplus \frac 
{1}{2}}$ & $\mathbf{\frac {3}{2}}$ & $\mathbf{2}$  &\textbf{Expt.} \\ 
\hline$m_\Delta $ & 
1055. & 1029. & 910. &$1232$ MeV \\ \hline
$\mu_{\Delta ^{++}}$ & 7.38 & 6.40 & 4.20 &$3.7-7.5$\\
\hline 
$\mu_{\Delta ^{+}}$& 3.02 & 2.73 & 2.01 &?\\
\hline 
$\mu_{\Delta ^{0}}$ & $-$1.33 &$-$0.94 & $-$0.19 &?\\
\hline 
$\mu_{\Delta ^{-}}$& $-$5.69 & $-$4.61 & $-$2.38 & ? \\
\hline
$\sqrt{\langle r^2_{E,I=0}\rangle }$ 
&0.91&0.87&0.72&? \\ \hline
\hline
\end{tabular}
\end{center}
\end{table} 
\setlength{\floatsep}{2cm}

\begin{table}
\caption{The predicted static $\Delta_{33}$-resonance observables for
different representations with fixed empirical values for the
$e=4.46$ and $f_\pi =59.8$~MeV (from nucleon observables $m_N =939$~MeV,
$\langle r^2_{E,I=0}\rangle ^{1/2}=0.72$~fm, representation $j=\frac 12
$ in Table~\protect\ref{nuc1}).}
\label{deltaseven}
\begin{center}
\begin{tabular}{|c|c|c|c|c|}
\hline \hline
$\mathbf{j}$ &$\scriptscriptstyle \mathbf{\frac 32 \oplus 1\oplus \frac
12}$ & $\mathbf{\frac 32}$ & $\mathbf{2}$  &\textbf{Expt.} \\
\hline$m_\Delta $ &
1008. & 974. & 809. &$1232$~MeV \\ \hline
$\mu_{\Delta ^{++}}$ & 6.05 & 5.15 & 3.00 &$3.7-7.5$\\
\hline
$\mu_{\Delta ^{+}}$& 2.63 & 2.36 & 1.63 &?\\
\hline
$\mu_{\Delta ^{0}}$ & $-$0.80 &$-$0.43 & 0.25 &?\\
\hline
$\mu_{\Delta ^{-}}$& $-$4.23 & $-$3.22 & $-$1.12 & ? \\
\hline
$m_\pi $& 104. & 172. & 438.&$138$~MeV \\
\hline
$\sqrt{\langle r^2_{E,I=0}\rangle }$
&0.84&0.79&0.62&? \\ \hline
\hline
\end{tabular}
\end{center}
\end{table}
\newpage

\begin{table}
\caption{The predicted static $\Delta_{33}$-resonance observables for
different representations with fixed empirical values for the
$e=4.24$ and $f_\pi =58.8$~MeV (from nucleon observables $m_N =939$~MeV,
$\langle r^2_{E,I=0}\rangle ^{1/2}=0.72$~fm, representation $j=1\oplus 
\frac 12\oplus \frac 12$ in Table~\protect\ref{nuc1}).}
\label{deltaeight}
\begin{center}
\begin{tabular}{|c|c|c|c|c|}
\hline \hline
$\mathbf{j}$ &$\scriptscriptstyle \mathbf{\frac 32 \oplus 1\oplus \frac
12}$ & $\mathbf{\frac 32}$ & $\mathbf{2}$  &\textbf{Expt.} \\
\hline$m_\Delta $ &
1040. & 1012. & 881. &$1232$~MeV \\ \hline
$\mu_{\Delta ^{++}}$ & 6.96 & 6.00 & 3.82 &$3.7-7.5$\\
\hline
$\mu_{\Delta ^{+}}$& 2.90 & 2.61 & 1.89 &?\\
\hline
$\mu_{\Delta ^{0}}$ & $-$1.16 &$-$0.78 & $-$0.05 &?\\
\hline
$\mu_{\Delta ^{-}}$& $-$5.23 & $-$4.17 & $-$1.98 & ? \\
\hline
$m_\pi $& 85.5 & 141. & 338.&$138$~MeV \\
\hline
$\sqrt{\langle r^2_{E,I=0}\rangle }$
&0.89&0.85&0.69&? \\ \hline
\hline
\end{tabular}
\end{center}
\end{table}

\begin{table}
\caption{The predicted static $\Delta_{33}$-resonance observables for
different representations with fixed empirical values for the
$e=3.76$ and $f_\pi =68.8$~MeV (from nucleon observables $m_\pi =138$~MeV,
$\langle r^2_{E,I=0}\rangle ^{1/2}=0.72$~fm, representation $j=1$ in 
Table~\ref{nuc6}).}
\label{deltatwo}
\begin{center}
\begin{tabular}{|c|c|c|c|c|}
\hline \hline
$\mathbf{j}$ &$\scriptscriptstyle \mathbf{\frac 32 \oplus 1\oplus \frac
12}$ & $\mathbf{\frac 32}$ & $\mathbf{2}$  &\textbf{Expt.} \\
\hline
$m_\Delta $ & 1360. & 1338. & 1245. &$1232$~MeV \\ \hline
$\mu_{\Delta ^{++}}$ & 8.14 & 7.24 & 5.28 &$3.7-7.5$\\
\hline
$\mu_{\Delta ^{+}}$& 3.14 & 2.86 & 2.23 &?\\
\hline
$\mu_{\Delta ^{0}}$ & $-$1.87 &$-$1.52 & $-$0.83 &?\\
\hline
$\mu_{\Delta ^{-}}$& $-$6.87 & $-$5.89 & $-$3.88 & ? \\
\hline
$m_\pi $& 65.3 & 106.5 & 234.&$138$~MeV \\
\hline
$\sqrt{\langle r^2_{E,I=0}\rangle }$
&0.84&0.81&0.72&? \\ \hline
\hline
\end{tabular}
\end{center}
\end{table}

\begin{table}
\caption{The predicted static $\Delta_{33}$-resonance observables for
different representations with fixed empirical values for the
$e=4.05$ and $f_\pi =63.1$~MeV (from nucleon observables $m_\pi =138$~MeV,
$\langle r^2_{E,I=0}\rangle ^{1/2}=0.72$~fm, representation $j=1 \oplus 
\frac 12\oplus \frac 12 $ in Table~\ref{nuc6}).}
\label{deltathree}
\begin{center}
\begin{tabular}{|c|c|c|c|c|}
\hline \hline
$\mathbf{j}$ &$\scriptscriptstyle \mathbf{\frac 32 \oplus 1\oplus \frac
12}$ & $\mathbf{\frac 32}$ & $\mathbf{2}$  &\textbf{Expt.} \\
\hline$m_\Delta $ &
1162. & 1137. & 1020. &$1232$~MeV \\ \hline
$\mu_{\Delta ^{++}}$ & 7.25 & 6.32 & 4.27 &$3.7-7.5$\\
\hline
$\mu_{\Delta ^{+}}$& 2.92 & 2.65 & 1.97 &?\\
\hline
$\mu_{\Delta ^{0}}$ & $-$1.40 &$-$1.03 & $-$0.32 &?\\
\hline
$\mu_{\Delta ^{-}}$& $-$5.72 & $-$4.71 & $-$2.62 & ? \\
\hline
$m_\pi $& 78.5 & 129. & 297.&$138$~MeV \\
\hline
$\sqrt{\langle r^2_{E,I=0}\rangle }$
&0.86&0.82&0.70&? \\ \hline
\hline
\end{tabular}
\end{center}
\end{table}

\begin{table}
\caption{The predicted static $\Delta_{33}$-resonance observables for
different representations with fixed empirical values for the
$e=4.37$ and $f_\pi =61.4$~MeV (from nucleon observables $g_A =1.26$,
$\langle r^2_{E,I=0}\rangle ^{1/2}=0.72$~fm, representation $j=\frac 12
$ in Table~\ref{nuc2}).}
\label{deltafour}
\begin{center}
\begin{tabular}{|c|c|c|c|c|}
\hline \hline
$\mathbf{j}$ &$\scriptscriptstyle \mathbf{\frac 32 \oplus 1\oplus \frac
12}$ & $\mathbf{\frac 32}$ & $\mathbf{2}$  &\textbf{Expt.} \\
\hline$m_\Delta $ &
1053. & 1021. & 865. &$1232$~MeV \\ \hline
$\mu_{\Delta ^{++}}$ & 6.18 & 5.28 & 3.19 &$3.7-7.5$\\
\hline
$\mu_{\Delta ^{+}}$& 2.64 & 2.37 & 1.67 &?\\
\hline
$\mu_{\Delta ^{0}}$ & $-$0.90 &$-$0.54 & 0.14 &?\\
\hline
$\mu_{\Delta ^{-}}$& $-$4.44 & $-$3.45 & $-$1.39 & ? \\
\hline
$m_\pi $& 99.7 & 165. & 409.&$138$~MeV \\
\hline
$\sqrt{\langle r^2_{E,I=0}\rangle }$
&0.83&0.79&0.62&? \\ \hline
\hline
\end{tabular}
\end{center}
\end{table}

\begin{table}
\caption{The predicted static $\Delta_{33}$-resonance observables for
different representations with fixed empirical values for the
$e=4.13$ and $f_\pi =58.9$~MeV (from nucleon observables $g_A =1.26$,
$\langle r^2_{E,I=0}\rangle ^{1/2}=0.72$~fm, representation $j=1
$ in Table~\ref{nuc2}).}
\label{deltafive}
\begin{center}
\begin{tabular}{|c|c|c|c|c|}
\hline \hline
$\mathbf{j}$ &$\scriptscriptstyle \mathbf{\frac 32 \oplus 1\oplus \frac
12}$ & $\mathbf{\frac 32}$ & $\mathbf{2}$  &\textbf{Expt.} \\
\hline$m_\Delta $ &
1064. & 1038. & 920. &$1232$~MeV \\ \hline
$\mu_{\Delta ^{++}}$ & 7.40 & 6.42 & 4.23 &$3.7-7.5$\\
\hline
$\mu_{\Delta ^{+}}$& 3.03 & 2.73 & 2.01 &?\\
\hline
$\mu_{\Delta ^{0}}$ & $-$1.35 &$-$0.96 & $-$0.21 &?\\
\hline
$\mu_{\Delta ^{-}}$& $-$5.72 & $-$4.65 & $-$2.43 & ? \\
\hline
$m_\pi $& 78.4 & 129. & 302.&$138$~MeV \\
\hline
$\sqrt{\langle r^2_{E,I=0}\rangle }$
&0.91&0.87&0.72&? \\ \hline
\hline
\end{tabular}
\end{center}
\end{table}

\begin{table}
\caption{The predicted static $\Delta_{33}$-resonance observables for
different representations with fixed empirical values for the
$e=4.20$ and $f_\pi =59.7$~MeV (from nucleon observables $g_A =1.26$,
$\langle r^2_{E,I=0}\rangle ^{1/2}=0.72$~fm, representation $j=1 \oplus 
\frac 12\oplus \frac 12$ in Table~\protect\ref{nuc2}).}
\label{deltasix}
\begin{center}
\begin{tabular}{|c|c|c|c|c|}
\hline \hline
$\mathbf{j}$ &$\scriptscriptstyle \mathbf{\frac 32 \oplus 1\oplus \frac
12}$ & $\mathbf{\frac 32}$ & $\mathbf{2}$  &\textbf{Expt.} \\
\hline$m_\Delta $ &
1062. & 1035. & 907. &$1232$~MeV \\ \hline
$\mu_{\Delta ^{++}}$ & 7.01 & 6.06 & 3.90 &$3.7-7.5$\\
\hline
$\mu_{\Delta ^{+}}$& 2.90 & 2.61 & 1.90 &?\\
\hline
$\mu_{\Delta ^{0}}$ & $-$1.21 &$-$0.83 & $-$0.10 &?\\
\hline
$\mu_{\Delta ^{-}}$& $-$5.32 & $-$4.27 & $-$2.10 & ? \\
\hline
$m_\pi $& 84.2 & 139. & 330.&$138$~MeV \\
\hline
$\sqrt{\langle r^2_{E,I=0}\rangle }$
&0.88&0.84&0.69&? \\ \hline
\hline
\end{tabular}
\end{center}
\end{table}
\chapter{Baryon densities for $B>1$ configurations}
\label{app3}
\begin{figure}
\begin{center}
\includegraphics*[scale=0.61]{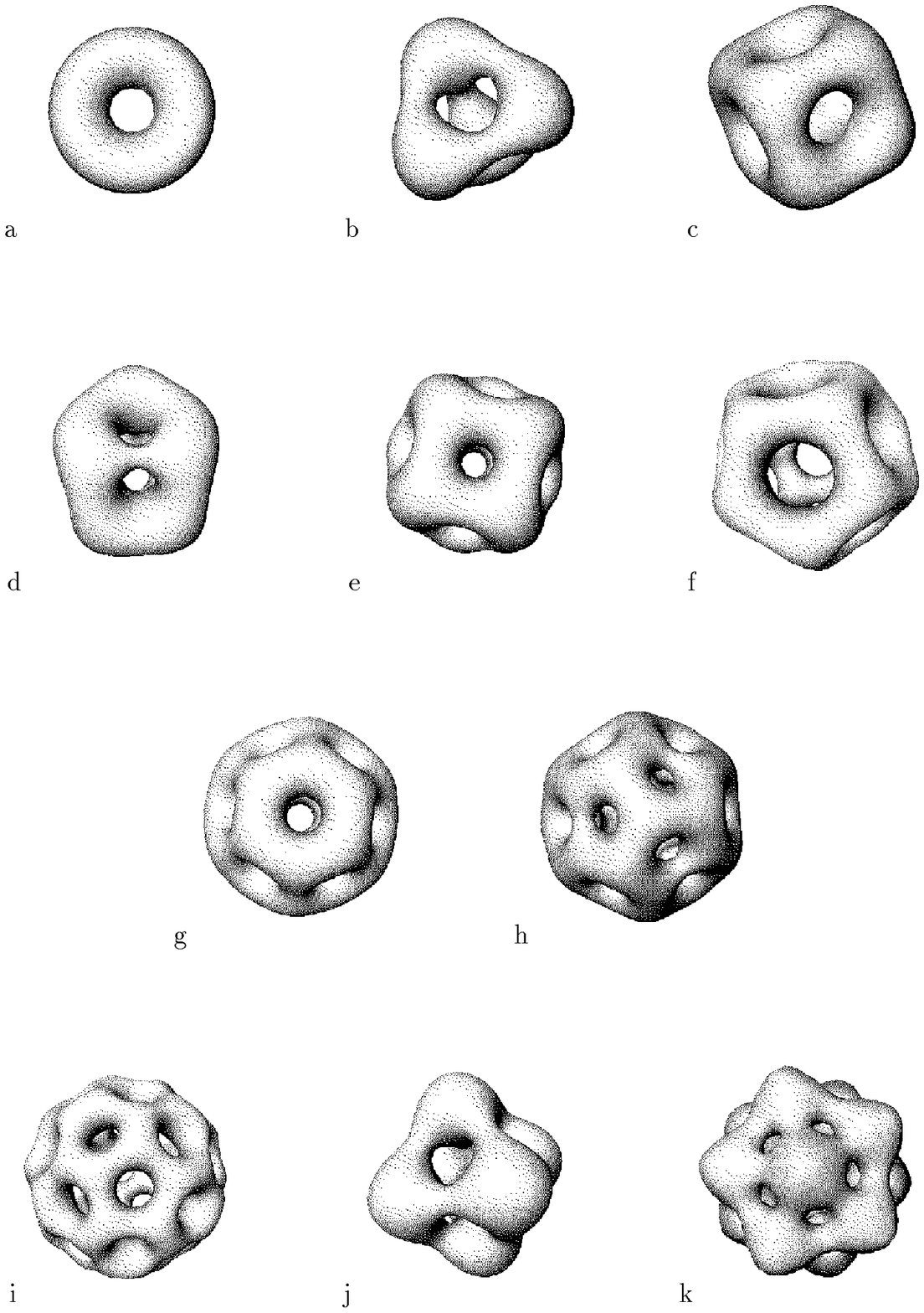}
\end{center}
\caption{Surfaces of constant baryon density for the following Skyrme 
fields~\cite{sutcliffe97b}:
a) $B=2$ torus 
b) $B=3$ tetrahedron
c) $B=4$ cube
d) $B=5$ with $D_{2d}$ symmetry
e) $B=6$ with $D_{4d}$ symmetry
f) $B=7$ dodecahedron
g) $B=8$  with $D_{6d}$ symmetry
h) $B=9$ with tetrahedral symmetry
i) $B=17$ buckyball
j) $B=5$ octahedron
k) $B=11$ icosahedron.}
\label{fig7}
\end{figure}
\penalty10000
\enlargethispage*{2cm}
%\end{document}

%\bibliography{skyrme}

\begin{thebibliography}{200}
\bibitem[1]{acus97}
{A.~Acus, E.~Norvai\v sas, and D.O.~Riska}.
\newblock {Stability and Representation Dependence of the Quantum Skyrmion}.
\newblock {\em {Phys.~Rev.}}, C57(5):2597--2604, 1998.

\bibitem[2]{acus96a}
{A.~Acus, E.~Norvai\v sas, and D.O.~Riska}.
\newblock {The Quantum Skyrmion in Representation of General Dimension}.
\newblock {\em Nucl. Phys. A}, 614:361--372, 1997.

\bibitem[3]{acus97a}
{A.~Acus and E.~Norvai\v sas}.
\newblock {Stability of SU(2) Quantum Skyrmion and Static Properties of
  Nucleons}.
\newblock {\em Lithuanian Journal of Physics}, 37(5):446--448, 1997.

\bibitem[4]{filipovbook}
{\cyrbib A.T.~Filippov}.
\newblock {\em {\cyritbib Mnogoliki\u i soliton}}, {\em {\cyrbib
Bibliotechka "Kvant" t.~48}}.
\newblock {\cyrbib Moskva, Nauka}, 1990.

\bibitem[5]{skyrme61a}
T.H.R. Skyrme.
\newblock A non-linear field theory.
\newblock {\em Proc.~Roy.~Soc.}, 260:127, 1961.

\bibitem[6]{skyrme62}
T.H.R. Skyrme.
\newblock A unified field theory of mesons and baryons.
\newblock {\em Nucl.~Phys.}, 31(4):556--569, 1962.

\bibitem[7]{rekalob}
{\cyrbib A.I.~Akhiezer, M.P. Rekalo}.
\newblock {\em {\cyritbib \`Elementarnye chasticy}}.
\newblock {\cyrbib Problemy Nauki i Tekhnicheskogo Progressa}. {\cyrbib Moskva,
  Nauka}, 1986.

\bibitem[8]{adkins83}
G.S.~Adkins, C.R.~Nappi, and E.~Witten .
\newblock {Static properties of nucleons in the Skyrme model}.
\newblock {\em Nuclear Physics}, B228:552--566, 1983.

\bibitem[9]{witten84}
{A.~Chodor, E.~Hadjimichael and C.~Tze}, editor.
\newblock {\em {Skyrmions and QCD; In:~Solitons in Nuclear and Elementary
  Particle Physics}}. World Scientific, 1984.
\newblock pp. 306-312.

\bibitem[10]{aitchison87}
{I.J.R.~Aitchison}.
\newblock {Effective Lagrangians and soliton physics I.~Derivative expansion,
  and decoupling}.
\newblock {\em Acta Physica Polonica}, B18(3):191--205, 1987.

\bibitem[11]{dorey95}
{N.~Dorey, M.P.~Mattis}.
\newblock {From Effective Lagrangians, to Chiral Bags, to Skyrmions with the
  Large-${N}_c$ Renormalization Group}.
\newblock {\em Phys. Rev.}, D52:2891--2914, 1995.

\bibitem[12]{mattisphd}
{M.P.~Mattis}.
\newblock {\em {Systematics of meson-skyrmion scattering}}.
\newblock PhD thesis, {Stanford Linear Accelerator Center, Stanford University,
  Stanford, California 94305}, 1986.

\bibitem[13]{mattis84}
{M.P.~Mattis and M.E.~Peskin}.
\newblock {Systematics of $\pi N$ Scattering in Chiral Soliton Models}.
\newblock {\em Phys.~Rev.}, D32(1):58--73, 1984.

\bibitem[14]{manton95}
{R.A.~Leese, N.S.~Manton, B.J.~Schroers}.
\newblock {Attractive Channel Skyrmions and the Deuteron}.
\newblock hep-ph/9502405.

\bibitem[15]{wambach97}
{Th.~Waindzoch and J.~Wambach}.
\newblock {Skyrmion dynamics on the unstable manifold and the nucleon-nucleon
  interaction}.
\newblock nucl-th/9705040.

\bibitem[16]{sutcliffe97c}
{R.A.~Battye and P.M.~Sutcliffe}.
\newblock {A Skyrme lattice with hexagonal symmetry}.
\newblock {\em Phys.~Lett.}, B416:385--391, 1998.
\newblock hep-th/9709221.

\bibitem[17]{carson91}
{L.~Carson}.
\newblock {B=3 nuclei as quantized multiskyrmions}.
\newblock {\em Phys.~Rev.~Lett.}, 66:1406, 1991.

\bibitem[18]{eisenberg93}
{G.~K\"albermann and J.M.~Eisenberg}.
\newblock {Three-nucleon interactions in the Skyrme model}.
\newblock {\em Phys.~Lett.}, B304:35--38, 1993.

\bibitem[19]{marleau90}
{L.~Marleau}.
\newblock {The Skyrme model and higher order terms}.
\newblock {\em {Phys.~Lett.}}, B235({1,2}):{141--146}, 1990.

\bibitem[20]{marleau92}
{L.~Marleau}.
\newblock {All-orders skyrmions}.
\newblock {\em {Phys.~Rev.}}, D45(5):1776--1781, 1992.

\bibitem[21]{walliser92}
{H.~Walliser}.
\newblock {The SU(n) Skyrme model}.
\newblock {\em Nucl.~Phys.}, A548:649--668, 1992.

\bibitem[22]{manton86}
{N.S.~Manton}.
\newblock {Skyrmions in flat space and curved space}.
\newblock KEK-86-7-463.

\bibitem[23]{sutcliffe97b}
{C.J.~Houghton, N.S.~Manton, and P.M.~Sutcliffe}.
\newblock Rational maps, monopoles and skyrmions.
\newblock {\em Nucl.~Phys.}, B510:507--537, 1998.
\newblock hep-th/9705151.

\bibitem[24]{mattis86}
{M.P.~Mattis}.
\newblock {Skyrmions and Vector Mesons}.
\newblock KEK~86-2-222.

\bibitem[25]{mattis88}
{M.~Mukerjee and M.P.~Mattis}.
\newblock {Skyrmions and Vector Mesons~II: Strange New Formulae}.
\newblock KEK~88-11-188.

\bibitem[26]{kaiser93}
{N.~Kaiser, Ulf-G.~Mei$\beta$ner}.
\newblock {On the axial charge in Skyrme models with vector mesons}.
\newblock {\em Phys.~Lett.}, B311:1--3, 1993.

\bibitem[27]{pari91}
{G.~Pari}.
\newblock {The role of the eta meson in the Callan-Klebanov approach to the
  Skyrme model}.
\newblock {\em Phys.~Lett.}, B261:347, 1991.

\bibitem[28]{rho90}
{M.~Rho, D.O.~Riska, and N.N.~Scoccola}.
\newblock {Charmed baryons as soliton-D meson bound states}.
\newblock {\em Phys.~Lett.}, B251:597--602, 1990.

\bibitem[29]{witten83}
{E.~Witten}.
\newblock {Global aspects of current algebra}.
\newblock {\em Nucl.~Phys.}, B223:422--432, 1983.

\bibitem[30]{rabinovici84}
{E.~Rabinovici, A.~Schwimmer, S.~Yankielowicz}.
\newblock {Quantization in the presence of Wess-Zumino terms}.
\newblock {\em {Nucl.~Phys.}}, B248:523--535, 1984.

\bibitem[31]{asano91}
{H.~Asano, H.~Kanada, and H.~So}.
\newblock {Quantization of the Non-linear Sigma Model and the Skyrme Model}.
\newblock {\em Phys.~Rev.}, D44:277--288, 1991.

\bibitem[32]{karliner86}
{M.~Karliner and M.P.~Mattis}.
\newblock {$\pi N,\ KN$, and $\bar KN$ scattering: Skyrme model versus
  experiment}.
\newblock {\em Phys.~Rev.}, D34(7):1991--2024, 1986.

\bibitem[33]{hayashi91}
{A.~Hayashi, S.~Saito, M.~Uehara}.
\newblock {Pion-nucleon scattering in the Skyrme model and the P-wave Born
  amplitudes}.
\newblock {\em Phys.~Rev.}, D43(5):1520--1531, 1991.

\bibitem[34]{braaten85}
{E.~Braaten, J.~P.~Ralston}.
\newblock {Limitations of semiclassical treatment of the Skyrme soliton}.
\newblock {\em Phys.~Rev.}, D31:598, 1985.

\bibitem[35]{hayashi92}
{A.~Hayashi, S.~Saito, M.~Uehara}.
\newblock {New formulation of pion-nucleon scattering and soft-pion theorems in
  the Skyrme model}.
\newblock {\em Phys.~Rev.}, D46(11):4856--4867, 1992.

\bibitem[36]{kostyuk95}
{\cyrbib A.~Kostyuk, A.~Kobushkin, N.~Chepilko, T.~Okazaki}.
\newblock {\cyrbib O korrektnosti perturbativnogo kvantovaniya
  dyxatel{\cprime}no\u i mody v modeli Skirma}.
\newblock {\em {\cyritbib Yadernaya Fizika}}, 58(8):1488--1491, 1995.

\bibitem[37]{sawada91}
{S.~Sawada and K.~Yang}.
\newblock {Stability of quantized chiral soliton with the Skyrme term}.
\newblock {\em Phys.~Rev.}, D44(5):1578--1584, 1991.

\bibitem[38]{weigel91}
{J.~Schechter and H.~Weigel}.
\newblock {Breathing mode in the SU(3) Skyrme model}.
\newblock {\em Phys.~Rev.}, D44(9):2916--2927, 1991.

\bibitem[39]{balakrishna92}
{B.S.~Balakrishna, V.~Sanyuk, J.~Schechter, and A. Subbaraman}.
\newblock {Cutoff quantization and the skyrmion}.
\newblock {\em Phys.~Rev.}, D45(1):344--351, 1992.

\bibitem[40]{dewittbook1}
{B.S.~DeWitt}.
\newblock {\em {Dynamical Theory of Groups and Fields}}.
\newblock Goron and Breach; New York, 1965.

\bibitem[41]{dewitt52}
{B.S.~DeWitt}.
\newblock {Point Transformations in Quantum Mechanics}.
\newblock {\em Phys.~Rev.}, 85(4):653--661, 1952.

\bibitem[42]{dewitt57}
B.S. DeWitt.
\newblock {Dynamical Theory in Curved Spaces. I. A Review of the Classical and
  Quantum Action Principles}.
\newblock {\em Rev.~Mod.~Phys.}, 29(3):377--397, 1957.

\bibitem[43]{cebula93}
{D.P.~Cebula, A.~Klein and N.R.~Walet}.
\newblock Quantization of the skyrmion.
\newblock {\em Phys.~Rev.}, D47(5):2113--2131, 1993.

\bibitem[44]{balachandranbook}
A.P.~Balachandran G.~Marmo B.S.~Skagerstam~A. Stern.
\newblock {\em Classical topology and quantum states}.
\newblock World Scientific, 1991.

\bibitem[45]{rho95}
{M.~Rho}.
\newblock {Compact star matter in chiral Lagrangians}.
\newblock {\em {Prog. Theor. Phys. Suppl.}}, 120:157--170, 1995.

\bibitem[46]{sondhi93}
{S.L.~Sondhi, A.~Karlhede, and S.A.~Kivelson}.
\newblock {Skyrmions and the crossover from the integer to fractional quantum
  Hall effect at small Zeeman energies}.
\newblock {\em Phys.~Rev.}, B47(24):16419--16426, 1993.

\bibitem[47]{fradkin88}
{E.~Fradkin and M.~Stone}.
\newblock {Topological terms in one- and two-dimensional quantum Heisenberg
  antiferromagnets}.
\newblock {\em Phys.~Rev.}, B38(10):7215--7218, 1988.

\bibitem[48]{fujii87}
{K.~Fujii, A.~Kobushkin, K.~Sato and N.~Toyota}.
\newblock {Skyrme-model Lagrangian in quantum mechanics: SU(2) case}.
\newblock {\em Phys. Rev.}, D35:1896--1907, 1987.

\bibitem[49]{norvaisas94}
{E.~Norvai\v sas, D.O.~Riska}.
\newblock {Representations of General Dimension for the Skyrme model}.
\newblock {\em Physics Scripta}, 50:634--638, 1994.

\bibitem[50]{mahankovbook}
{\cyrbib V.G.~Makhan\cprime kov, Yu.P.~Rybakov, V.I.~Sanyuk}.
\newblock {\cyritbib Model{\cprime } Skirma i Solitony v Fizike Adronov}.
\newblock {\cyrbib Dubna}, 1989.

\bibitem[51]{balachandran86}
{A.P.~Balachandran}.
\newblock {Skyrmions}.
\newblock In {M.J.~Bowick, F.~G\"ursey}, editor, {\em {High Energy Physics
  1985: vol.~I}}, pages 1--81. World Scientific, 1986.

\bibitem[52]{holzwarth86}
{G.~Holzwarth and B.~Schwesinger}.
\newblock {Baryons in the Skyrme model}.
\newblock {\em Rep.~Prog.~Phys.}, 49:825--871, 1986.

\bibitem[53]{zahed86}
{I.~Zahed and G.E.~Brown}.
\newblock {The Skyrme Model}.
\newblock {\em Physics Reports (Review Section of Physics Letters)},
  142(1{\&}2):1--102, 1986.

\bibitem[54]{rajaramanbook}
R.~Rajaraman.
\newblock {\em {Solitons and Instantons -- An Introduction to Solitons and
  {Instantons} in Quantum Field Theory}}.
\newblock North-Holland Personal Library, Amsterdam, 1982.

\bibitem[55]{rebbibook}
C.~Rebbi and G.~Soliani.
\newblock {\em Solitons and Particles}.
\newblock World Scientific, Singapore, 1984.

\bibitem[56]{fadeev76}
L.D. Faddeev.
\newblock {Some comments on the many-dimensional solitons}.
\newblock {\em Letters in Mathematical Physics}, 1:289--293, 1976.

\bibitem[57]{derrick64}
G.H. Derrick.
\newblock Comments on nonlinear wave equations on models of elementary
  particles.
\newblock {\em J.~Math.~Phys.}, 5:1252--1254, 1964.

\bibitem[58]{matuzeviciusbook}
A.~Matuzevi\v cius.
\newblock {\em Topologija}.
\newblock Mokslas, 1982.

\bibitem[59]{williams70}
J.G. Williams.
\newblock Topological analysis of a nonlinear field theory.
\newblock {\em J.~Math.~Phys.}, 11(8):2611--2615, 1970.

\bibitem[60]{shashinbook}
{\cyrbib Yu.A.~Shashkin}.
\newblock {\em {\cyritbib Nepodvizhnye tochki}}.
\newblock {\cyrbib Populyarnye lekcii po matematike}. {\cyrbib Moskva, Nauka},
  1989.

\bibitem[61]{belavin75}
A.A. Belavin and A.M. Polyakov.
\newblock {Metastable states of Two-Dimensional Isotropic Ferromagnets}.
\newblock {\em JETP Lett.}, 22:245--247, 1975.

\bibitem[62]{hawkingbook}
S.W. Hawking.
\newblock {\em {A Brief History of Time}}.
\newblock Bantam Books, 1990.

\bibitem[63]{meissner86}
{U.-G.~Meissner and I.~Zahed}.
\newblock {Skyrmions in the Presence of Vector Mesons}.
\newblock {\em {Phys.~Rev.~Lett.}}, 56(10):1035--1038, 1986.

\bibitem[64]{jain89}
{P.~Jain, J.~Schechter, and R.~Sorkin}.
\newblock {Quantum stabilization of the Skyrme soliton}.
\newblock {\em Phys.~Rev.}, D39(3):998--1001, 1989.

\bibitem[65]{kobushin95}
{\cyrbib A.P.~Kostyuk, A.P.~Kobushkin, N.M.~Chepilko, T.~Okazaki}.
\newblock {\cyrbib Kvantovye solitony neline\u ino\u i $\sigma$-modeli s
  narushenno\u i kiral{\cprime}no\u i simmetrie\u i}.
\newblock {\em {\cyritbib Yadernaya fizika}}, 58(8):1482--1487, 1995.

\bibitem[66]{bhaduri90}
{R.K.~Bhaduri and A.~Suzuki}.
\newblock {Quantum stabilization of the chiral soliton}.
\newblock {\em Phys.~Rev.}, D41(3):959--963, 1990.

\bibitem[67]{iwasaki89}
{M.~Iwasaki and H.~Ohyama}.
\newblock {Profile function of chiral quantum baryon}.
\newblock {\em Phys.~Rev.}, D40(9):3125--3126, 1989.

\bibitem[68]{rajaraman86}
{R.~Rajaraman, H.M.~Sommermann, J.~Wambach, and H.W.~Wyld}.
\newblock {Stability of the rotating skyrmion}.
\newblock {\em Phys.~Rev.}, D33(1):287--289, 1986.

\bibitem[69]{witten79}
E.~Witten.
\newblock {Baryons in the $1/N$ Expansion}.
\newblock {\em Nucl.~Phys.}, B160:57--115, 1979.

\bibitem[70]{thooft74}
{G.~'t~Hooft}.
\newblock {A Two-dimensional Model for Mesons}.
\newblock {\em Nucl.~Phys.}, B75:461--470, 1974.

\bibitem[71]{skyrme61b}
T.H.R. Skyrme.
\newblock Particle states of a quantized meson field.
\newblock {\em Proc.~Roy.~Soc.}, 262:237--245, 1961.

\bibitem[72]{skyrme58}
T.H.R. Skyrme.
\newblock {A non-linear theory of strong interactions}.
\newblock {\em Proc.~Roy.~Soc.}, 247:260--278, 1958.

\bibitem[73]{thooft76}
{G.~`t~Hooft}.
\newblock {Symmetry breaking through Bell-Jackiw anomalies}.
\newblock {\em Phys.~Rev.~Lett.}, 37:8, 1976.

\bibitem[74]{crewther77}
{R.J.~Crewther}.
\newblock {Chirality selection rules and the U(1) problem}.
\newblock {\em Phys.~Lett.}, 70B:349, 1977.

\bibitem[75]{ecker98}
{G.~Ecker}.
\newblock {Chiral Symmetry}.
\newblock hep-ph/9805500, 1998.

\bibitem[76]{adkins86}
{G.S.~Adkins}.
\newblock {Rho mesons in the Skyrme model}.
\newblock {\em Phys.~Rev.}, D33(1):193--197, 1986.

\bibitem[77]{weinberg97}
{S.~Weinberg}.
\newblock {Effective Field Theories in Large $N$ Limit}.
\newblock hep-th/9706047, 1997.

\bibitem[78]{mattis89}
{M.~Mattis and E.~Braaten}.
\newblock {Hadron scattering in the large-$N_c$ limit as a problem in linear
  algebra}.
\newblock {\em Phys.~Rev.}, D39(9):2737--2750, 1989.

\bibitem[79]{mattis89a}
{R.D.~Amado, M.~Oka, M.P.~Mattis}.
\newblock {$1/N_c$ corrections to $\pi$-nucleon scattering relations in chiral
  soliton models}.
\newblock {\em Phys.~Rev.}, D40(11):3622--3626, 1989.

\bibitem[80]{mattis90}
{P.B.~Arnold and M.P.~Mattis}.
\newblock {Summing Graphs in Large-$N_c$ Quantum Hadrodynamics}.
\newblock {\em Phys.~Rev.~Lett.}, 65(7):831--834, 1990.

\bibitem[81]{bidenkharnbook}
{L.C.~Biedenharn, J.D.~Louck}.
\newblock {\em {Angular Momentum in Quantum Physics}}, volume~8 of {\em
  Encyclopedia of Mathematics and its Applications}.
\newblock Addison-Wesley, 1981.

\bibitem[82]{varshalovichbook}
{\cyrbib D.A.~Varshalovich, A.H.~Moskal\"ev, V.K.~Khersonski\u i}.
\newblock {\em {\cyritbib Kvantovaya teoriya uglovogo momenta}}.
\newblock {\cyrbib Nauka }, 1975.

\bibitem[83]{donoghue84}
{J.F.~Donoghue, E.~Golowich, and B.R.~Holstein}.
\newblock {Predicting the proton mass from $\pi\pi$ scattering data}.
\newblock {\em Phys.~Rev.~Lett.}, 53:747, 1984.

\bibitem[84]{jain86}
{P.~Jain, R.~Johnson, and J.~Schechter}.
\newblock {Constraints on bag formation from the scalar sector}.
\newblock {\em Phys.~Rev.}, D35(7):2230--2237, 1986.

\bibitem[85]{koch97}
{V.~Koch}.
\newblock {Aspects of Chiral Symmetry}.
\newblock nucl-th/9706075, 1997.

\bibitem[86]{elliottbook1}
{J.P.~Elliott and P.G.~Dawber}.
\newblock {\em {Symmetry in Physics: principles and simple applications}},
  volume~1.
\newblock The Macmillan Press Ltd, London, 1979.

\bibitem[87]{sohnius85}
M.F. Sohnius.
\newblock Introducing supersymmetry.
\newblock {\em Phys.~Rep.}, 128:39, 1985.

\bibitem[88]{gieres98}
C.~Lucchesi F.~Gieres, M.~Kibler and O.~Piguet.
\newblock {\em Symmetries in Physics}.
\newblock Editions Fronti\`eres, 1998.
\newblock hep-th/9712154.

\bibitem[89]{kaulfuss85}
{U.B.~Kaulfuss and U-G. Meissner}.
\newblock {Deformation effects in the skyrmion-skyrmion interaction}.
\newblock {\em Phys.~Rev.}, D31(11):3024--3026, 1985.

\bibitem[90]{kopeliovic87}
{\cyrbib V.B.~Kopeliovich, B.E.~Shtern}.
\newblock {\cyrbib \`Ekzoticheskie solitony v modeli Skirma}.
\newblock {\em {\cyritbib Pis{\cprime}ma v Zh\`ETF}}, {\bf 45}(4):165--168, 
1987.

\bibitem[91]{verbaarshot}
J.J.M. Verbaarshot.
\newblock Axial symmetry of bound barion-number two soliton of the {S}kyrme
  model.
\newblock {\em Phys. Lett.}, B195(2):235--239, 1987.

\bibitem[92]{sutcliffe97a}
{R.A.~Battye and P.M.~Sutcliffe}.
\newblock Symmetric skyrmions.
\newblock {\em Phys.~Rev.~Lett.}, 79:363--367, 1997.
\newblock hep-th/9705151.

\bibitem[93]{walet91}
{N.R.~Walet, G.~Do~Dang, and A.~Klein}.
\newblock Theory of large-amplitude collective motion applied to the structure
  of ${}^{28}{Si}$.
\newblock {\em Phys.~Rev.}, C43:2254, 1991.

\bibitem[94]{sugano71}
R.~Sugano.
\newblock {On consistency between Lagrange and Hamilton formalisms in quantum
  mechanics}.
\newblock {\em Prog.~Theor.~Phys.}, 46(1):297--307, 1971.

\bibitem[95]{kimura71}
T.~Kimura and R.~Sugano.
\newblock {On consistency between Lagrangian and Hamilton formalisms in quantum
  mechanics. II}.
\newblock {\em Prog.~Theor.~Phys.}, 47(3):1004--1025, 1971.

\bibitem[96]{kimura72}
T.Ohtani T.~Kimura and R.~Sugano.
\newblock {On the consistency between Lagrangian and Hamiltonian formalisms in
  quantum machanics. III}.
\newblock {\em Prog.~Theor.~Phys.}, 48(4):1395--1407, 1972.

\bibitem[97]{kimura71a}
T.~Kimura.
\newblock {On the Quantization in Non-Linear Theories}.
\newblock {\em Prog.~Theor.~Phys.}, 46(4):1261--1277, 1971.

\bibitem[98]{sugano72}
{T.~Ohtani and R.~Sugano}.
\newblock {Variation Principle for Non-Linear Lagrangian in Quantum Mechanics}.
\newblock {\em Prog.~Theor.~Phys.}, 47(5):1704--1713, 1972.

\bibitem[99]{sugano73a}
R.~Sugano.
\newblock {Schwinger's Variation Principle by Means of Q-Number Variation for
  Non-Linear Lagrangian}.
\newblock {\em Prog.~Theor.~Phys.}, 49(4):1352--1361, 1973.

\bibitem[100]{dyson90}
H.P. Noyes.
\newblock {Comment on "Feynman's proof of the Maxwell equations" by
  F.J.~Dyson}.
\newblock {\em Am.~J.~Phys.}, 58:209--211, 1990.

\bibitem[101]{hojman91}
{S.A.~Hojman and L.C.~Shepley}.
\newblock {No Lagrangian? No quantization!}
\newblock {\em J.~Math. Phys.}, 32:142--146, 1991.

\bibitem[102]{klauder98}
J.R. Klauder.
\newblock Metrical quantization.
\newblock quant-ph/9804009, 1998.

\bibitem[103]{born26}
{M.~Born, W.~Heisenberg, and P.~Jordan}.
\newblock {Zur Quantenmechanik II}.
\newblock {\em Z. Phys.}, 35:557--615, 1926.

\bibitem[104]{dirac58}
P.A.M. Dirac.
\newblock {\em The Principles of Quantum Mechanics}.
\newblock Clarendon Press, Oxford, 4th edition, 1958.

\bibitem[105]{schrodinger26}
{E.~Schr\"odinger}.
\newblock {Quantisierung als Eigenwertproblem I}.
\newblock {\em Ann.~Phys.}, 79:361--376, 1926.

\bibitem[106]{feynman48}
R.P. Feynman.
\newblock Space-time approach to non-relativistic quantum mechanics.
\newblock {\em Rev.~Mod.~Phys.}, 20:367--387, 1948.

\bibitem[107]{stuckens86}
{C.~Stuckens and D.H.~Kobe}.
\newblock {Quantization of a particle with a force quadratic in the velocity}.
\newblock {\em Phys.~Rev.}, A34(5):3565--3567, 1986.

\bibitem[108]{crewther95}
{R.J.~Crewther}.
\newblock Introduction to quantum field theory.
\newblock In {\em Seventh Physics School --- Statistical Mechanics and Field
  Theory}. World Scientific, 1995.
\newblock hep-th/9505152.

\bibitem[109]{pauli33}
W.~Pauli.
\newblock {\em General principles of quantum mechanics}.
\newblock Springer, Berlin and Heidelberg, 1980.

\bibitem[110]{berezinbook}
{\cyrbib F.A.~Berezin}.
\newblock {\em {\cyritbib Metod vtorichnogo kvantovaniya}}.
\newblock {\cyrbib Moskva, Nauka}, 1986.

\bibitem[111]{tdleebook}
T.D. Lee.
\newblock {\em Particle physics and introduction to field theory}.
\newblock Harwood academic publishers, 1981.

\bibitem[112]{faddeev88}
{L.~Faddeev and R.~Jackiw}.
\newblock {Hamiltonian Reduction of Unconstrained and Constrained Systems}.
\newblock {\em Phys.~Rev.~Lett.}, 60(17):1692--1694, 1988.

\bibitem[113]{balachandran92}
{A.P.~Balachandran}.
\newblock {Gauge Symmetries, Topology and Quantisation}.
\newblock SU-4240-506, 1992.
\newblock {Lectures delivered in the Summer Course on "Low Dimensional Quantum
  Field Theories for Condensed Matter Physicists", International Centre for
  Theoretical Physics, Trieste, 24 August to 4 September, 1992}.

\bibitem[114]{diracbook}
P.A.M. Dirac.
\newblock {\em {Lectures on quantum mechanics}}.
\newblock {Yeshiva Univeristy, New York}, 1964.

\bibitem[115]{gitmanbook}
{\cyrbib D.M.~Gitman, I.V.~Tyutin}.
\newblock {\em {\cyritbib Kanonicheskoe kvantovanie pole\u i so svyazyami}}.
\newblock {\cyrbib Nauka}, 1986.

\bibitem[116]{witten83a}
{E.~Witten}.
\newblock {Current algebra, baryons, and quark confinement}.
\newblock {\em Nucl.~Phys.}, B223:433--444, 1983.

\bibitem[117]{bogolubov50}
{\cyrbib N.N.~Bogolyubov}.
\newblock {\cyrbib Ob odno\u i novo\u i forme adiabatichesko\u i teorii
  vozmushcheni\u i v zadache o vzaimode\u istvii chastic s kvantovym polem}.
\newblock {\em {\cyritbib Ukr.~Mat.~Zhurn.}}, 2(2):3--24, 1950.
\newblock {\cyrbib (Izbrannye trudy, t.2, Kiev;"Naukova dumka", 1970,
  499-520)}.

\bibitem[118]{lin70}
{H.E.~Lin, W.C.~Lin, and R.~Sugano}.
\newblock {On velocity-dependent potentials in quantum mechanics}.
\newblock {\em Nucl.~Phys.}, B(16):431--449, 1970.

\bibitem[119]{flandersbook}
H.~Flanders.
\newblock {\em {Differential Forms with Applications to Physical Sciences}}.
\newblock Dover Publications, Inc., New York, 1989.

\bibitem[120]{sugano73}
{T.~Ohtani and R.~Sugano}.
\newblock {Q-Number Variational Method for Non-Linear Lagrangian in Quantum
  Mechanics}.
\newblock {\em Prog.~Theor.~Phys.}, 50(5):1715--1728, 1973.

\bibitem[121]{kimura73}
T.~Kimura.
\newblock {Note on Quantum Form of Non-Linear Lagrangian}.
\newblock {\em Prog.~Theor.~Phys.}, 50:1769--1771, 1973.

\bibitem[122]{kiang69}
{D.~Kiang, K.~Nakazawa, and R.~Sugano}.
\newblock {Velocity-dependent potentials in Heisenberg picture}.
\newblock {\em Phys.~Rev.}, 181(4):1380--1382, 1969.

\bibitem[123]{jain90a}
{P.~Jain}.
\newblock {Static properties of the nucleon as a quantum stabilized soliton}.
\newblock {\em Phys.~Rev.}, D41(11):3527--3530, 1990.

\bibitem[124]{meissner87}
{U-G.~Meissner, N.~Kaiser and W.~Weise}.
\newblock {Nucleons as Skyrme solitons with vector mesons: electromagnetic and
  axial properties}.
\newblock {\em Nucl.~Phys.}, A466:685, 1987.

\bibitem[125]{riska90}
{E.M.~Nyman and D.O.~Riska}.
\newblock {Low-energy properties of baryons in the Skyrme model}.
\newblock {\em Rep.~Prog.~Phys.}, 53:1137--1181, 1990.

\bibitem[126]{ascher81}
{U.~Ascher, J.~Christiansen, and R.D.~Russell}.
\newblock {Collocation Software for Boundary-Value ODEs}.
\newblock {\em {ACM Trans.~Math.~Softw.}}, 7(2):209--222, 1981.

\bibitem[127]{pdg94}
{Particle Data Group}.
\newblock {\em Phys.~Rev.}, D(50):1173, 1994.

\bibitem[128]{bosshard91}
{A.~Bosshard, C.~Amsler, M.~D\"{o}beli, M.~Doser, M.~Schaad, J.~Riedlberger,
  P.~Truu\"{o}l, J.A.~Bistirlich, K.M.~Crowe, S.~Ljungfelt, C.A.~Meyer,
  B.~van~den~Brandt, J.A.~Konter, S.~Mango, D.~Renker, J.F.~Loude,
  J.P.~Perroud, R.P.~Haddock, and D.L.~Sober}.
\newblock {Analyzing power in pion-proton bremsstrahlung, and the
  $\Delta^{++}(1232)$ magnetic moment}.
\newblock {\em Phys.~Rev.}, D44(7):1962--1974, 1991.

\bibitem[129]{xenlibook}
{\cyrbib G.~Frau\`enfel{\cprime}der, \`E.~Khenli}.
\newblock {\em {\cyritbib Subatomnaya fizika}}.
\newblock {\cyrbib Mir, Moskva}, 1979.

\bibitem[130]{bander84}
{M.~Bander and F.~Hayot}.
\newblock {Instability of rotating chiral solitons}.
\newblock {\em Phys.~Rev.}, D30:1837, 1984.

\bibitem[131]{li87}
{B-A.~Li, K-F.~Liu, M-M.~Zhang}.
\newblock {Semiclassical skyrmion equation of motion}.
\newblock {\em Phys.~Rev.}, D35(5):1693--1697, 1987.

\bibitem[132]{adkins84}
{G.S.~Adkins and C.R.~Nappi}.
\newblock {The Skyrme model with pion masses}.
\newblock {\em Nucl.~Phys.}, B233:109, 1984.

\bibitem[133]{fujii92}
{K.~Fujii and N.~Ogawa}.
\newblock {Quantization of Chiral Solitons in Collective-Coordinate Approach}.
\newblock {\em Prog.~Theor.~Phys.~Suppl.}, (109):1--17, 1991.

\bibitem[134]{krupovnickas97}
{T.~Krupovnickas}.
\newblock {\it Solitoninio sprendinio ie\v skojimas remiantis supaprastintu 
Skyrme modeliu}.
\newblock {Vilniaus universitetas, teorin\.es fizikos katedra, kursinis
  darbas}, 1997.

\end{thebibliography}
%\bibliographystyle{amsplain}
%\bibliographystyle{plain}

\chapter*{Colophon}
The manuscript was prepared in \LaTeXe, a standard markup format, written 
by Leslie Lamport in \TeX\ typesetting language, created by 
{\it Donald E.~Knuth.}\/
The text is typeset in Computer Modern Roman font, combined with Computer 
Modern Bold Extended Roman and Computer Modern Italic. 
Mathematical formulas were typeset using symbols and macros provided by 
\verb|amsmath| package for \AmS-\LaTeX, distributed by 
{\it American Mathematical Society}\/ (AMS). Additional mathematical font 
shapes for Calligraphic, Doublestroke and Gotish  fonts are provided by 
the packages \verb|calrsfs| (AMS), \verb|dsfont| ({\it Olaf Kummer}) and 
\verb|goth| ({\it Yannis Haralambour}) respectively. 
Computer Modern font family was designed for 
use with \TeX\ by {\it D.E.~Knuth.} 

Feynman diagrams (Fig.~\ref{gluoncorrection} and Fig.~\ref{gluonvertex}) 
were drawn using \verb|FEYNMAN| package by {\it M.J.S.~Levine.}\/ Other
graphics in this manuscript were drawn in 
{\sc PostScript} graphic programming language of {\it Adobe Systems Inc.}\/ 
and included in the \TeX\ output using standard \LaTeXe\ `Graphics Bundle' 
package designed by {\it David Carlisle Sebastian Rahtz.}\/ 
The output was converted 
to {\sc PostScript} for printing using {\sc DviPS} driver 
program by {\it Tomas Rokicki.}

\end{document}